\documentclass{aa}
\usepackage{natbib}
\usepackage{amssymb,verbatim}
\usepackage{amsmath}
\usepackage{bigints}
\usepackage{graphicx}
\usepackage{appendix}
\usepackage{subfig}
\usepackage{color}
\usepackage{lineno}

\begin{document}

\title{Metal enrichment: the apex accretor perspective}
\author{S. Molendi\inst{1}, S. Ghizzardi\inst{1}, S. De Grandi\inst{2}, M. Balboni\inst{1,3}, I. Bartalucci\inst{1}, D. Eckert\inst{4,1}, F. Gastaldello\inst{1}, L. Lovisari\inst{1}, G. Riva\inst{1,5} and M. Rossetti\inst{1}} 

\offprints{S. Molendi \email{silvano.molendi@inaf.it}}

\institute{
INAF - IASF Milano, via A. Corti 12 I-20133 Milano, Italy 
\and
INAF - Osservatorio Astronomico di Brera, via E. Bianchi 46, I-23807 Merate (LC), Italy
\and
Dipartimento di Scienza e Alta Tecnologia, Universit\`a dell’Insubria, Via Valleggio 11, I-22100 Como, Italy
\and
Department of Astronomy, University of Geneva, ch. d'Ecogia 16, 1290 Versoix, Switzerland
\and
Dipartimento di Fisica, Universit\`a degli Studi di Milano, Via G. Celoria 16, I-20133 Milano, Italy
}
\date{\today}
\abstract
{}
{The goal of this work is to devise a description of the enrichment process in large-scale structure that explains the available observations and makes predictions for future measurements.}
{We took a spartan approach to this study, employing observational results and algebra to connect stellar assembly in star-forming halos with metal enrichment of the intra-cluster and group medium.} 
{On one hand, our construct is the first to provide an explanation for much of the phenomenology of metal enrichment in clusters and groups. It sheds light on the lack of redshift evolution in metal abundance, as well as the small scatter of metal abundance profiles, the entropy versus abundance anti-correlation found in cool core clusters, and the so-called Fe conundrum, along with several other aspects of cluster enrichment.
On the other hand, it also allows us to infer the properties of other constituents of large-scale structure. We find that
 gas that is not bound to halos must have a metal abundance similar to that of the ICM and only about one-seventh to one-third of the Fe in the Universe is locked in stars. A comparable amount is found in gas in groups and clusters and, lastly and most importantly, about  three-fifths of the total Fe is contained in a tenuous warm or hot gaseous medium in or between galaxies. 
 We point out that several of our results follow from two critical but well motivated assumptions: 1) the stellar mass in massive halos is currently underestimated and 2) the adopted Fe yield is only marginally consistent with predictions from synthesis models and SN rates. 
 }
{One of the most appealing features of the work presented here is that it provides an observationally grounded construct where vital questions on chemical enrichment in the large-scale structure can be addressed. 
We hope that it may serve as a useful baseline for future works.      
}

\titlerunning{Clusters as apex accretors}

\authorrunning{Molendi et al.}

\maketitle
\nolinenumbers
\section{Introduction}\label{sec:intro}

Over the last two decades, we have accumulated a wealth of observational constraints on the enrichment process in large-scale structure. We have measured Fe abundance profiles in the hot gas in clusters, the so called intracluster medium (ICM), and (to a lesser extent) in the hot gas in groups, namely the intragroup medium  \citep[IGrM,][and refs. therein]{Gasta:2021,Mernier:2022}. We have measured how the metal abundance in the ICM varies with cosmic time \citep[e.g.,][]{Ettori:2015,McDonald:2016,Liu:2020}. We have estimated abundance ratios between different elements for the core and circum regions of mostly relaxed low mass clusters \citep[e.g.,][]{Mernier:2017}.

The key question we address in this paper is whether we can come up with a description of enrichment in large-scale structure that explains the wealth of available observations and possibly makes predictions on future measurements.
In undertaking this study, we must begin by making a connection with the stellar assembly process.  
The metals we detect in the hot gas halos of massive systems have been produced in stars and it is only by linking 
star formation with enrichment that we are able to gain an understanding of the process.
The available measurements provide important constrains that can help us in making the connection. 
Metal abundances, in the outer regions\footnote{Within the scope of this paper, outer regions are defined by the relation $R \gtrsim 0.3 R_{500} $, where $R_{500}$ is the radius within which the mean density is 500 times the critical density.} of systems with halo masses of $M_{\rm h} \gtrsim 5 \times 10^{13}$ M$_\odot$, appears to be largely independent of mass, redshift, and radius (see \citealt{Gasta:2021,Mernier:2022}, for the group and cluster radial profiles, and \citealt{McDonald:2016} for the redshift dependence). Indeed, a case can be made for a "universal" abundance, with all systems investigated thus far, from local groups to distant clusters, which is consistent with featuring the same metal abundance in their outer regions. We note that no other ICM/IGrM observable displays a behavior that is anywhere as self-similar as the metal abundance.
Moreover, while self-similarity in the radial profiles of astrophysical quantities (e.g., pressure or entropy) requires the application of a renormalization process (referred to as scaling), no such operation is needed for the metal abundance.
 This is all the more surprising since self-similarity is the hallmark of scale-free gravitational processes \citep{Kaiser:1986,Evrard:1991} and the enrichment process is, by its very nature, non-gravitational. Thus, for metal abundance, self-similarity must somehow be achieved not in spite of non gravitational processes but because of them. 
In this work, we consider where this feedback driven self-similarity comes from and what mechanism produces constant and low scatter abundance profiles in clusters and groups. 

The observed self-similarity of metal abundance profiles suggests the processes at play can be modeled in simple terms.
Thus, we  address the issue of cluster enrichment with a spartan approach, using mostly observational results and algebra (we do indulge in the occasional bit of calculus here and there). More specifically, we make a connection between stellar assembly in dark matter (DM) halos and observed properties of the metal abundance of the hot gas in massive systems. 
In the age of peta byte simulations, such an approach may be viewed with skepticism. 
However, we begin by pointing out that metals are produced in stars, mostly supernovae (SNe), and, as such, they are the result of feedback processes occurring on scales that are many orders of magnitude smaller than those captured by simulations.
Typically, interactions occurring on these scales are not described in terms of elementary physical processes, they are introduced through semi-analytical recipes \cite[e.g.,][and refs therein]{Biffi:2018}. 
In light of these considerations, an approach such as the one followed here is highly complementary. 
One of the difficulties related to simulation-based studies  is that it can be very challenging to extricate results based on well understood physical laws embedded in the simulation from others that arise from sub grid recipes. This is particularly true when investigating metals whose synthesis occurs at subgrid scales. As can be easily understood, this is not the case for the approach we take here. The arguments we make and the equations we use here lead to predictions that will either be confirmed or disproved, leaving little doubt as to what works and what does not. In keeping with this approach, we also refrain from using simulation-based results to guide or justify our choices, in those instances where feedback plays a key role. 

To help readers navigate through the many ramifications of the paper, we provide a rather detailed description of its structure. We start off with a brief review  of the literature on baryon assembly and derive a simple description of star formation and enrichment from the point of view of massive systems, which we refer to as ``apex" accretors. We highlight how the bulk of the stars in these systems are synthesized in smaller halos which are later accreted onto the more massive ones (Sect. \ref{sec:ass}).  Our next step is to make a connection between star-forming halos and apex accretors (see Sect. \ref{sec:connect}). This allows us to perform an assessment of the efficiency with which stars produce metals
(see Sect. \ref{sec:coutskirts}).
By framing the similarity of metal abundance between  galaxy groups and clusters within an evolutionary scenario, we infer that the large reservoir of gas outside halos is not pristine, but enriched in metals to a degree similar to the one in massive halos (see Sect. \ref{sec:goutskirts}). 
In Sect. \ref{sec:scatter}, following a similar approach, we show how the small scatter in metal abundance in clusters originates from the large ratio in mass between accretor and accreted. In Sect. \ref{sec:revol}, by noting the concomitant lack of redshift evolution  in metal abundance and stellar fraction, in groups and clusters, from $z \sim 1.3$, we expose the tight connection between the stellar assembly and enrichment processes across cosmic time. In Sect. \ref{sec:antico}, we propose an explanation for the entropy versus abundance anti-correlation found in cool core clusters. In Sect. \ref{sec:stratif}, we  further explore the connection between chemo and thermo-dynamic properties of groups and clusters. In Sect. \ref{sec:ratios}, we propose an explanation for the lack of gradients in abundance ratios observed in cluster radial profiles.
In Sect. \ref{sec:missing}, we provide a solution to the long standing "missing stellar mass" problem in clusters; namely, that the measured Fe mass is significantly larger than the stellar mass required to synthesize it. 
In Sect. \ref{sec:budget}, we take advantage of the census of metals we have made in massive systems to present a metal budget for the Universe. 
In Sect. \ref{sec:prospects}, we discuss possible developments, emphasizing the role played by future high resolution and wide field-of-view experiments, such as those on XRISM \citep{Tashiro_XRISM:2018} and ATHENA \citep{Nandra_Athena:2013}. Finally, in Sect. \ref{sec:summary}, we provide a  summary of our main findings.

Throughout the paper, we assume a $\Lambda$ cold dark matter cosmology with $H_0 = $ 70 km s$^{-1}$ Mpc$^{-1}$, $\Omega_M = 0.3,$ and $\Omega_\Lambda = 0.7$.  We also adopt solar abundances from \cite{Asplund:2009}. Across the scope of this work, the terms "metal abundance" and "Fe abundance" are to be considered interchangeable (unless otherwise stated).

\section{Baryon assembly}\label{sec:ass}
 Here, we provide a brief review of the literature, our goal is to motivate the simplified enrichment model discussed at the end of the section. Since metals are produced in stars, we shall start by looking at the stellar mass function and how it connects to the dark matter dominated halo assembly process. We then move onto the issue of metal production and dispersal. Finally, we make use of the material summarized in previous subsections to construct a description of stellar formation and metal enrichment processes from the point of view of massive systems.

\subsection{Stellar mass function }

The galaxy stellar mass function (GSMF), as measured by many observers (see \citealt{Weaver:2022} for a recent example and \citealt{Behroozi:2019} for a compilation), features a characteristic stellar mass of $ \sim 10^{11} $M$_\odot$. Galaxies appear to have considerable difficulty in growing beyond it and, quite importantly, this finding seems to be independent of redshift  up to $z\sim 2$.
The break in the GSMF is associated with quenching. It is expected to begin at an earlier time in more massive systems \citep{Dekel:2006}. For instance, \citet[][see their Fig.13]{Behroozi:2019}, have found that star formation is largely stopped at $z\sim 2$ for massive halos, $M_{\rm h}>10^{13}$ M$_\odot$, and at $z\sim 0.5$ for $M_{\rm h} > 10^{12}$M$_\odot$.

\subsection{Linking star formation to halo assembly}\label{sec:sm-hm}
Over the last decade, considerable progress has been made in linking observations of galaxy stellar mass and star formation rates to dark matter (DM) halos across cosmic history.
A rich and diverse data set, combined with simulation results, have been used to provide a comprehensive description of how stellar mass, $M_\star$, is assembled on different scales and at different times \citep[e.g.,][]{Leauthaud:2012,Coupon:2012,Behroozi:2013,Moster:2013,Coupon:2015,Cowley:2018,Behroozi:2019,Legrand:2019,Girelli:2020,Shuntov:2022}.  
Here, we briefly review some of the most salient features, keeping in mind that our focus is on massive systems, $M_{\rm h}>10^{13}$M$_\odot$, and late times, $z<0.1$.

\begin{figure}
        \centerline{\includegraphics[angle=0,width=8.8cm]{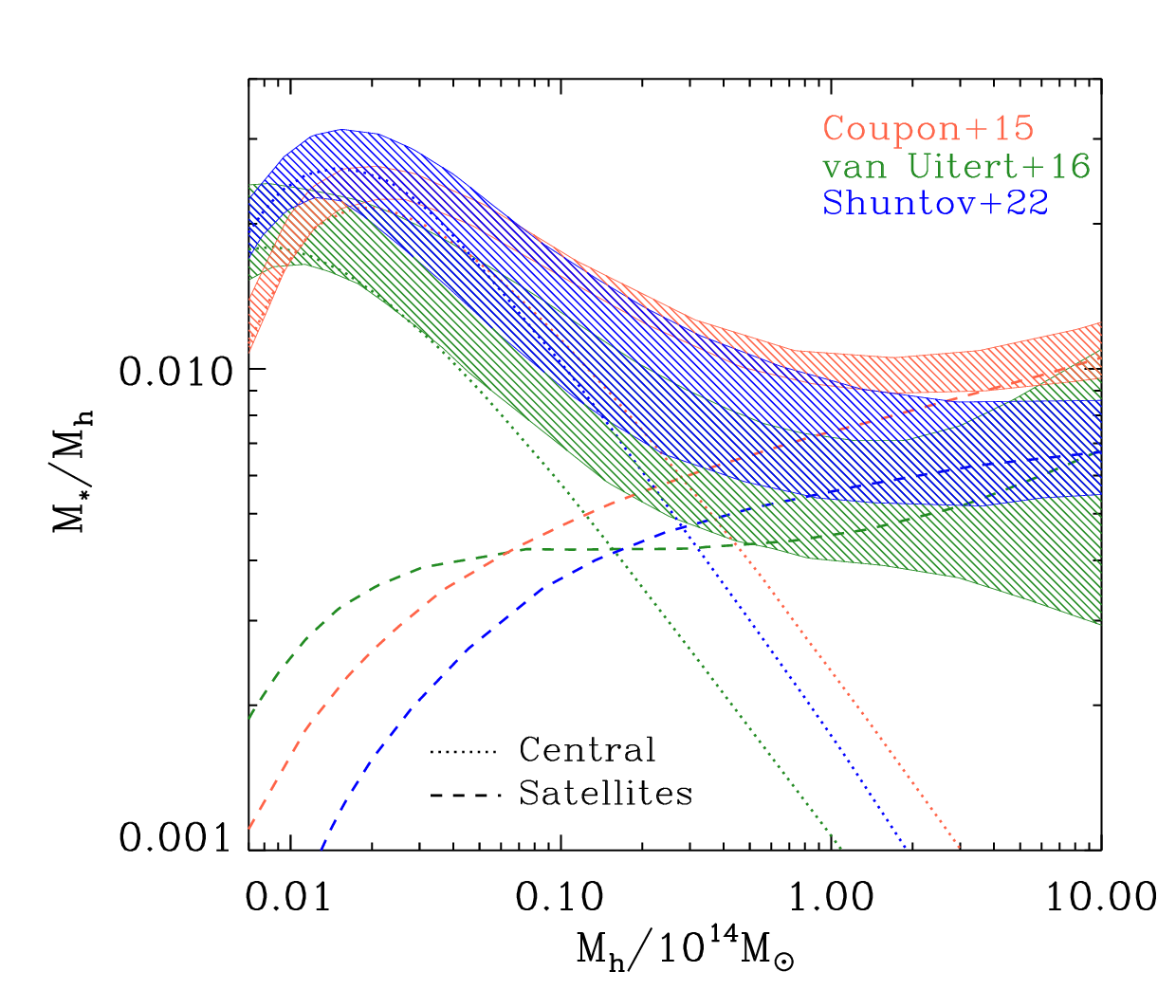}}
        \caption{Stellar mass over halo mass as a function of halo mass. Shaded regions indicate total stellar fractions, dotted and dashed lines central and satellite contributions respectively. Results from \cite{Coupon:2015},
                \cite{vanUitert:2016} and \cite{Shuntov:2022}  are reported in red, green and blue respectively. 
        }
        \label{fig:shmr}
\end{figure}

Matching DM-dominated halos with galaxies has been achieved by different methods: abundance matching \citep{Behroozi:2010}, halo occupation density \citep{Leauthaud:2012}, and empirical matching  \citep{Behroozi:2019}. Abundance matching performs a match of observed galaxies with simulated DM halos, the latter two making use of auto and cross-correlation functions.
All these methods broadly converge on a stellar-to-halo mass relation (SHMR) characterized by a peak at $M_{\rm h} \sim 10^{12} $M$_\odot$, with  stellar mass over halo mass, $M_\star/M_{\rm h}$, decreasing both at smaller and larger halo masses.   This scenario appears to change only moderately with cosmic time, for $z<4$ \citep[e.g.,][]{Leauthaud:2012,Coupon:2012,Behroozi:2013,Moster:2013,Coupon:2015,Cowley:2018,Behroozi:2019,Legrand:2019,Girelli:2020,Shuntov:2022}.
In Fig. \ref{fig:shmr}, we show the stellar-to-halo mass relation from three different low-redshift samples. In all cases, we see the stellar mass over halo mass, $M_\star/M_{\rm h}$, peaks at $ M_{\rm h} \sim 10^{12} $M$_\odot$ and declines both at lower and higher halo masses.

From the SHMR, \citet{Behroozi:2013,Behroozi:2019} and \citet{Shuntov:2022}, under the reasonable assumption that baryonic mass accretion rate is proportional to the DM accretion rate \citep[see][]{vandeVoort:2011,Wright:2020,Mitchell:2022}, found that star formation efficiency (SFE), defined as the ratio of star formation rate to baryon accretion rate, depends only weakly on cosmic time for $z<4$. In other words the bulk of star formation occurs in a narrow halo mass range  around $M_{\rm h} \sim 10^{12}$ M$_\odot$ and this does not change much with cosmic time.

An important distinction ought to be made between in situ stellar assembly, namely, stars that are produced within the halo, and ex situ assembly, namely, stars that are produced in another halo and are later accreted onto the massive halo under consideration. 
Generally, in situ assembly operates at early times in central galaxies, while ex situ assembly is associated with the late times infall of satellite galaxies. We note that infall may lead to some in situ star formation, therefore, not all stellar mass in satellites is associated with ex situ assembly. 
Since,  for $ M_{\rm h} > $ a few $ 10^{12}$M$_\odot$, star formation is largely quenched, systems that evolve well beyond such a mass (i.e., groups and clusters of galaxies) increase their stellar mass by accreting it as they accrete other types of matter (baryonic or otherwise). As shown in Fig. \ref{fig:shmr},  \citet{Coupon:2015}, \citet{vanUitert:2016}, and \citet{Shuntov:2022} all found that for $M_{\rm h} \gtrsim 3\times 10^{13}$ M$_\odot$, the accreted stellar mass becomes dominant over the one synthesized within the halo itself \citep[see also Fig.27 in][]{Behroozi:2019}.
On the scale of massive clusters ($M_{\rm h} \sim 10^{15}$ M$_\odot$), the stellar mass synthesized  within the halo is no more than a few percent of the accreted one (again, see Fig. \ref{fig:shmr}).

\subsection{Metal production and dispersal}\label{sec:metal}

All elements heavier than H,  excluding  He and (in part) Li, are produced in stars, with the bulk coming from supernovae explosions. We refer to \citet{Hoyle:1946} for a seminal work and \citet{Rauscher:2011} for a more recent review.
Essentially all core collapse supernovae (SNcc) explode within a few tens of Myr of their formation (and therefore closely track the star formation process). Moreover, about half type Ia supernovae (SNIa) explode within less than 1 Gyr \citep{Maoz:2017,Freundlich:2021}. Given the relative contribution of SNIa and SNcc to Fe production, it can be shown that  more than 90\% of the Fe is produced within $\sim$ 3 Gyr 
\footnote{In this paper we do not consider enrichment from an early stellar population,
characterized by a significantly different initial mass function (IMF), so called Pop III stars. The possible role of this population in the enrichment of the ICM has been explored by a few authors \cite{Bregman:2010,Lowenstein13,Blackwell:2022}.}.

A detailed characterization of the enrichment process of the gas in star forming halos cannot be achieved through observations; indeed, the gas distributed outside the galaxy, but within the halo, the so called circumgalactic medium \citep[CGM, see][for a recent review]{Tumlinson:2017}
is very hard to detect, let alone characterize \citep[e.g.,][]{Comparat:2022}. We consider what we can assume on the basis of our relatively scarce knowledge and elementary considerations.
First, we know that the same process that injects metals in the CGM also injects energy, which eventually leads to the ejection of a part of the CGM from the DM halo. We prudently assume that the process of mixing and ejection each operate on its own timescale. 
If the mixing timescale is much shorter than the ejection timescale, the gas ejected from the CGM will have the same metallicity of the one that is not ejected.
If the mixing timescale is much longer than the ejection timescale, the ejected gas will have a much lower metallicity of the one that is retained.
These two extreme scenarios lead to significantly different enrichment scenarios. 
In the first case, metals are shared equally between the two components, whereas in the second case, they are retained by the non-ejected gas.
Of course, if the timescales are comparable, we will end up with an intermediate solution where the metal abundance of the ejected gas will be smaller by some multiplicative factor of that of the non ejected one.
For the time being, we leave this unknown factor as a free parameter in our model and we  return to it in Sect. \ref{sec:goutskirts}.

\subsection{Stellar mass assembly \& enrichment from an apex accretor perspective}\label{sec:model}

In this section, we provide a description of the stellar mass assembly and enrichment process from the point of view of systems that are positioned at the vertex of the accretion chain.  We often refer to them as  ``apex accretors", although they are more widely known as clusters of galaxies. 

The findings summarized in the previous subsections suggest that a relatively simple description can be attempted. 
We can envisage two major modes or phases of stellar mass assembly in groups and clusters:
assembly from star formation within the forming halo (in situ) and from galaxies outside the halo (ex situ). 
The first mode and phase (in situ) dominates at early times when the core of the structure is being assembled. 
The second, at later times, when the halo grows by accreting less massive systems (ex situ). This phase occurs when star formation within the progenitor halo has been mostly quenched 
and stellar assembly is associated to the infall of galaxies onto the halo.
We note that some early mode star formation will continue around the central galaxy.
As we follow the structure formation process up in mass, most of the halos are accreted. 
Clusters act as apex accretors, that is, they accrete without being accreted, with only a small fraction of stars assembled within the progenitor halo and the bulk accreted from other halos \citep[e.g.,][]{Coupon:2015,Behroozi:2019,Shuntov:2022}.

Halos that evolve well beyond $10^{12}$M$_\odot$, experience a decrease in star formation efficiency. For large halo masses, synthesis of new stars (in situ star formation) rapidly falls off and the total (in situ plus ex situ) SHMR flattens out (see Fig. \ref{fig:shmr}). This can be understood if we consider that at the high-mass end,  the stellar
assembly process is nothing more than a transfer of stellar mass from smaller to larger halos with no further synthesis.  
We note that while there is a broad agreement on the shape of the stellar fraction, $M_*/M_{\rm h}$, as a function of halo mass, $M_{\rm h}$, there are also differences; for example, the  reduction in $M_*/M_{\rm h}$ with increasing halo mass appears to be significantly larger in \cite{Shuntov:2022} than in \cite{Coupon:2015}. Details can also be seen in Fig. \ref{fig:shmr}.

The gas enrichment process can be described through two major modes and phases echoing star formation.
In the first  mode or phase (in situ), metals synthesized within stars are expelled via feedback mechanism and mixed 
in with gas bound to the halo.
In the second mode or phase (ex situ), accreted sub-halos donate their pre-enriched gas. We note that some in situ enrichment is likely to continue around the central galaxy well into times dominated by ex situ  formation.

For halos that evolve well beyond $10^{12}$M$_\odot$, the decrease in star formation efficiency will necessarily result in a decrease in overall metallicity up to a few $10^{13}$M$_\odot$\footnote{We use the term "overall" to denote the mean abundance  over all baryons irrespective of their physical state.}.  For larger halo masses, the synthesis of new stars (in situ star formation) rapidly falls off (see Fig. \ref{fig:shmr}), as does the fraction of accreted gas not associated to halos  \citep[see][and refs. therein]{Eckert:2021}. 

The two modes or phases of enrichment, as we have identified them, differ in some crucial aspects.
In the first mode or phase, the gas that is expelled by galaxies finds itself inside the forming halo and does not experience an accretion shock. Its entropy is raised only through feedback mechanisms.
In the second mode or phase, the gas finds itself outside the forming halo and experiences an accretion shock when it eventually falls in the halo. Its entropy is raised by feedback mechanisms and gravitational heating, with the latter providing the dominant contribution for sufficiently massive halos ($M_{\rm h} > $ a few $ 10^{13}$ M$_\odot$). 
This has important consequences: shock heating of the gas in the second mode guarantees that essentially all of this gas, whatever its original physical state, ends up in the hot phase, be it the IGrM or the ICM.

\section{Connecting star forming halos to apex accretors}\label{sec:connect}

Metals are mostly produced in stars in halos with $M_h \sim 10^{12}$M$_\odot$; conversely, gas abundances  
are measured in halos of $M_h \sim 10^{15}$M$_\odot$, which are sufficiently massive to feature a baryon fraction that is close to the cosmic one \citep[see][and refs. therein]{Eckert:2021}.
In this section, we make a connection between these two mass scales.

\subsection{Massive halos}\label{sec:coutskirts}

 We start by connecting the  metal abundance measurements of the ICM of massive systems, as reported in \citet{Ghizzardi:2021},  
with  the stellar assembly and enrichment scenario that we sketch in the previous section.
As a first step, we derived a prediction for the metal abundance and compared it with the measured value. To this end, 
we made use of the Fe yield, $\mathcal{Y}_{\rm Fe}$, introduced in \cite{Greggio:2011} and \cite{RA14}.
It is defined as the total Fe mass, which, for massive systems, is the sum of the iron mass locked in stars, $M^{\star}_{\rm Fe}$,  and the iron mass in the ICM, $M^{\rm ICM}_{\rm Fe}$, divided by the stellar mass, $M_{\star}(0)$, that produced the iron :

\begin{equation}
        \mathcal{Y}_{\rm Fe} = { {M^{\star}_{\rm Fe} + M^{\rm ICM}_{\rm Fe}} \over  M_{\star}(0)} .
        \label{eq:y}
\end{equation}

Since stars suffer significant mass loss,  $M_{\star}(0)$ is related to the present mass in stars, $M_{\star}$, via the relation $M_{\star}(0) = r_o M_{\star}$, where $r_o$ is the return factor, see \citet{RA14} and refs. therein for further details.
We can rewrite Eq. \ref{eq:y} in a slightly different form:
        
\begin{equation}
        \mathcal{Y}_{{\rm Fe}} = {Z^{\star}_{{\rm Fe}}  \over r_{o}} \Biggl( 1 + {  Z^{\rm ICM}_{{\rm Fe}} \over Z^{\star}_{{\rm Fe}}} {M_{{\rm ICM}} \over  M_{\star}} \Biggr) ,
        \label{eq:y2}
\end{equation}
where we have expressed Fe masses as Fe abundances times ICM or stellar masses: 
$M^{\star}_{{\rm Fe}} =  Z^{\star}_{{\rm Fe}} M_{\star}$, 
$M^{\rm ICM}_{{\rm Fe}} =  Z^{\rm ICM}_{{\rm Fe}} M_{{\rm ICM}}  $,
note that $Z^{\star}_{{\rm Fe}}$ and $Z^{\rm ICM}_{{\rm Fe}}$ are respectively the mean stellar and ICM Fe abundances for the halo and  $M_{{\rm ICM}}$ refers to the total gas mass bound to the halo. 

 Next we solve the equation for $Z^{\rm ICM}_{{\rm Fe}}$ and rewrite $M_{\rm ICM}$ as $M_b - M_{\star}$, where $M_b$ is the total baryon mass\footnote{We are assuming that the mass lost through feedback effects can be neglected at the massive cluster scale, we revisit this assumption in the next section when we move to less massive systems.}, as follows:

\begin{equation}
        Z^{\rm ICM}_{{\rm Fe}} =   \Big(   r_{o} \mathcal{Y}_{{\rm Fe}}  - Z^{\star}_{{\rm Fe}} \Big)  {M_{\star} \over  M_{\rm b}} \, {1\over 1 - {M_{{\star}} \over  M_{\rm b}}} \, .
        \label{eq:zgas}
\end{equation}
Finally, we express abundances and yield in solar units and rewrite $M_{\star} /  M_{\rm b}$ as $f_{\star} /  f_{\rm b}$, where $f_{\star}$ and $f_{\rm b}$, defined as $f_{\star} \equiv  M_{\star} / M_{\rm h}$ and  $f_{b} \equiv  M_{\rm b}  /  M_{\rm h}$, are, respectively, the stellar and baryon fraction and $ M_{\rm h}$ is the halo mass:

\begin{equation}
        Z^{\rm ICM}_{{\rm Fe},\odot} =  \Big(   r_{o} \mathcal{Y}_{{\rm Fe},\odot} - Z^{\star}_{{\rm Fe},\odot}  \Big)  {f_{\star} \over  f_{\rm b}} \, {1\over 1 - {f_{{\star}} \over  f_{\rm b}}} \, .
        \label{eq:zgas2}
\end{equation}
To gain some insight into this expression, we can think of  $r_{o} \mathcal{Y}_{{\rm Fe}}  - Z^{\star}_{{\rm Fe}}$ as a stellar fraction to ICM metal abundance conversion factor.

\begin{figure}
        \centerline{\includegraphics[angle=0,width=8.8cm]{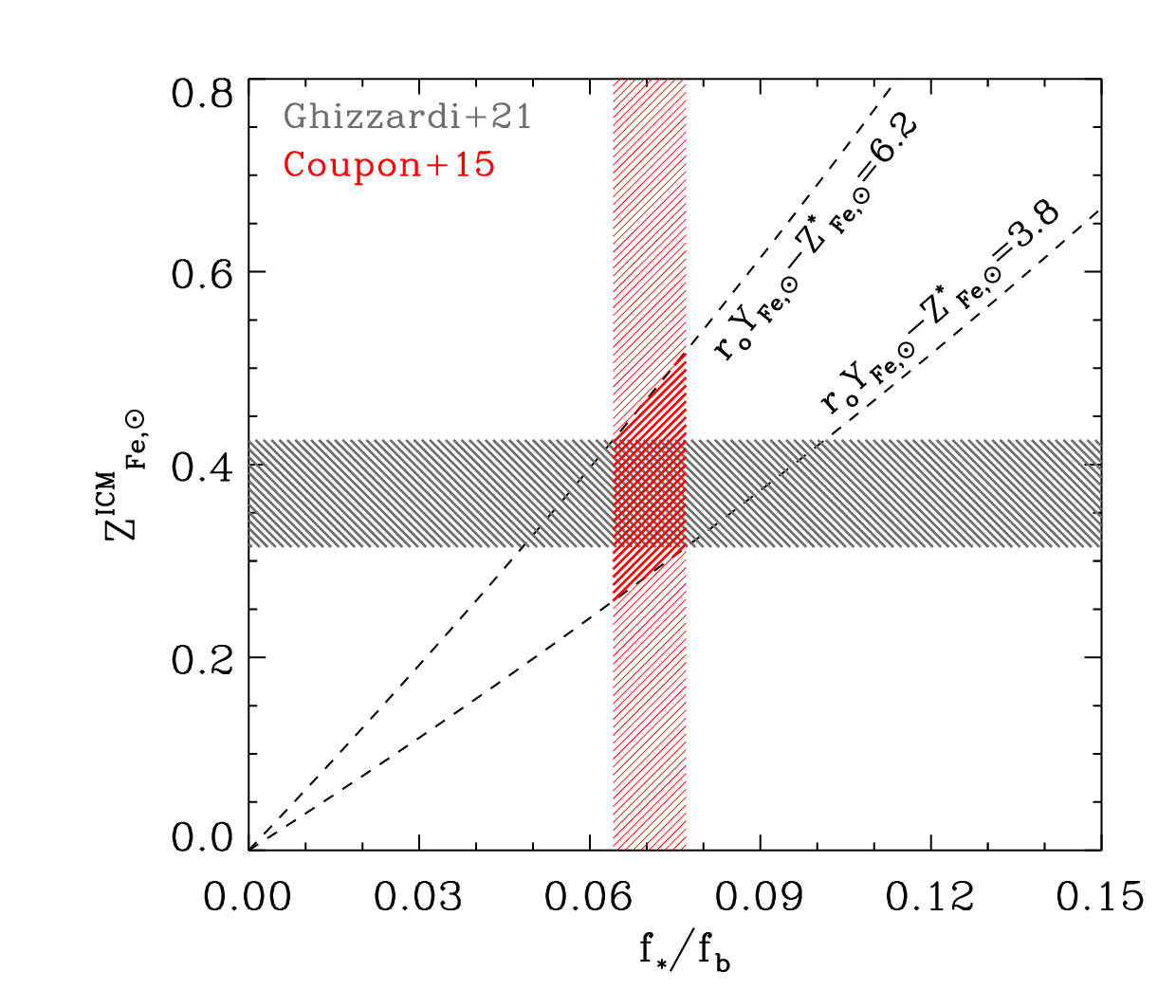}}
        \caption{ICM Fe abundance as a function of stellar over baryon fraction.
                The red shaded region show ${f_{\star} / f_{\rm b}}$ values for massive systems as reported in \cite{Coupon:2015}. The gray shaded  region represents the Fe abundance measurements for massive clusters \citep{Ghizzardi:2021}.
                The dashed lines trace the minimum and maximum values of $r_{o} \mathcal{Y}_{{\rm Fe},\odot} - Z^{\star}_{{\rm Fe},\odot }$ for which the ICM abundance, predicted from ${f_{\star} / f_{\rm b}}$, through Eq. \ref{eq:zgas2},  agree with the measured one. A deeper red is adopted to highlight the concordance region between predicted and measured quantities.
        }
        \label{fig:z_fe_vs_ms_mb}
\end{figure}

We  estimate the ICM Fe abundance from Eq. \ref{eq:zgas2} by taking values of $f_\star$ at the high mass end ($M_{\rm h} \sim 10^{15}$M$_\odot$) from one of the works reported in Fig. \ref{fig:shmr}, namely \cite{Coupon:2015}\footnote{One could alternatively measure $f_\star$ directly from a sample of clusters. We have pursued this approach in a previous paper \citep{Ghizzardi:2021}, and return to it in Sect.\ref{sec:missing}}.
For the baryon fraction, we assume, $f_{\rm b}= 0.16$  \citep[see][]{Coupon:2015,Eckert_non_th_XCOP:2019,Shuntov:2022}. Errors on $f_{\rm b}$ are neglected as it always appears in combination with $f_{{\star}}$, which is characterized by much larger uncertainties. 
The predicted ICM Fe abundance is compared to the one measured in \cite{Ghizzardi:2021}, see their Sect. 3.5, more specifically we make use of the mass weighted Fe abundance within $R_{500}$ averaged over the full sample and the associated scatter\footnote{Note that the value reported here has been converted from the \cite{AG:1989} system used in \cite{Ghizzardi:2021} to the \cite{Asplund:2009} system adopted here.}. 
In Fig. \ref{fig:z_fe_vs_ms_mb}, we provide a graphical representation of the comparison, as we can see, by imposing that the predicted Fe abundance match the observed one, we restrict the $r_{o} \mathcal{Y}_{{\rm Fe},\odot} - Z^{\star}_{{\rm Fe},\odot }$ factor in the range of $3.8 <r_{o} \mathcal{Y}_{{\rm Fe},\odot} - Z^{\star}_{{\rm Fe},\odot } < 6.2$.

It is enlightening to compare our estimate of the  $r_{o} \mathcal{Y}_{{\rm Fe},\odot} - Z^{\star}_{{\rm Fe},\odot }$ factor with ones derived from stellar synthesis models and SN rates. Without going into too much detail,  $\mathcal{Y}_{{\rm Fe},\odot}$ is computed as the product of the Fe mass produced per SN explosion and the number of SN events per unit mass of gas turned into stars, both contributions from Ia and  CC SN are considered as they are of the same order. 
Following \cite{Ghizzardi:2021b}, who made use of work by \cite{RA14} and \cite{Freundlich:2021}, we estimated $ \mathcal{Y}_{{\rm Fe},\odot}< 3.0$. We assumed $1/r_{o}$ to be between 0.58 and 0.70, where the former and latter values have been obtained assuming a top heavy and a Salpeter initial mass function (IMF) respectively (see \citealt{Maraston:2005} and \citealt{RA14} for details). For the stellar abundance, we assumed $ Z^{\star}_{{\rm Fe},\odot} = 1.2\pm 0.1$  \citep[see][]{Gallazzi:2014,Zahid:2017,Saracco:2023}.
From these estimates, we derived $r_{o} \mathcal{Y}_{{\rm Fe},\odot} - Z^{\star}_{{\rm Fe},\odot } < 4.1$. The two values of the 
$r_{o} \mathcal{Y}_{{\rm Fe},\odot} - Z^{\star}_{{\rm Fe},\odot }$ factor (the one coming from application of Eq. \ref{eq:zgas2} and the one derived from stellar synthesis models) are both poorly constrained, the former features larger values than the latter, there is however a small region of overlap for $3.8 <r_{o} \mathcal{Y}_{{\rm Fe},\odot} - Z^{\star}_{{\rm Fe},\odot } < 4.1$.
  
From the constraint of $3.8 < r_{o} \mathcal{Y}_{{\rm Fe},\odot} - Z^{\star}_{{\rm Fe},\odot}< 6.2 $, which is based on the use of Eq. \ref{eq:zgas2}, and the one on the stellar Fe abundance discussed above, $ Z^{\star}_{{\rm Fe},\odot} = 1.2\pm 0.1$, we  derived bounds on the fraction of Fe mass in stars, defined as:  $ M^{\star}_{\rm Fe}/ M^{\rm b}_{\rm Fe}$, where $M^{\rm b}_{\rm Fe}$ is the total Fe mass produced by $r_{o}M_{\star}$; namely, the Fe mass associated to all baryons.
As we can  see:
\begin{equation}
     {  M^{\star}_{\rm Fe} \over  M^{\rm b}_{\rm Fe} } =   {Z^{\star}_{{\rm Fe},\odot} \over r_{o} \mathcal{Y}_{{\rm Fe},\odot}} \, ,
        \label{eq:m_fe_s_m_fe_t}
\end{equation}
from which we get: $0.15 < M^{\star}_{\rm Fe} /  M^{\rm b}_{\rm Fe} < 0.25$. Thus, by relating the observed ICM Fe abundance with the stellar mass fraction from the SHMR, we derive that the amount of Fe locked in stars is roughly bound between 1/7 and 1/4 of the total Fe. It is worth pointing out that, in light of the overlap between the different estimates of $r_{o} \mathcal{Y}_{{\rm Fe},\odot} - Z^{\star}_{{\rm Fe},\odot}$ discussed above, the upper bound on ${  M^{\star}_{\rm Fe} / M^{\rm b}_{\rm Fe} }$ is consistent with an independent evaluation based on stellar synthesis models and SN rates. It should also be noted that a similar, albeit somewhat smaller, value of ${  M^{\star}_{\rm Fe} /  M^{\rm b}_{\rm Fe} }$ has been derived by direct measurement of stellar mass and ICM metal abundance \citep{Ghizzardi:2021}, we return to this difference in Sect. \ref{sec:missing}.
Finally, we note that corroborating evidence that the bulk of metals are ejected from the galaxies they are produced in comes from work on star forming galaxies \citep[see][]{Peeples:2014,Sanders:2023}.

\begin{figure}
        \centerline{\includegraphics[angle=0,width=8.8cm]{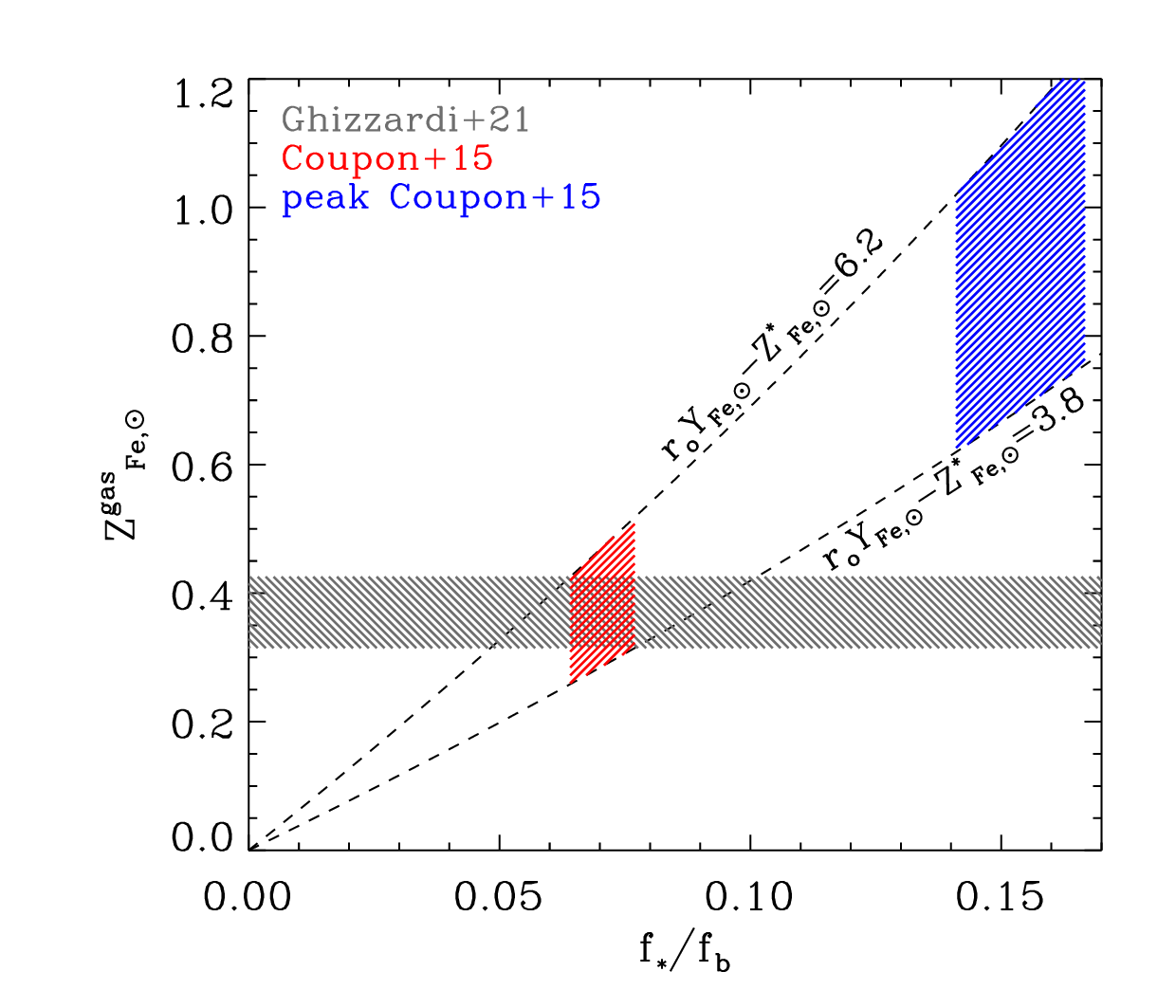}}
        \caption{Gas Fe abundance as a function of stellar over baryon fraction.
                The shaded gray region represents the Fe abundance measurements for massive clusters. The red shaded region shows where ${f_{\star} / f_{\rm b}}$ values for massive systems, as reported by \cite{Coupon:2015},  are consistent with the measured Fe abundance. The shaded blue region shows  how constraints on $r_{o} \mathcal{Y}_{{\rm Fe},\odot} - Z^{\star}_{{\rm Fe},\odot }$ allow us to estimate Fe abundance for $\sim 10^{12}$M$_\odot$ halos from the value of the peak stellar fraction, as measured by \cite{Coupon:2015}.
        }
        \label{fig:z_fe_vs_ms_mb_csp}
\end{figure}
As discussed in Sect. \ref{sec:ass}, the bulk of star formation occurs in halos with masses $\sim 10^{12}$M$_\odot$, and does not depend strongly on redshift, the galaxies hosted by these halos are later accreted by more massive systems and make up virtually all the stellar mass in apex accretors.  The upshot is that constraints on the  $r_{o} \mathcal{Y}_{{\rm Fe},\odot} - Z^{\star}_{{\rm Fe},\odot }$ factor, derived at the high mass end, can be applied to the $\sim 10^{12}$M$_\odot$ mass range, because the process that is being described is essentially the same. Indeed the central galaxies, that build up the bulk of their stellar mass when residing in $\sim 10^{12}$M$_\odot$ halos, are later accreted and end up as satellites in massive, $\sim 10^{15}$M$_\odot$, halos. 
Thus, the  $r_{o} \mathcal{Y}_{{\rm Fe},\odot} - Z^{\star}_{{\rm Fe},\odot }$ factor, despite being estimated at cluster scales,  can be thought of as a mean stellar fraction to gas\footnote{Although we are making use of Eq.\ref{eq:zgas2}, we describe the predicted quantity as gas abundance rather than ICM abundance because the prediction is being made at the $ 10^{12}$M$_\odot$ mass scale.} metal abundance conversion factor, where the  averaging is over the halos ingested by the apex accretor. There are two rather important consequences that follow.
The first is that the conclusion that the bulk of metals are to be found in gas does  not apply to massive accretors alone but to the Universe as a whole. We elaborate further on this point in Sects. \ref{sec:goutskirts}, \ref{sec:scatter} and Sect \ref{sec:budget}.
The second is that we can place some interesting constraints on the $\sim 10^{12}$M$_\odot$ mass scale.
In Fig. \ref{fig:z_fe_vs_ms_mb_csp} we see that by making use of the range  $3.8 <r_{o} \mathcal{Y}_{{\rm Fe},\odot} - Z^{\star}_{{\rm Fe},\odot } < 6.2$ the peak $f_{\star}$ measured by \cite{Coupon:2015}, at $ M_{\rm h} \sim 10^{12}$M$_\odot$, leads to a gas abundance in the range 0.65-1.2 Z$_\odot$, in $\sim 10^{12}$M$_\odot$ halos. This implies an Fe dilution, i.e. a reduction in Fe abundance, of about 1.5-4.0, when going from  $\sim 10^{12}$M$_\odot$ to $\sim 10^{15}$M$_\odot$ halos. Note that, unlike the case of massive systems, where essentially all baryons are within the halo, for $\sim 10^{12}$M$_\odot$ halos, the fraction of baryons lost through feedback effects is expected to be large \cite[see][and refs. therein]{Tumlinson:2017}. Thus, the Fe gas abundance is intended as the mean abundance extended to all gas accreted on the halo, of which a substantial part will have been ejected. We revisit this issue in Sects. \ref{sec:goutskirts},  \ref{sec:scatter}, and \ref{sec:budget}.

\begin{figure}
        \centerline{\includegraphics[angle=0,width=8.8cm]{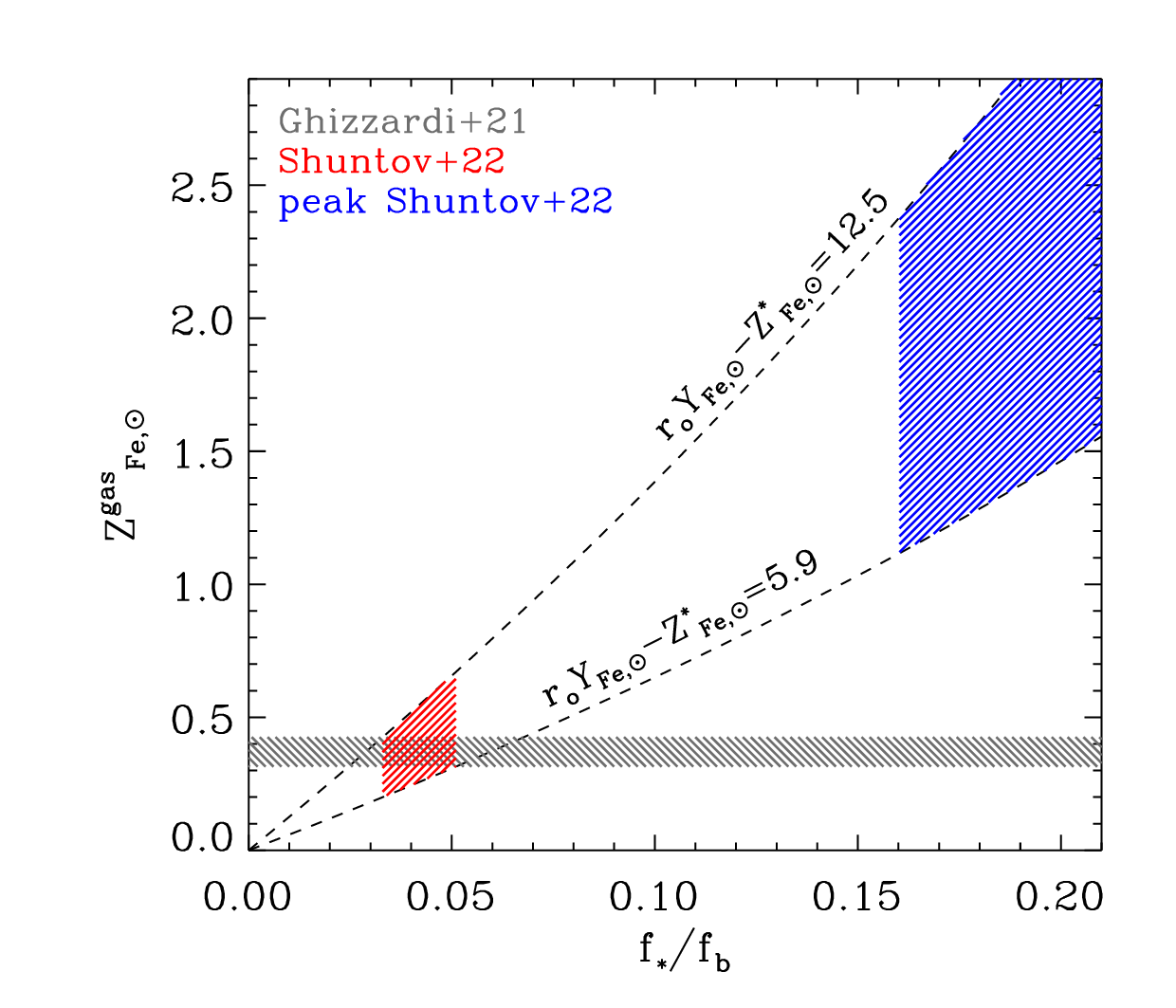}}
        \caption{Gas Fe abundance as a function of stellar over baryon fraction. The shaded gray region represents the Fe abundance measurements for massive clusters. The red shaded region shows where ${f_{\star} / f_{\rm b}}$ values for massive systems, as reported by \cite{Shuntov:2022}, are consistent with Fe abundance measurements.  The shaded blue region shows  how constraints on $r_{o} \mathcal{Y}_{{\rm Fe},\odot} - Z^{\star}_{{\rm Fe},\odot }$ allow us to estimate Fe abundance for $\sim 10^{12}$M$_\odot$ halos from the value of the peak stellar fraction, as measured by \cite{Shuntov:2022}.
        }
        \label{fig:z_fe_vs_ms_mb_sp}
\end{figure}
It is instructive to perform the same exercise depicted in Figs. \ref{fig:z_fe_vs_ms_mb} and \ref{fig:z_fe_vs_ms_mb_csp}  using the measurements reported in \cite{Shuntov:2022}. As we can see in Fig. \ref{fig:z_fe_vs_ms_mb_sp}, this leads to $5.9 <r_{o} \mathcal{Y}_{{\rm Fe},\odot} - Z^{\star}_{{\rm Fe},\odot } < 12.5$.
From this, using Eq. \ref{eq:m_fe_s_m_fe_t} and assuming, as done above, that the stellar Fe abundance is constrained between 1.1 $Z_\odot$ and 1.3 $Z_\odot$, we assess $0.08 < M^{\star}_{\rm Fe} /  M^{\rm b}_{\rm Fe} < 0.18$. In other words, the fraction of Fe in stars is even smaller than the one based on \cite{Coupon:2015} estimates of $f_{\star}$. From  the stellar fraction, $f_{\star} $, measured by \cite{Shuntov:2022}  at $\sim 10^{12}$M$_\odot$, we can estimate a gas abundance in the range 1.1-3.0 Z$_\odot$, in $\sim 10^{12}$M$_\odot$ halos and an Fe dilution of at least a factor of 2.6, when going from  $\sim 10^{12}$M$_\odot$ to $\sim 10^{15}$M$_\odot$ halos.

There are other estimates of the SHMR in the literature. \citet{Zu:2015} and \citet{Leauthaud:2012} derive values of $f_{\star}$ at  $\sim 10^{15}$M$_\odot$ that are about 20\% and 65\% higher, respectively, when compared to the one in \citet{Coupon:2015}. Clearly such values would lead to a reduction in the $r_{o} \mathcal{Y}_{{\rm Fe},\odot} - Z^{\star}_{{\rm Fe},\odot }$ factor, modest in the former case and substantial in the latter.  However, we are reluctant to make use of these estimates because the  associated SHMRs appear to be significantly offset from other measurements. For example, at  $\sim 10^{12}$M$_\odot$, $f_{\star}$ is (respectively) about two and three times higher than more recent estimates \cite[see][Fig. 34 for a compilation]{Behroozi:2019}. 
Another estimate of the SHMR is provided by \cite{vanUitert:2016}, see Fig.\ref{fig:shmr}. 
We do not use it   its large uncertainty makes it consistent with  $M_{\star}/M_h$ estimated by \cite{Coupon:2015} and \cite{Shuntov:2022}, thereby providing much weaker constraints on the $r_{o} \mathcal{Y}_{{\rm Fe},\odot} - Z^{\star}_{{\rm Fe},\odot }$ factor. 

\begin{figure}
        \hspace{-0.3cm}
        {\includegraphics[angle=0,width=9.3cm]{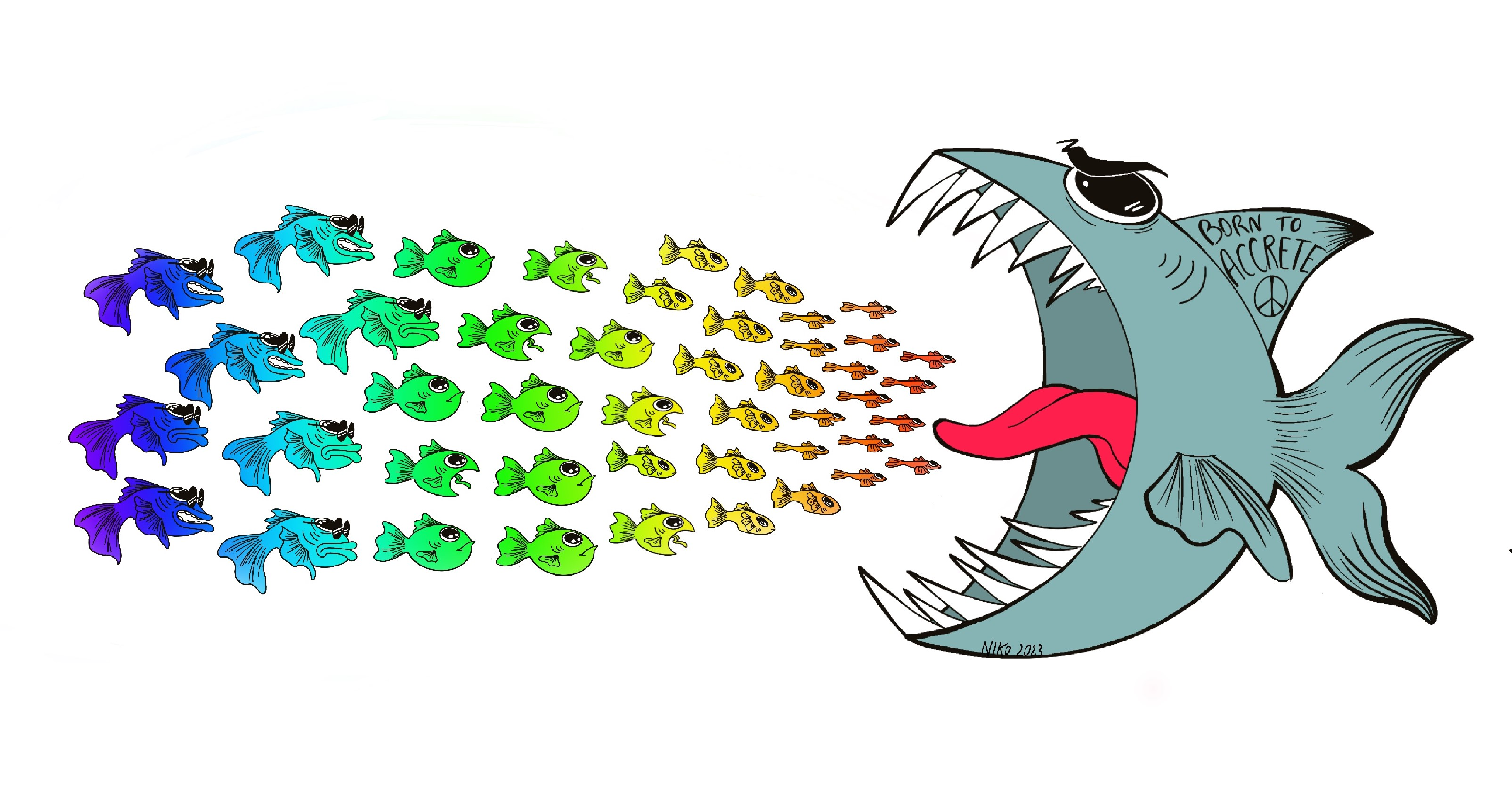}}
        \caption{
                Cartoon representation of apex accretor and sub-units undergoing accretion. The color (metal abundance) of the sub-units varies with size (halo mass), with the smaller being redder (metal richer) and the larger bluer (metal poorer). Note: the apex accretor features a color (metal abundance) that is a "mean"  of the sub-units' colors (metal abundances).  
        }
        \label{fig:fish}
\end{figure}

Despite the substantial quantitative difference between metal abundances we estimate for  $M_{h} \sim 10^{12}$M$_\odot$ halos from Fig. \ref{fig:z_fe_vs_ms_mb_csp} and Fig. \ref{fig:z_fe_vs_ms_mb_sp}, the decrease in $f_{\star}$, with increasing halo mass, is common to \cite{Coupon:2015} and \cite{Shuntov:2022} and, for that matter, to \citet{vanUitert:2016} (see Fig. \ref{fig:shmr}), \citet{Leauthaud:2012} and \citet{Zu:2015}. It is this decrement that ensures that gas in halos at the low mass end ($\sim 10^{12}$M$_\odot$) will always be  richer in metal than at the high-mass end ($\sim 10^{15}$M$_\odot$). To understand the reason for this in a simple way it is best to think of the metal abundance of massive halos as a weighted mean of the metal abundance of gas previously accreted and enriched in less massive halos, ranging from $\sim 10^{12}$M$_\odot$ to  a few $10^{13}$M$_\odot $, which is later reaccreted onto the halo under consideration. Halos at the low-mass end ($\sim 10^{12}$M$_\odot$) will contribute metal richer gas than halos at the high-mass end (a few $10^{13}$M$_\odot $). A somewhat unorthodox depiction of this process is presented in Fig. \ref{fig:fish}.
        
\subsection{Intermediate halos}\label{sec:goutskirts}

Iron abundance is known to a lesser extent in less massive than in more massive systems. Indeed, it has long been understood that measuring Fe in cooler objects presents greater challenges than in hotter ones. For one thing the L-shell blend measurements are more prone to systematic errors than those based on the K$\alpha$ line \citep[e.g.,][]{Buote:2000,MG:2001}.
In spite of these limitations, current measurements suggest that, to first order, group and poor cluster abundance profiles are flat, with a mean value similar to the one found in massive clusters,  and a  scatter larger than that found in more massive systems \citep[see][for a recent review]{Gasta:2021}. However, in the absence of a systematic study like the one performed on massive clusters, it is difficult to say how much of the difference between groups and clusters is due to real dissimilarities in the objects or to difficulties in the analysis.
 
When viewed within a structure formation scenario, the similarity in metal abundance between massive clusters and groups is surprising. Indeed, as halos evolve from the group to the massive cluster scale, they increase their gas mass by more than an order  of magnitude and accrete a large amount of potentially pristine gas, yet they appear to retain essentially the same Fe abundance.
To address this issue, we need to construct a model that predicts how metal abundance in the hot gas varies, as we go from the group to the cluster mass scale. We do this by including, in  Eq. \ref{eq:y2},  a term that accounts for the Fe mass expelled from halos through feedback:

\begin{equation}
        \mathcal{Y}_{{\rm Fe}} = {Z^{\star}_{{\rm Fe}}  \over r_{o}} \Biggl( 1 + {  Z^{\rm gas}_{{\rm Fe}} \over Z^{\star}_{{\rm Fe}}} {M_{{\rm gas}} \over  M_{\star}} + {  Z^{\rm m}_{{\rm Fe}} \over Z^{\star}_{{\rm Fe}}} {M_{{\rm m}} \over  M_{\star}}\Biggr) ,
        \label{eq:y_m}
\end{equation}
where $M_{\rm m}$ is the mass of the "missing" gas, i.e. the gas that has been ejected from the halo and $Z^{\rm m}_{{\rm Fe}}$ is its mean Fe abundance. As done in Sect. \ref{sec:coutskirts}, we solve for the gas abundance\footnote{Note that we have substituted the term ``ICM" adopted in  Sect. \ref{sec:coutskirts}, e.g. Eq. \ref{eq:y2}, with the more generic ``gas" as its application is here extended to lower mass systems.}, $Z^{\rm gas}_{{\rm Fe}}$:
\begin{equation}
        Z^{\rm gas}_{{\rm Fe},\odot} =  \Big(   r_{o} \mathcal{Y}_{{\rm Fe},\odot} - Z^{\star}_{{\rm Fe},\odot}  - Z^{\rm m}_{{\rm Fe},\odot} {f_{\rm m} \over  f_{\star}} \Big)  {f_{\star} \over  f_{\rm b}} \, {1\over 1 - {f_{{\rm m}} \over  f_{\rm b}} - {f_{{\star}} \over  f_{\rm b}}} \, ,
        \label{eq:zgas_m}
\end{equation}
where $f_{\rm m} = M_{\rm m} / M_{\rm h}$.
We assume $f_{\rm b}=0.16$ and derive $f_{\rm m}$ by imposing that the fraction of mass in the stellar, gas and missing components add up to the baryon fraction:
\begin{equation}
        f_{\rm b} = f_{\rm m} + f_{\rm gas} + f_{\star} \, ,
        \label{eq:f_m}
\end{equation}
where $f_{\rm gas}$ is adopted from \cite{Eckert:2021}, with a small modification. We normalize at $10^{15}  M_\odot $ rather than at  $10^{14}  M_\odot $, to allow for a larger scatter at lower masses\footnote{The use of the $\mp$ symbol instead of $\pm$ is intentional.}:
\begin{equation}
        f_{\rm gas} =  0.113^{\times 1.05}_{/1.05} \, \times \, \Big(  { M_{\rm h} \over 10^{15} M_\odot }\Big)^{0.22\mp 0.04}  \, .
        \label{eq:f_gas}
\end{equation}
 The stellar fraction, $f_{\star}$, is taken from \cite{Coupon:2015}, see Fig. \ref{fig:shmr}, and a 10\% uncertainty is assumed. As discussed in Sect.\ref{sec:coutskirts}, the term $r_{o} \mathcal{Y}_{{\rm Fe},\odot} - Z^{\star}_{{\rm Fe},\odot}$ is assumed to be independent of halo mass and will take on values derived at $M_{\rm h} = 10^{15} M_\odot$ (see Figs. \ref{fig:z_fe_vs_ms_mb} and \ref{fig:z_fe_vs_ms_mb_sp}). 

We consider three different values for the metal abundance of the ejected gas:  $Z^{\rm m}_{\rm Fe} = 0$, this represents the case where the missing gas is metal free
because its passage through less massive halos has not led to metal enrichment, see Sect. \ref{sec:metal};  $Z^{\rm m}_{\rm Fe,\odot} = Z^{\rm ICM}_{\rm Fe,\odot}$, this represents the case where the missing gas has previously been enriched in lower mass halos to an abundance similar to the  one found in the hot gas of  massive halos and $Z^{\rm m}_{\rm Fe,\odot} = 2 Z^{\rm ICM}_{\rm Fe,\odot}$, where the missing gas has previously been enriched in lower mass halos to an abundance twice that found in the hot gas of  massive halos. The third option, while difficult to justify, is, as we shall soon see, quite insightful. 

Having written $f_{\rm gas}$, $f_{\star}$ as a function of the halo mass, we can use Eq.\ref{eq:zgas_m} to express $Z^{\rm gas}_{{\rm Fe},\odot}$ as a function of $M_{\rm h}$. 
\begin{figure}
        \centerline{\includegraphics[angle=0,width=8.8cm]{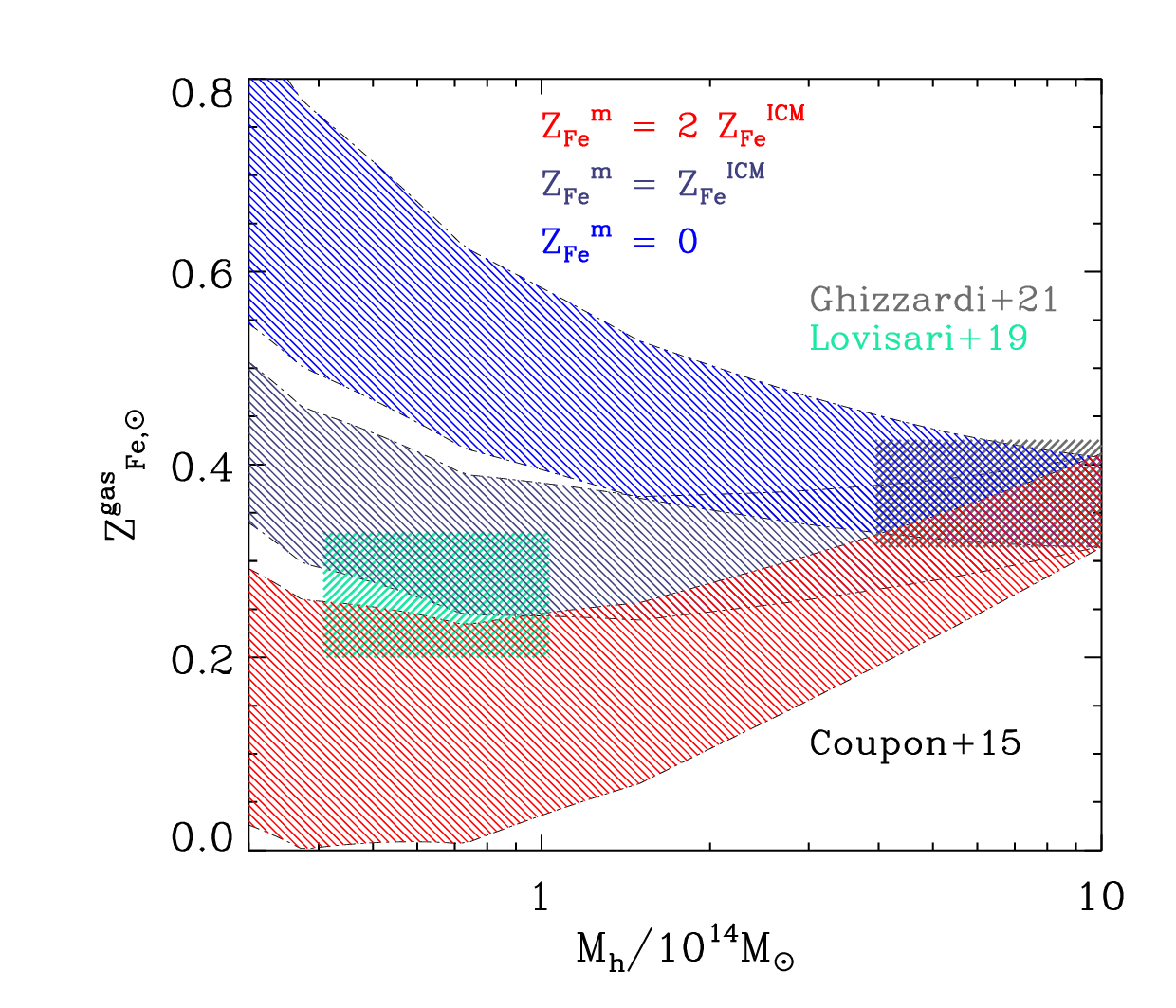}}
        \caption{Fe abundance of hot gas in massive halos as a function of halo mass. Blue, gray, and red shaded regions represent predicted abundances for different values of the metallicity of the missing gas, they have all been derived using the SHMR reported in \cite{Coupon:2015}. Green and gray rectangles are  measured abundances for groups and massive clusters, respectively.   
        }
        \label{fig:z_fe_vs_mh_cou}
\end{figure}
In Fig. \ref{fig:z_fe_vs_mh_cou} we compare the metal abundance predicted by Eq. \ref{eq:zgas_m}, with measurements in clusters \citep{Ghizzardi:2021} and groups \citep{Lovisari:2019}. At $M = 10^{15}$M$_\odot$,
$f_{\rm m} \simeq 0 $,  Eq. \ref{eq:zgas_m} reduces to Eq. \ref{eq:zgas} and all 3 choices of $Z^{\rm m}_{\rm Fe,\odot}$ lead to the same estimate for $Z^{\rm ICM}_{\rm Fe,\odot}$, which is consistent by construction, see Fig. \ref{fig:z_fe_vs_ms_mb}, with the measured one. As we move to lower halo masses, $f_{\rm m}$ increases and the three cases separate out. As easily understandable, if the missing gas is metal free, its accretion has the sole effect of diluting the metal content of hot halos and, as we move from lower to higher halo masses, the hot gas abundance decreases. If the missing gas has a metal abundance similar to that found in massive clusters, the predicted metal abundance of the hot gas will not vary much with halo mass. Finally, if most of the metals are in the missing gas, its accretion will lead to an increase in the metal abundance of the hot gas. As can be seen, the similarity between the abundance of the hot gas in groups and massive clusters implies that the missing gas cannot be pristine, a substantial,  most likely dominant part,  must have been previously accreted by less massive halos. 
The comparison described here can be used to derive a crude estimate of the abundance of the missing gas. 
By gradually varying $Z^{\rm m}_{\rm Fe,\odot}$, we identify values for which the predicted region intersects the measured group region. In doing so we find:
\begin{equation}
     0.25 \lesssim Z^{\rm m}_{\rm Fe,\odot} \lesssim 0.75 \, .
        \label{eq:z_m}
\end{equation}

We have investigated the dependence of this result on the specific choice of SHMR.  
\begin{figure}
        \centerline{\includegraphics[angle=0,width=8.8cm]{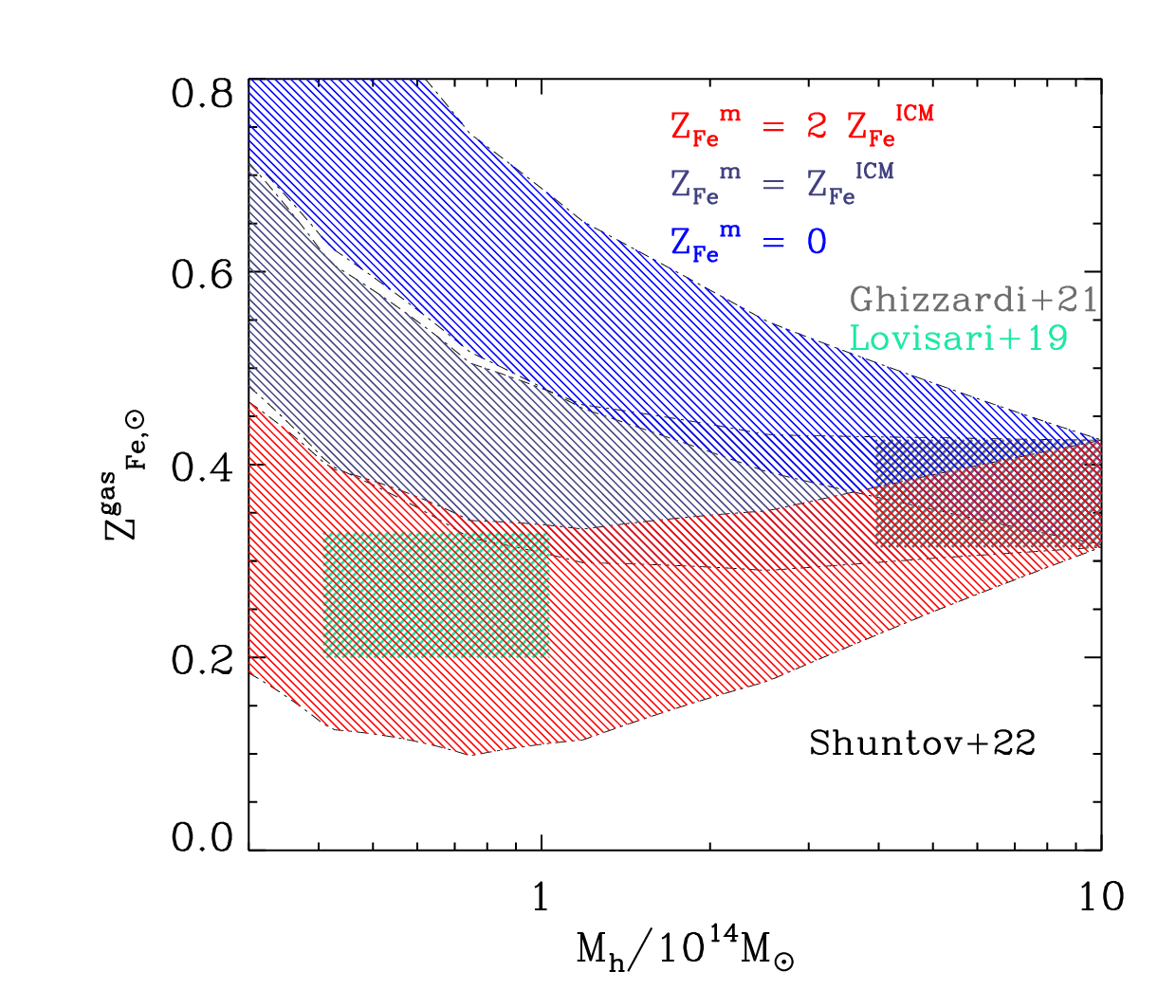}}
        \caption{Fe abundance of hot gas in massive halos as a function of halo mass. Predicted values are  computed as in  Fig. \ref{fig:z_fe_vs_mh_cou} except for the adopted SHMR which, in this case, is taken from \cite{Shuntov:2022}. 
        }
        \label{fig:z_fe_vs_mh_shu}
\end{figure}
In Fig. \ref{fig:z_fe_vs_mh_shu} we show the Fe abundance of hot gas in massive halos predicted using the \cite{Shuntov:2022} SHMR rather than the \cite{Coupon:2015} adopted in Fig. \ref{fig:z_fe_vs_mh_cou}. As we can see, the main result, namely,  the impossibility for the missing gas to be metal-free, remains unchanged. Indeed, the very high Fe yield required to reproduce the observed Fe abundance in massive clusters leads to an even larger discrepancy between predicted and measured Fe abundance at the group scale. The inescapable conclusion is that gas in the CGM must have had the time to be significantly enriched before it was ejected from the potential well of the dark matter halo. 

The argument we have made and the constrain we have derived on the missing gas metallicities are based on the assumption that the metal abundance of the  groups that evolve in the low redshift clusters we observe, is 
reasonably approximated by the metal abundance of low redshift groups. There are good reasons to expect this to be the case.
The enrichment process is intimately related to star formation and SHMR is essentially unchanged since $z\sim 3$ \citep[see][Figs. 34 and 35]{Behroozi:2019}.  Furthermore, at the cluster scale, we have direct observational evidence that the metal abundance of the ICM does not change at least out to $z\sim 1.5$ (see  \citealt{Baldi:2012}, \citealt{Ettori:2015}, \citealt{McDonald:2016}, \citealt{Mantz:2017}, \citealt{Liu:2020}, and \citealt{Flores:2021}, as well as Sect. \ref{sec:revol}). 

Looking beyond formulae and figures, the argument made here is a very simple one: the only way we can retain the same abundances, while increasing the mass of the halo by more than an order of magnitude, is by requiring that the accreted gas, much of which previously resided outside halos, be contaminated with metals in roughly the same proportion as the gas already bound to the halo. To the best of our knowledge, this is the first time that observational constraints on the metal abundance of the unbound gas are presented. It is certainly true that simulations predict the unbound gas to to be significantly contaminated by metals \citep[e.g.][]{Artale:2022,Mitchell:2022} and that the flat abundance profiles in cluster outskirts suggest a similar scenario \citep{Werner:2013}. It is also evident that our approach is an indirect one and that constraints are quite loose, as pointed out above the only thing we can say is that the unbound gas is not pristine and that its abundance is "similar" to that measured in groups and clusters. However, in the absence of any other measurement, we believe this estimate to be of some value. Indeed, in a much cited review,  \citet{Tumlinson:2017} identify the question of whether metals are retained by the CGM or leave the halo altogether as an important one. The arguments presented here provides a first answer. It is also worth mentioning that the method we have adopted can in principle be used to provide stronger constraints. If (or more hopefully, when) better measurements of the gas fraction, $f_{\rm gas}$, and hot gas metal abundance, $Z^{\rm ICM}_{\rm Fe,\odot}$, as functions of the halo mass, $M_{\rm h}$, and redshift  become available, it will be possible to derive improved measurements of the unbound gas metal abundance.

In Sect. \ref{sec:coutskirts} we inferred that most of metals in the Universe are in gas rather then locked in stars, we can go a little further by saying that part of that metal rich gas is bound to DM halos and part of it is not. We revisit this point in Sect. \ref{sec:budget}.

\subsection{Scatter}\label{sec:scatter}
Another important result from the analysis of the sample of massive clusters, is that the scatter around the mean abundance value is small. In  \cite{Ghizzardi:2021} we derived a total scatter of $\sim$ 15\% on the mass weighted abundance within $R_{500}$. We did not provide an estimate of the intrinsic scatter because a sizable fraction of the total scatter is likely associated with systematic errors which we could not quantify. Under such circumstances, all that can be said is that the intrinsic scatter is smaller than the total scatter; thus, this  becomes the starting point of the present analysis.  
Starting from Eq. \ref{eq:zgas2}, we connect the intrinsic scatter in metal abundance in massive clusters with that in the 4 variables  on the right hand side of the equation, namely:  $r_{o} \mathcal{Y}_{\rm Fe,\odot}$, $Z^{\star}_{\rm Fe,\odot}$, $f_{\star}$, and  $f_{\rm b}$. To this end we apply the standard error propagation technique: compute first-order derivatives with respect to the four variables and express the square of the standard deviation on the abundance, $\sigma_{Z^{\rm ICM}_{\rm Fe,\odot}}$,  as the sum of the squares of the standard deviations of the four variables weighted by the square of the respective first-order derivatives (note: by doing so we neglect the covariance and we return to this point later). After some algebra we find:
\begin{multline}
        {\sigma_{Z^{\rm ICM}_{\rm Fe,\odot}} \over Z^{\rm ICM}_{\rm Fe,\odot}} =  \Biggl[ \Big(  { \sigma_{ r_{o} \mathcal{Y}_{\rm Fe,\odot}}  \over r_{o} \mathcal{Y}_{\rm Fe,\odot} - Z^{\star}_{\rm Fe,\odot}  } \Big)^2  +  \Big(  { \sigma_{Z^{\star}_{\rm Fe,\odot}} \over r_{o} \mathcal{Y}_{\rm Fe,\odot} - Z^{\star}_{\rm Fe,\odot}  } \Big)^2  +  \\ \Big(  { \sigma_{f_{\star}}  \over f_{\star}  ( {1 - {f_{{\star}} \over  f_{\rm b}}}) } \Big)^2  +  \Big(  { \sigma_{f_{\rm b}}  \over f_{\rm b}  ( {1 - {f_{{\star}} \over  f_{\rm b}}}) } \Big)^2 \Biggr]^{1/2} \, .
\label{eq:sigma}
\end{multline}
We now investigate how the upper limit on the left hand side of Eq.\ref{eq:sigma}:
\begin{equation}
    {\sigma_{Z^{\rm ICM}_{\rm Fe,\odot}} \over Z^{\rm ICM}_{\rm Fe,\odot}} < 0.15  \, ,
\end{equation}   
derived in \cite{Ghizzardi:2021}, impacts on terms on the right hand side. The equation tells us that, neglecting covariance, each and every term on the right hand side has a scatter that is limited by the scatter of the term on the left hand side. For any given right hand side term, the maximum scatter is obtained when scatters on the other  three terms are imposed to be equal to zero.
 Let us consider the first term, if all others are set to zero, the upper limit on the term on the left hand side applies to this term as well, in mathematical form:
\begin{equation}
        { \sigma_{ r_{o} \mathcal{Y}_{\rm Fe,\odot}}  \over r_{o} \mathcal{Y}_{\rm Fe,\odot} - Z^{\star}_{\rm Fe,\odot}  } < 0.15   \, .
\end{equation}   
Since the denominator is smaller than  $r_{o} \mathcal{Y}_{\rm Fe,\odot}$, this implies that the relative scatter on $r_{o} \mathcal{Y}_{\rm Fe,\odot}$ will be smaller than 0.15. The question of by how much depends on the exact value taken on by $r_{o} \mathcal{Y}_{\rm Fe,\odot} - Z^{\star}_{\rm Fe,\odot}$. In the case of limiting values 3.8 and 6.2  (presented in Fig. \ref{fig:z_fe_vs_mh_cou}) and assuming the range of values 1.1 to 1.3 for $Z^{\star}_{{\rm Fe},\odot}$ ( discussed in Sect. \ref{sec:coutskirts}), we find 
$ {\sigma_{ r_{o} \mathcal{Y}_{\rm Fe,\odot}}  / r_{o} \mathcal{Y}_{\rm Fe,\odot}} < 0.13$. 
If we apply the same argument to the second term, we find that the relative scatter on $Z^{\star}_{\rm Fe,\odot}$ is bound by an upper limit that is larger than 0.15, because the denominator is larger than $Z^{\star}_{\rm Fe,\odot}$. As before, an estimate can be achieved from the limiting values 3.8 and 6.2 on $r_{o} \mathcal{Y}_{\rm Fe,\odot} - Z^{\star}_{\rm Fe,\odot}$ and the 1.1, 1.3 range for $Z^{\star}_{{\rm Fe}\odot}$. We estimate a much weaker constraint, ${ \sigma_{Z^{\star}_{\rm Fe,\odot}} / Z^{\star}_{\rm Fe,\odot}  } < 0.7$, this is a direct consequence of the fact that most of the Fe is in the gas and not locked in stars, thus even a large scatter in the distribution of $Z^{\star}_{\rm Fe,\odot}$ will have a modest impact on $\sigma_{Z^{\rm ICM}_{\rm Fe,\odot}} / Z^{\rm ICM}_{\rm Fe,\odot}$.
A certain degree of correlation might be present between the first two terms. 
We may imagine that the total production of Fe, encoded in the first term, is correlated with the metallicity of stars described by the second term. For example, halos where overall metal production is larger could feature metal richer stars and vice versa.  In such a case, the small scatter on $Z^{\rm ICM}_{\rm Fe,\odot}$ may result from large, but co-varying, scatters in  $r_{o} \mathcal{Y}_{\rm Fe,\odot}$ and  $Z^{\star}_{\rm Fe,\odot}$. Fortunately, we have an independent estimate of the scatter on $Z^{\star}_{\rm Fe,\odot}$  of $\sim 15\%$ \citep[see][]{Gallazzi:2014,Zahid:2017,Saracco:2023}, this suggests that any covariance between $r_{o} \mathcal{Y}_{\rm Fe,\odot}$ and  $Z^{\star}_{\rm Fe,\odot}$ provides a contribution to the scatter in either  $r_{o} \mathcal{Y}_{\rm Fe,\odot}$ or $Z^{\star}_{\rm Fe,\odot}$  that is smaller than our current upper limits.

In the case of the third term, assuming scatter in all other terms is zero, we find  $\sigma_{f_{\star}} / f_{\star}  < 0.15 $, the same holds true for the fourth term: $\sigma_{f_{\rm b}}  / f_{\rm b} < 0.15$. As pointed out earlier, this analysis does not account for covariance between the variables. However, it is not difficult to imagine that a halo where baryons are accreted in quantities larger than average may end up producing more stars than average and, of course, vice versa.
In such a case, the small scatter on $Z^{\rm ICM}_{\rm Fe,\odot}$ may result from larger but co-varying scatters in $f_\star$ and  $f_{\rm b}$. Fortunately, we have an independent estimate of the upper limit of the scatter on  $f_{\rm b}$ from XCOP \citep{Eckert_non_th_XCOP:2019}: $\sigma_{f_{\rm b}}  / f_{\rm b} < 0.18$, this tells us that any covariance between $f_\star$ and  $f_{\rm b}$ is related to a scatter in either $f_\star$ and  $f_{\rm b}$ that is significantly smaller than our current upper limits.  

It is interesting to compare our upper limit on  scatter on $f_{\star}$ with measurements available in the literature. \cite{Chiu:2018} derive an intrinsic scatter of $\sim70$\%. We suspect this may result from an underestimation of systematic errors on stellar mass calculation, possibly arising from a disparity of treatment between objects coming from different samples. Interestingly, \cite{Andreon:2012}, with a smaller and more homogeneous sample, ends up with an upper limit on intrinsic scatter similar to ours, namely: 15\%.
    
 There are at least three reasons why having a small scatter is important.
 First, it provides a strong justification for our approach, which describes the enrichment process, that is without doubt characterized by many fascinating and complicated details, through simple averages over large populations.
 Second, at present, several of the average quantities, $r_{o} \mathcal{Y}_{\rm Fe,\odot} $, $f_\star$, and so on,  are poorly constrained, more accurate measurements can lead us to a clearer picture of the enrichment process, even within a model as simple as this one. 
 Finally, the average values of  $r_{o} \mathcal{Y}_{\rm Fe,\odot} $ and $f_\star $ that we derived bear strong connections with average properties of the halos, where the bulk of these stars and metals are created. In other words, it may be possible to work our way back and constrain CGM properties starting from apex accretors. We return to this point in  Sect. \ref{sec:budget}. 
 
Although our analysis of the scatter on $Z^{\rm ICM}_{\rm Fe,\odot}$ allows us to derive constraints on important quantities, it does not answer the question of why the scatter is small.
As previously discussed, the vast majority of metals in the ICM, or more generally in clusters, is synthesized in halos of mass of roughly $10^{12}$M$_\odot$, similarly the bulk of the gas in the ICM was originally accreted in similar, perhaps a little more massive, halos.
As we have seen when discussing groups, the metal rich gas expelled by small size halos is later re-accreted by massive ones.
By taking the ratio of halo masses between massive clusters($\sim 10^{15}$M$_\odot$) and star forming halos ($\sim 10^{12}$M$_\odot$ to a few $\sim 10^{13}$M$_\odot$), we see that the re-accreted gas in each massive cluster must have been enriched in several hundred smaller halos. Furthermore, its mean metal abundance, which is essentially the ICM abundance measured beyond the core region, will feature a scatter that is about 20-30 times smaller than that characterizing the individual halos, where the enrichment process occurred. From this argument and the 15\% upper limit on the intrinsic scatter on   $Z^{\rm ICM}_{\rm Fe,\odot}$, we can work out that the dispersion on the metal abundance in the population of halos later accreted by massive clusters cannot be larger than  $\sim$ 250\%. This admittedly poor constraint could be improved upon through more precise measurements of the metallicity in the ICM of massive clusters.

In simple terms, the reason why scatter on $Z^{\rm ICM}_{\rm Fe,\odot}$ is small is because it arises from the averaging of hundreds of independent enrichment events (see Fig. \ref{fig:fish} for an evocative representation or look up ``central limit theorem" on wikipedia for a more theoretical perspective). Of course, as we move from the massive cluster scale to the poor cluster and even more the group scale, the number of enrichment events become fewer and the averaging effect will be reduced. Thus, at the group scale, the scatter should increase by a factor of a few with respect to what is measured in massive clusters. This is a simple prediction that future observations might be able to test (more on this is Sect. \ref{sec:prospects}).

\subsection{Redshift evolution}\label{sec:revol}
Early results by \cite{Balestra:2007} were indicative of a substantial decrease in metal abundance at high redshift.
These findings were not confirmed by later measurements. The results from  \cite{Baldi:2012,Ettori:2015,McDonald:2016,Mantz:2017,Liu:2020,Flores:2021}  are all consistent, with no evolution of metal abundance in core excised clusters (see more in Table \ref{tab:slopes}).
The broad range in mass and redshift (Table \ref{tab:slopes}) suggests these systems have formed and evolved over different cosmic times. 
Low-redshift ($z\sim 0.01-0.1$) and low-mass ($M_{\rm h}\sim 10^{14}$M$_\odot$) systems are young, with about half of their mass having been accreted since $z=1$, conversely massive ($M_{\rm h} \sim 10^{15}$M$_\odot$) high-redshift ($z\sim 1$) systems must have formed much earlier, about half of their mass has been accreted before $z=2$, yet they appear to have very similar abundances.
This requires the enrichment process, which for these core excised measurements is dominated by ex-situ enrichment, to be similar in a broad redshift range, at least up to 2, when the age of the Universe was $\sim 1/4$ of what it is now, and in an equally substantial mass range: $10^{14}-10^{15}$M$_\odot$.

We go on to examine how the lack of redshift evolution for metal abundance can be accommodated within the baryon enrichment picture we have outlined in this paper.
As discussed in Sect. \ref{sec:coutskirts}, for massive halos ($M_{\rm h} \gtrsim  5 \times 10^{13}$M$_\odot$) in the local Universe, metal abundance outside the core is determined by 
an averaging process: less massive halos,  where star formation efficiency peaks, $M_{h} \sim 10^{12}$M$_\odot$, contribute metal richer gas than more massive systems, $M_{h} \sim 10^{13}$M$_\odot$, where star formation is less efficient.
As pointed out by several authors \citep{Behroozi:2013,Behroozi:2019,Legrand:2019,Shuntov:2022} the dependence of star formation efficiency  on halo mass described above, is not limited to the local Universe but actually extends back in time to  $z \sim 4$. This suggests that the enrichment process of the hot gas in massive halos should be similar up to relatively high redshifts $z \sim 2$: this is indeed what we observe.  
\begin{table}
        \centering
        \caption{Redshift dependence of metal abundance and stellar mass fraction}
        \resizebox{\columnwidth}{!}{%
                \begin{tabular}{|c| r | r |c |r|}       
                        \hline
                        &       &                   &          &   \\
                        $z$ range  &   $M_{500}$ range & $R_{500}$ range  & $\gamma^{\mathrm{^{a}}}$ & Ref.      \\
                        &$10^{14}$M$_\odot$ &                  &                             &         \\
                        \hline
                        $<1.2$         &        $1< M < 10$     & $R >0.4$                &  $-0.26\pm 0.61$      &  Et15 \\
                        $<1.3$         &        $3< M < 20 $    & $0.15 < R < 0.5$ &  $-0.41\pm 0.25$      &  Mc16 \\
                        $<0.8$         &   $2< M < 20 $    & $0.5 < R < 1.0 $ &  $-0.30\pm 0.91$      &  Ma17 \\
                        $<1.0$         &  $1 < M < 20 $    & $ R < 1.0 $      &  $-0.28\pm 0.31$      &  Li20 \\
                        $<1.7$         &  $1 < M < 20 $    & $0.3 < R < 1.0 $ &  $-0.5^{+0.4}_{-0.3}$ &  Fl21 \\
                        \hline
                        $<1.3$         & $0.4< M < 20 $    & $ R < 1.0 $      &  $+0.05\pm 0.27$      &  Ch18 \\
                        \hline
                \end{tabular}%
        }
        \begin{list}{}{}
                \item[Notes.]
                $\mathrm{^{(a)}}$ slope defined from relation: $ X \propto (1+z)^{\gamma}$, where $X$ is $Z^{\rm ICM}_{\rm Fe}$ for the first 4 entries and $f_\star$ for the last.
                \item[References.] (Et15) \cite{Ettori:2015}; (Mc16) \cite{McDonald:2016}; (Ma17) \cite{Mantz:2017}; (Li20) \cite{Liu:2020}; (Fl21) \cite{Flores:2021}; (Ch18) \cite{Chiu:2018}.
        \end{list}
        \label{tab:slopes}
\end{table}

Striking confirmation of this picture comes from observations at longer wavelengths. As discussed in Sects. \ref{sec:coutskirts} and \ref{sec:goutskirts} (and highlighted in Eqs. \ref{eq:zgas2} and \ref{eq:zgas_m}) in the simple model we propose here, the metal abundance of the hot gas in groups and clusters should  be proportional to the stellar fraction found in these systems. Thus, if the metal abundance is independent of redshift and mass, the same should hold true for the stellar fraction, $f_\star$. This is indeed what is observed: \cite{Chiu:2018} find that $f_\star \propto (1+z)^{0.05\pm 0.27}$, for $5\times 10^{13}$M$_\odot <M_{500} < 2 \times 10^{15}$M$_\odot$ and $z<1.3$ (see also their Fig. 6 and Table \ref{tab:slopes}).

In simple words we may conclude that just as the lack of redshift dependence for metal abundance argues in favor of an enrichment process that is self similar in a broad redshift range, the lack of redshift dependence for the stellar mass fraction argues in favor of a stellar assembly process that is self-similar in a broad redshift range.

The findings reported in this section suggest we may be on the verge of a shift in paradigm; thus far, the accepted interpretation for abundance measurements in the outskirts has been that: ”enrichment must have occurred early on in the proto-cluster phase” \citep[e.g.,][and refs. therein]{Gasta:2021}. Comparisons of low-redshift  and low-mass systems with high-redshift and high-mass systems may force us to abandon this conclusion for an even stronger one, positing that: “enrichment efficiency is essentially the same from $z \sim 2$ to today”. A more quantitative assessment than the one presented here can come from follow-up works on a dedicated sample.

\subsection{Entropy vs. abundance anti-correlation}\label{sec:antico}

In this subsection, we provide the first explanation for the entropy versus abundance anti-correlation observed in clusters and groups. To do this, we must first take a step back and address issues related to the physics of the ICM.

Over the last two decades, we have accumulated a considerable body of evidence that favors suppression of thermal conduction, (see  \citealt{Molendi:2023} for a recent review). 
For example, in the case of substructures falling into A2142 \citep{Eckert:2014,Eckert:2017} and Hydra A \citep{Degrandi:2016} and in the case of coronae \citep{Sun:2007}, the difference in temperature between the infalling structure or the corona and its environment is accompanied by a difference in metal abundance suggesting that the same mechanism inhibiting conduction is also operating on mixing. 
The evidence for suppression of mixing is perhaps not as strong, however, some of the cases that have been used to argue in favor of suppression of conduction can also be used to claim inhibition of mixing. Moreover, recent work \citep{Zhuravleva:2019} finds evidence for the suppression of a directly related quantity, namely, the viscosity.
Assuming thermal conduction and mixing are inhibited \citep[see][and refs. therein]{Molendi:2023}, the ICM stratifies according to entropy and metals do not diffuse much.

Under the circumstances described above, we expect the two phase enrichment process discussed in Sect. \ref{sec:model} to leave an imprint on the ICM and IGrM; this  will mainly be visible on the two variables that are most affected by feedback, that is, metallicity and entropy. The lower entropy  gas 
expelled via the first mode or phase in the progenitor
will be concentrated at the center. Its abundance will be high because it is enriched at a time when the stellar fraction is at its peak and, to a lesser extent, because star formation in the central galaxy continues, albeit at a reduced rate, well after the major star assembly phase is over.

Moreover, as the progenitor halo grows, the low entropy high metallicity gas at its center is  augmented through donations from sub-halos whose cores survive disruption onto the major halo\footnote{The reason for the survival of the metal rich low entropy core, which makes up no more than a few percent of the total gas in structure, is that the energy associated with the free fall of the subhalo is channeled into  the heating and stripping of its outer regions.}. This process has been documented in clusters. In  A2142 \citep[see][]{Eckert:2014} and Hydra A \citep[see][]{Degrandi:2016} we observe subhalos whose metal rich and low entropy cores have survived intact down to $R_{500}$. It is also consistent with predictions from hydro-dynamical simulations, \citep[see][]{AM06,Sheardown:2018}. More specifically Figs.10 and 15 of \citet{Sheardown:2018} show that, while the bulk of the gas is stripped from the infalling structure, its low entropy core can be preserved and end up in the core of the host halo.
 It is also worth noting that, in light of its nature, the donation process should also play a role in the formation of cluster BCGs. This has indeed been observed, in a recent paper, \citet{Kluge:2023}, by comparing the properties of BCGs with those of other luminous ellipticals, find that the former show excess emission at large radii, which they explain through a process that is very similar to the one we have just outlined.  Infalling massive galaxies that, thanks to their deep potential wells, survive the merger process, are stripped once they reach the cluster center, where they contribute to the build up of the BCG. 

Further out, accreted gas is shock heated, its entropy increasing as halo mass rises. The metallicity of this gas is determined by enrichment occurring in the first phase in smaller halos later accreted by the one under consideration. As pointed out in Sect. \ref{sec:coutskirts}, the abundance of this gas is a weighted mean of the abundance of the gas enriched in halos ranging from $\sim 10^{12}$ to a few $10^{13}$ M$_\odot$, with the former contributing metal richer gas than the latter. Thus, the abundance of this gas will be lower than that of the gas bound to the central galaxy. These processes  lead to a scenario where the low entropy central gas is metal richer than the higher entropy gas located further out.

While a detailed quantification of the metallicity gradient between the low entropy and high entropy gas is hard to make, a rudimental calculation can be attempted. The metallicity of the high entropy gas has been discussed  in Sect. \ref{sec:coutskirts} and illustrated in Figs. \ref{fig:z_fe_vs_ms_mb} and \ref{fig:z_fe_vs_ms_mb_sp}. There we showed that the mean metal abundance of massive clusters, which is almost indistinguishable from that measured beyond the core and circum core regions \citep[see][]{Ghizzardi:2021}, can be reproduced by the $f_\star/f_{\rm b}$ ratio of massive halos, for a range of values of the  $r_{o} \mathcal{Y}_{{\rm Fe},\odot} - Z^{\star}_{{\rm Fe},\odot }$ factor. Based on the arguments presented here, we can estimate the metal abundance of the core to be roughly associated with the $f_\star/f_{\rm b}$ ratio of less massive halos ($\sim 10^{12}$ M$_\odot$) and the same range for the $r_{o} \mathcal{Y}_{{\rm Fe},\odot} - Z^{\star}_{{\rm Fe},\odot }$ factor. The predicted metal abundance of the gas in such halos is shown in Figs. \ref{fig:z_fe_vs_ms_mb_csp} and \ref{fig:z_fe_vs_ms_mb_sp}, in the form of blue shaded regions, for the case of $f_\star/f_{\rm b}$ estimated by \cite{Coupon:2012} and \cite{Shuntov:2022} respectively. In the former case the metal abundance is in the range $0.65< Z^{\rm ICM}_{\rm Fe,\odot}< 1.2$ and in the latter $1.1< Z^{\rm ICM}_{\rm Fe,\odot}< 3.0$. 
\begin{figure}
        \centerline{\includegraphics[angle=0,width=8.8cm]{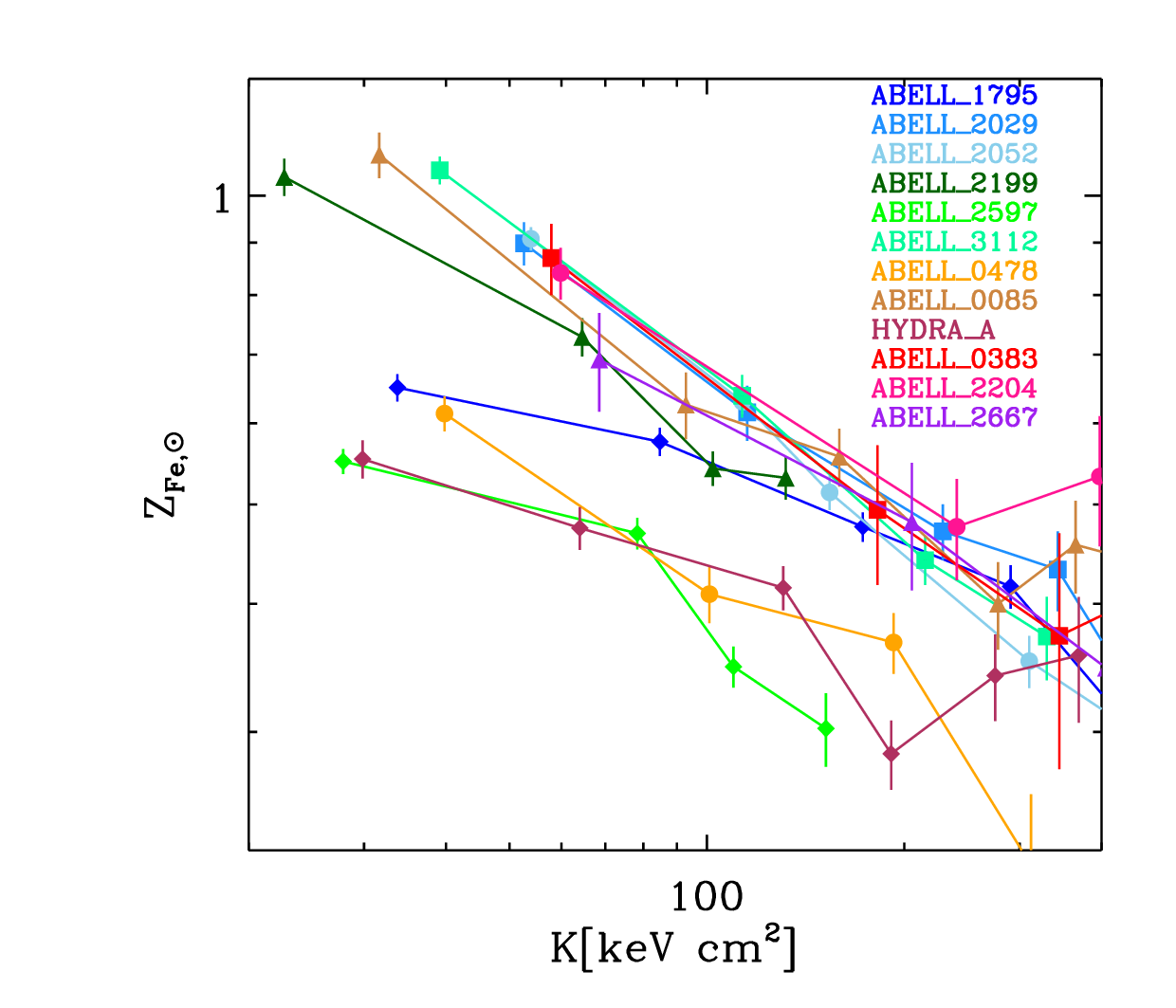}}
        \caption{Fe abundance versus entropy for cool core clusters. Abundance measurements are taken from \cite{Leccardi:2010}, they have been converted to \cite{Asplund:2009} solar abundances. Entropy measurements come from the ACCEPT archive of Chandra data \citep{Cavagnolo:2009}. The sample comprises the 12 cool core systems identified in \cite{Leccardi:2010} for which adequate entropy measurements could be found in ACCEPT.
        }
        \label{fig:z_vs_k}
\end{figure}
 
It is interesting to compare these considerations with actual measurements in groups and clusters.  Indeed, whenever an abundance gradient is measured, it is accompanied by a variation in entropy. The most striking anti-correlations \citep[e.g][]{Degrandi:2004,Rossetti:2010,Ghizzardi:2014} are found in a  subclass of groups and clusters known as cool cores \citep[CC, see][for a definition]{Molendi:2001}, which feature a strong central gradient in entropy and abundance.
In Fig. \ref{fig:z_vs_k}, we report the abundance versus entropy curves for a sample of 12 CC clusters, the anti-correlation is observed in each and every system (see the figure caption for more details).  Moreover, the abundances associated with the lowest entropy regions of CC systems, see Fig. \ref{fig:z_vs_k}, are in broad agreement  with those predicted  from values of $f_\star/f_{\rm b}$  at peak star formation efficiency. 
We note that the entropy versus abundance  anti-correlation is also found in so called cool core remnants \citep[CCR, see Fig. 3 in ][]{Rossetti:2010}, where the gradients are not as strong. As we discussed in a recent publication \citep{Molendi:2023}, merging systems, which do not show evidence for such gradients,  may well settle back to a configuration characterized by the entropy versus abundance anti-correlation once the disruption caused by the merger subsides. 

From data reported in Figs. 13 and 14 of  \citet{Ghizzardi:2021}, we estimate that, for XCOP CC clusters, the excess metal mass in the core exceeds by a factor of  5-10 what would be expected if only the progenitor contributed to it. From the same data, we infer that the iron mass in the core is  only a few percent of the  Fe mass integrated within $R_{500}$. 
We conclude that the donation process we discussed earlier in this section plays a dominant role in the enrichment of the core and a very modest one in that of the cluster. 
In light of the very crude nature of our quantitative estimates we defer a more detailed comparison with observations to future work.

\subsection{Entropy stratification}\label{sec:stratif}

Having reviewed how the gas outside the core and circum core regions has a metal abundance that depends weakly if at all with the mass (see Sects. \ref{sec:coutskirts} and \ref{sec:goutskirts}) and the redshift (Sect. \ref{sec:revol}) of the halo, we attempt to connect these properties with the thermodynamic structure of the ICM. 

In massive systems, gas undergoing accretion is shock heated to the virial temperature of the halo, as the halo mass increases so does the virial temperature. The gas stratifies according to entropy with the gas accreted earlier located closer to the center and the gas accreted later further out.
If conduction and mixing are heavily suppressed, as discussed in Sect. \ref{sec:antico}, then the stratification may well persist and even survive major merger events, as it very likely does in the core \cite[see][]{Molendi:2023}. If this is indeed the case, then we can look at abundance radial profiles very much like geologists look at layers of rock. More to the point, the absence of substantial radial abundance gradients beyond core regions in massive clusters would result from the lack of a halo mass and redshift dependence of metal abundance, see Sect. \ref{sec:revol}. 

\subsection{Abundance ratios}\label{sec:ratios}
Currently, evidence points to a lack of variation in abundance ratios as we move from the central BCG dominated region to more external regions; the most constraining measurements come from the Si/Fe ratio \citep[e.g.][]{Mernier:2017}, see also \cite{Biffi:2018}. Since $\alpha$ elements should be mostly produced by SNcc and Fe should be primarily synthesized in SNIa \cite[e.g][and refs. therein]{DGM:2009}, a radial abundance gradient could be interpreted as evidence of a different contribution of SNIa and SNcc to core and outskirts. However, as pointed out in Sect. \ref{sec:metal},  the bulk of SNIa explosions occur within a few Gyr of their formation. Thus metal enrichment in star forming halos should be characterized by a roughly constant $\alpha$ over Fe ratio.
Under these circumstances, the lack of abundance ratio gradients is not very surprising. We expect it to  extend to larger radii where measurements are currently either unavailable or unconstraining. 

There is one simple prediction we can make on the basis of our model. Since the metal mass at the center originates from a much smaller number of star forming halos (see Sect. \ref{sec:antico}) then in the outskirts (see Sect. \ref{sec:scatter}) we expect the scatter in abundance ratios to decrease as we move out from the core.

\section{The missing stellar mass problem}\label{sec:missing}

Over the last decade there have been several attempts to take a census of metals in galaxy clusters \citep[i.e.][]{Lowenstein13,RA14,Ghizzardi:2021}. In all instances, the
Fe mass measured in the ICM has been found to be in excess of what could be produced by the cluster stellar population.
While for the earlier results the problem could be ascribed to systematics in the estimate of the Fe mass \citep[see][]{Molendi:2016}, the thorough work presented in \cite{Ghizzardi:2021}, where stellar masses and ICM metal abundances were consistently and homogeneously measured out to a well defined radius in a representative sample of massive systems, is much harder to explain away. 

An important point is that  the stellar mass fraction can vary substantially from object to object: \cite{Chiu:2018} estimate an intrinsic scatter of $\sim$ 70\% on  $f_\star$, 
suggesting that the lack of stellar mass could different substantially from object to object. 
However, other work  on a smaller sample with homogeneously measured stellar masses \citep{Andreon:2012} suggests  it could be much smaller: $\sigma_{f_\star}/f_\star \sim 0.15$.
Moreover,  as discussed in Sect.\ref{sec:scatter}, the combination of limits on scatter on the ICM metallicity, ${\sigma_{Z^{\rm ICM}_{\rm Fe,\odot}} / Z^{\rm ICM}_{\rm Fe,\odot}} < 0.15$, and on baryon fraction $\sigma_{f_b}/f_b < 0.18$, suggests the scatter in $f_\star$ should be of the same order,  $\sigma_{f_b}/f_b  \lesssim 0.2$. These contradictory results can be reconciled if we postulate that stellar mass estimates in clusters are characterized by large systematic uncertainties. This is quite  likely the case: in Sect.4.2 of \cite{Ghizzardi:2021} we performed a detailed comparison of stellar masses reported in \cite{Lin:2012,Chiu:2018} and \cite{vdB15} finding a systematic discrepancy of $\sim$50\%-60\%
between the latter and the former two.  
In light of these considerations we do not address the issue of scatter in our proposed solution of the so called Fe ``conundrum".

Working within the enrichment framework presented in this paper, we start off  by rewriting Eq. \ref{eq:y2} in a slightly modified form:
\begin{equation}
        \mathcal{Y}_{\rm Fe,\odot} = { 1 \over r_o }\Bigl({Z^{\star}_{{\rm Fe,\odot}} + Z_{{\rm Fe}\odot}^{\rm gas}  {f_{\rm gas} \over  f_{\rm \star}}}\Bigr) \, .
        \label{eq:y2b}
\end{equation}
Next, we take  available measurements of  the dependence of  $f_{\rm gas}$ and $f_{\rm \star}$ on the halo mass, and plug them into Eq. \ref{eq:y2b}.
We perform our estimate at on overdensity of 500 and assume a power-law dependence of  $f_{\rm gas,500}$ and $f_{\rm \star,500}$ on $M_{500}$.
For the stellar fraction: for the high mass range $10^{14}< M_{500} < 10^{15}$M$_\odot$, we consider more direct measurements \citep{Eckert:2016,Chiu:2018,Ghizzardi:2021}; for the low mass range $10^{13}$M$_\odot < M_{500} < 10^{14}$M$_\odot$, we make use of \citet{Leauthaud:2012,Coupon:2015,Behroozi:2019}, which are based on  the SHMR, see Sect. \ref{sec:sm-hm} for details. This leads to a parametrization of the form: 
\begin{equation}
        f_{\rm \star,500} = (1.1 \times 10^{-2})^{\times 1.3}_{/1.3}  \times \Bigl({M_{500} \over 10^{14} \rm M_\odot}\Bigr)^{-0.18\pm 0.02} \, .
        \label{eq:fstar} 
\end{equation}
\begin{figure}
        \centerline{\includegraphics[angle=0,width=8.8cm]{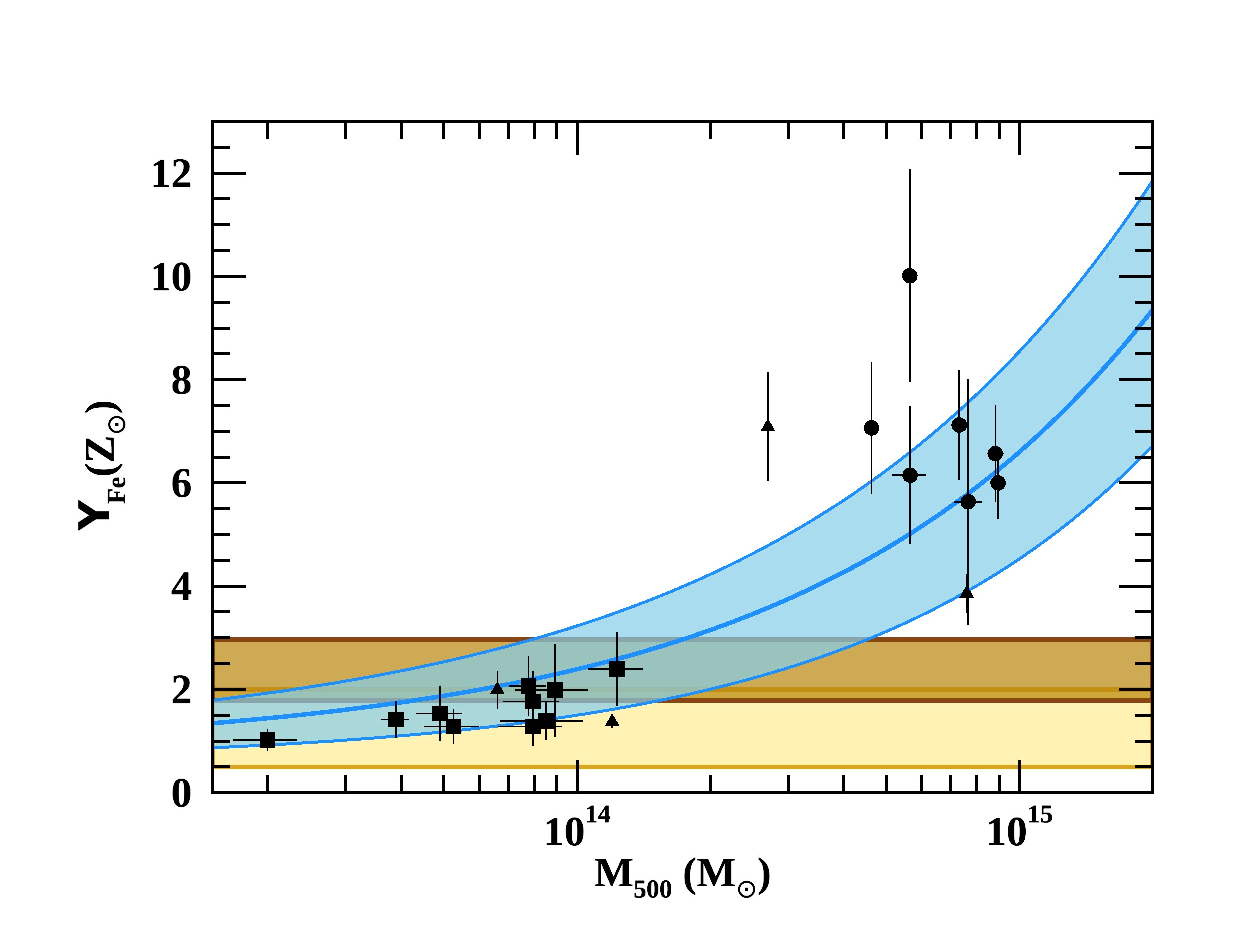}}
        \caption{Iron yield vs. halo mass, data points are shown as filled symbols \citep[see][for details]{Gasta:2021}. The blue thick line represents the relation reported in Eq. \ref{eq:y3} and the lighter shaded blue region associated uncertainties. Predicted yields for two different estimates of SN rates \citep[see][]{RA14,Freundlich:2021} are shown as yellow and brown shaded regions, respectively.
        }
        \label{fig:yield_vs_mh}
\end{figure}
For the gas fraction, $f_{\rm gas,500}$, we adopt the parametrization  based on \cite{Eckert:2021} and presented in Eq. \ref{eq:f_gas}. As in Sect. \ref{sec:coutskirts}: for the stellar metal abundance we assume $Z^{\star}_{{\rm Fe,\odot}} = 1.2\pm 0.1 $, for the abundance of the hot gas $ Z_{\rm Fe,\odot}^{\rm gas} = 0.37\pm 0.05$, and for the return factor we take the value $1/r_o = 0.64\pm0.06$.
Inserting all these elements into Eq. \ref{eq:y2b}, we come up with an equation relating  $\mathcal{Y}_{\rm Fe,\odot,500}$ with  $M_{500}$, which is well fit by the following formula:

\begin{equation}
 \mathcal{Y}_{\rm Fe,\odot} =  (0.79^{+0.14}_{-0.19}) \; + \; (1.6\pm 0.7) \Bigl({M_{500} \over 10^{14} \rm M_\odot }\Bigr)^{0.56^{-0.04}_{+0.08}}  \, .
        \label{eq:y3}
\end{equation}

It is only fair to point out that a similar parametrization has been proposed several years ago by \citet{RA14}. The main difference is that the above authors go through the $r$-band luminosity, rather than the stellar mass, however their end result, see their Fig. 2, is not at all unlike ours. 

In Fig. \ref{fig:yield_vs_mh}, we report the relation described in Eq. \ref{eq:y3}. As we can see, the yield estimated following the parametrization described above is broadly consistent with currently available direct measurements. At the high mass end both direct measurements and estimates  based on  Eq. \ref{eq:y3} are in excess of yield expectations based on SN rates \citep[see][]{RA14,Freundlich:2021} highlighted by the yellow and brown shaded regions.

The successful outcome of the exercise we have just performed provides much needed insight into the nature of the Fe conundrum, in simple terms: the discrepancy between the yield measured from clusters with the one predicted from stellar synthesis models  and SN rates (as well as the agreement for groups), must be connected to how the observed distribution of stars and hot gas varies as a function of halo mass.  
With this consideration in mind, we address the Fe conundrum within the framework of the model we sketched in Sect. \ref{sec:model}. The first point we need to make is that, on the basis of the discussion presented in Sect.  \ref{sec:coutskirts} (see Sect. \ref{sec:budget} for a more detailed version), we do not expect $\mathcal{Y}_{\rm Fe,\odot}$ to vary with halo mass. Thus, the behavior reported in Eq.\ref{eq:y3} and graphed in Fig. \ref{fig:yield_vs_mh} is likely the result of some observational bias.

During the first phase or mode, stellar mass is assembled from stars forming inside the halo (in situ), in the case of the progenitors of massive systems this stellar mass will end up mostly in the BCG or brightest group galaxy (BGG), there is considerable observational evidence that is consistent with this scenario \citep[e.g.,][]{Edwards:2020,Chu:2021,Chu:2022}.
During the second phase or mode, stellar mass is accreted in the form of galaxies from less massive halos falling into the main halo (ex situ). Part of this stellar mass will be stripped and will contribute to diffuse emission, so called intra cluster light \citep[ICL, see][for a recent review]{Contini:2021}, and part will remain in the form of satellite galaxies.
Within the first mode or phase the stellar mass rapidly builds up at early times until quenching stops almost completely further growth. This is confirmed from observations where the stellar mass is found to grow with halo mass up to  $M_\star \sim 2-3\times 10^{11}$ M$_\odot$ roughly corresponding to $M_{h} \sim 10^{13}$ M$_\odot$ and rapidly saturates there-after \citep[e.g.][see also Fig.\ref{fig:shmr}]{Coupon:2015,Behroozi:2019}. The contribution of accreted stellar mass  becomes progressively more important with growing halo mass and, starting from $\sim  3 \times 10^{13}$ M$_\odot$, is the dominant contribution to stellar mass within the halo \citep[e.g.][]{Leauthaud:2012,Coupon:2015,Shuntov:2022}.

Since the process by which stellar mass is assembled in the two modes is different, we do not expect them to follow the same radial distribution. Indeed, the stellar mass profile associated to the BCG is much more centrally concentrated than that of satellite galaxies, or of the ICL, \citep[e.g.][see their Fig. 12]{Zhang_ICL_2019}.
A consequence of the observational facts  outlined above is that the fraction of stellar mass within a given overdensity, say $f_{\star,500}$, differs significantly for halos of different mass. Within less massive halos ($10^{13}-10^{14}$ M$_\odot$), $f_{\star,500}$ will be larger than in more massive ones ($M_{h} > 10^{14}$M$_\odot$), this is actually consistent with a wealth of observational data \citep[e.g.][ and refs. therein, see also their Fig. 12]{Eckert_non_th_XCOP:2019}. It should also be mentioned that in situ and ex situ components are likely to suffer differently from detection biases. Indeed, the much more diffuse distribution of the latter (this is particularly true for ICL) favors flux underestimation. The ensuing differential bias likely contributes to the steepening of the observed $f_{\star}$ versus $M_{h}$ relation.

For the hot gas in groups and clusters, the situation is  the opposite, more massive systems feature more centrally peaked gas distribution than less massive ones. We note that outside the core ($R > 0.25 R_{500}$), this is true for both CC and non cool core (NCC) systems  \citep{Pratt:2022}. This is because, in lower mass systems, AGN feedback, contending with a shallower potential well, can redistribute gas to significantly larger radii and, perhaps, expel it altogether. 
As a consequence of this,  the fraction of gas within a given overdensity, say $f_{gas,500}$, will differ significantly for halos of different masses. Within less massive halos, $10^{13}-10^{14}$ M$_\odot$, $f_{gas,500}$  will be smaller than in more massive ones, $M_{h} > 10^{14}$ M$_\odot$. This is actually consistent with a wealth of observational data  \citep[e.g.][ and refs. therein, see also their Fig. 7]{Eckert:2021}.
Thus, by computing $\mathcal{Y}_{\rm Fe}$ within a fixed overdensity, we do not sample the same fraction of gas and stellar mass and the difference depends on the mass scale and specific overdensity at which $\mathcal{Y}_{\rm Fe}$ is computed. 
This is in agreement with work by \cite{Sarkar:2022}, who find that, for a small sample of groups, the measured yield increases by about 1/3 when computation is extended from $R_{500}$ to $R_{200}$.

The negative slope of the  $f_{\star,500}$ versus  $M_{500}$ relation, based on direct estimates of the stellar fraction in massive halos, see Eq. \ref{eq:fstar}, is at variance with the one derived from the SHMR, see Fig. \ref{fig:shmr}, \cite{Coupon:2015} and \citet{Shuntov:2022}. The latter is consistent with being flat for halo masses larger than $\sim 10^{14}$ M$_\odot$.
If we assume that the correct dependence of $f_{\star,500}$ on $M_{500}$ is the one found from SHMR
and that the bias on the  stellar mass fraction derived from direct observations in groups and clusters is  zero somewhere between $3\times 10^{13}$ M$_\odot$ and $  3\times 10^{14}$ M$_\odot$, we come up with a correction to $f_{\star,500}(10^{15}{\rm M_\odot})$ of $(10^{\times 3}_{/3})^{0.18} \sim 1.5^{\times 1.2}_{/1.2}$. If we apply this correction factor to  the  $f_{\star,500}(10^{15}{\rm M_\odot})$ estimated in Eq. \ref{eq:fstar}, we get  $f^{\rm c}_{\star}(10^{15}{\rm M_\odot})/f_{\rm b} = 0.064^{\times 1.4}_{/1.4}$,  which is in agreement with the $f_{\star}(10^{15}{\rm M_\odot})/f_{\rm b}$ derived by \citet{Coupon:2015}, see also Fig. \ref{fig:z_fe_vs_ms_mb}. Plugging $f^{\rm c}_{\star}(10^{15}{\rm M_\odot})$ into Eq. \ref{eq:y2b}, we get  a corrected yield, $\mathcal{Y}^{\rm c}_{\rm Fe,\odot}=3-6$. 

\begin{figure}
        \centerline{\includegraphics[angle=0,width=8.8cm]{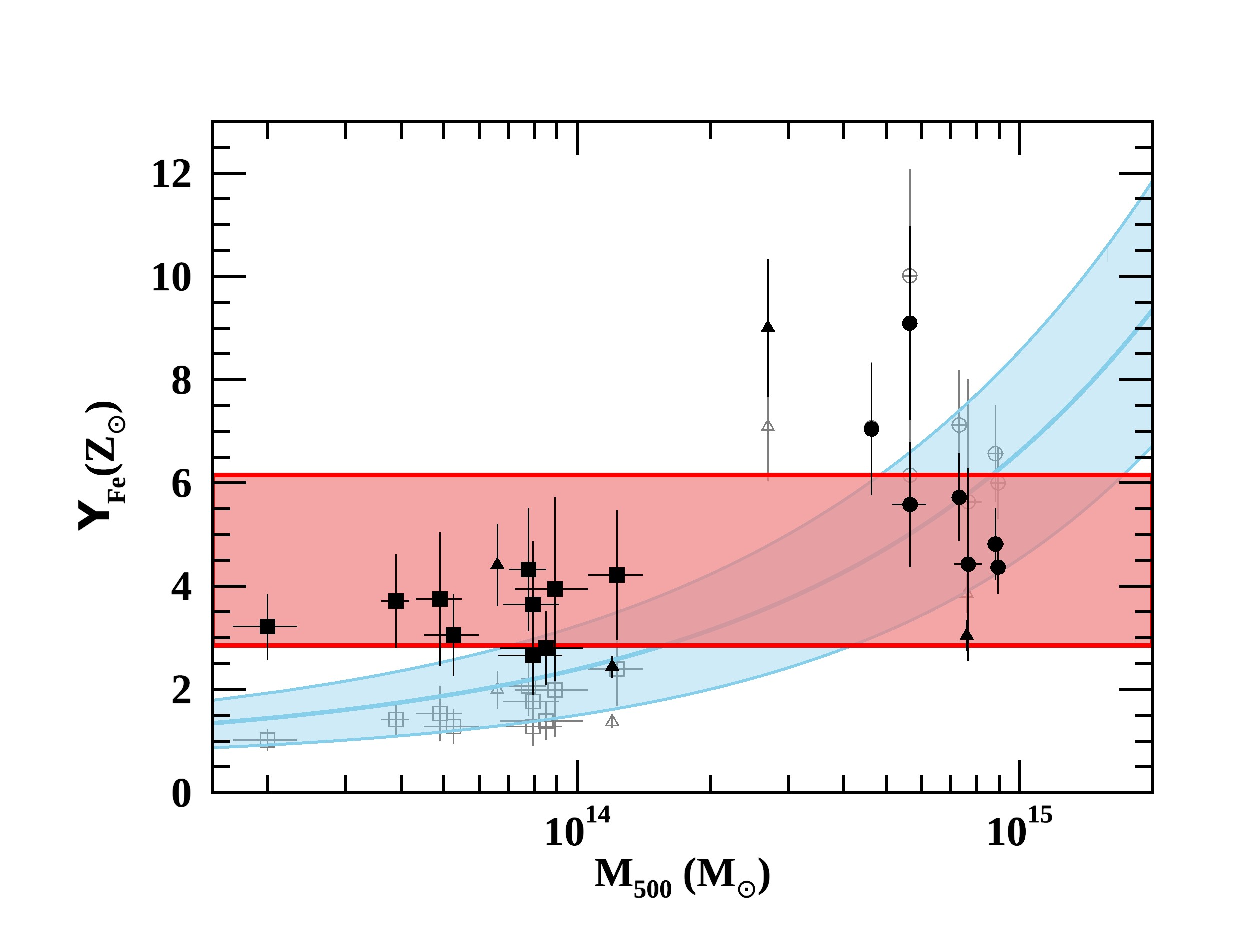}}
        \caption{Corrected iron yield vs. halo mass. Original data points and yield vs. halo mass relation are shown  as semi-transparent. Filled data points and region are obtained by correcting the original ones as  described in the text. 
        }
        \label{fig:yield_vs_mh_cor}
\end{figure}

$\mathcal{Y}^{\rm c}_{\rm Fe,\odot}$ has been obtained by applying a correction to the stellar fraction, at a mass scale where corrections to the gas fraction can be neglected. In accord with the assumption discussed earlier in this section and in Sects.\ref{sec:coutskirts} and \ref{sec:budget} that the yield  does not vary with halo mass, we take $\mathcal{Y}^{\rm c}_{\rm Fe,\odot}$ as an estimate of the true yield. We show it in Fig. \ref{fig:yield_vs_mh_cor} as a shaded horizontal region along with corrected data points obtained by applying the same correction to the original ones. Note that, for low mass systems, the change from measured to corrected yield derives mostly from having substituted the observed gas fraction with the baryon fraction.  
The solution we propose is based on the \citet{Coupon:2015} stellar fraction, a similar one can be derived from the \citet{Shuntov:2022} stellar fraction. The major difference between the two is that the latter implies an even larger yield and significant tension with estimates based on synthesis models and SN rates.

In summary, we have come up with a self consistent picture where the conundrum is solved if the following  conditions are met:
1) about one-third of the total stellar mass in massive halos has thus far eluded detection;
2) the gas mass fraction of less massive halos ($<$ a few $10^{14}$ M$_\odot$) is significantly underestimated; and
3) the yield is only marginally consistent with predictions from synthesis models and SN rates, and is largely independent of halo mass.
While it can be argued that there is already substantial evidence for the second point, the same cannot be said for points one and three. 
In simpler terms, our proposal for this conundrum will stand as an applicable solution if our assumptions are addressed through further observational work and verified.  

\section{A metal budget for the Universe}\label{sec:budget}

Here we put together several results disseminated through Sects. \ref{sec:connect} and \ref{sec:missing} to provide a coherent picture of the distribution of metals in the Universe. As pointed out in \citet{Peeples:2014}, the major problem in carrying out an estimate of the metal budget for the Universe is the uncertainty on the quantity of metals that galaxies produce. We approach this issue in Sect. \ref{sec:coutskirts}, by deriving an estimate of the Fe yield, or more specifically of the $r_{o} \mathcal{Y}_{{\rm Fe},\odot} - Z^{\star}_{{\rm Fe},\odot }$ factor, in clusters of galaxies, where we have a reliable estimate of both the Fe mass and the stellar mass that has been responsible for its production. Despite the relatively large uncertainties in the estimated factor, it allows us to infer that no more than $\sim 1/4$ of the Fe synthesized in stars currently resides in them. By making use of the description of the assembly and enrichment process provided in Sect. \ref{sec:ass}, we  extend this result to the Universe. Since this is a rather important point, we go over the argument once more and elaborate upon it. The $r_{o} \mathcal{Y}_{{\rm Fe},\odot} - Z^{\star}_{{\rm Fe},\odot }$ factor, which can be thought of as a stellar fraction to gas metal abundance conversion factor, see Eq. \ref{eq:zgas2}, is estimated in clusters of galaxies, but, since virtually the entire stellar mass in clusters has been synthesized in smaller halos, it describes the Fe production efficiency in the latter. More specifically, it is a mean of the values that it takes on in the many galaxies currently residing in clusters. One could argue that the distribution of subhalos that are accreted by the apex accretor is significantly different from the distribution of halos in the Universe and that this might, at least in principle, lead to a biased estimate of the conversion factor derived from clusters. However, this is not the case, indeed the shape of the sub-halo mass function \citep[see][and refs. therein]{Grylls:2019} is quite similar to that of the halo mass function (HMF) \citep[e.g.][]{Tinker:2008}. 

 Galaxies in clusters are mostly passive and have not produced metals for some time, conversely, those found in lower density environments continue to synthesize heavy elements today. If, for some reason, the efficiency characterizing metal production depends on cosmic time, the value we estimate in clusters might be biased. 
As shown in \citet[][see their Fig. 16]{Behroozi:2019}, star formation peaks at earlier times in more massive halos, however significant contribution to star formation occurs over a broad range in cosmic time in all environments that provide a substantial contribution to the overall stellar mass in the current universe, suggesting that late contributions even if characterized by a different efficiency will provide a modest contribution. Indeed only about one-third of the total stellar mass in the Universe is synthesized at $z<1$, \citep[][see their Fig.11]{Madau:2014}.

Having proved to ourselves, and hopefully to our readers, that our measurement of the yield in clusters provides a reasonable estimate for the one in the Universe, we proceed to discuss the budget. We start by defining, $Z^{\rm b}_{\rm Fe}$, the  Fe abundance averaged over all baryons within a given halo: $Z^{\rm b}_{\rm Fe} \equiv M^{\rm b}_{\rm Fe} / M_{\rm b}$. It is easy to see that:

\begin{equation}
        Z^{\rm b}_{\rm Fe,\odot}(M_{\rm h}) = r_{o}(M_{\rm h}) \mathcal{Y}_{\rm Fe,\odot}(M_{\rm h}) \, {f_\star(M_{\rm h})  \over f_{\rm b}} \, .
        \label{eq:z_b} 
\end{equation}
To estimate the metal abundance of the Universe we average $Z^{\rm b}_{\rm Fe,\odot}$ over the distribution of mass in the Universe:
\begin{equation}
        \langle Z^{\rm b}_{\rm Fe,\odot} \rangle \; =  \; { \bigintssss Z^{\rm b}_{\rm Fe,\odot}(M_{\rm h}) \; M_{\rm h} \; {{\rm d}n / {\rm d}M_{\rm h}}  \; {\rm d}M_{\rm h} \over  \bigintssss  M_{\rm h} \; {{\rm d}n / {\rm d}M_{\rm h}}  \; {\rm d}M_{\rm h}} \, ,
        \label{eq:z_u} 
\end{equation}
where, ${\rm d}n / {\rm d}M_{\rm h}$ is the HMF. 
By substituting Eq. \ref{eq:z_b} into \ref{eq:z_u} and assuming the dependence of $r_{o}(M_{\rm h}) \mathcal{Y}_{\rm Fe,\odot}(M_{\rm h})$ on halo mass, if any, is modest, we get:

\begin{equation}
        \langle Z^{\rm b}_{\rm Fe,\odot} \rangle  \; =  \; { r_{o} \mathcal{Y}_{\rm Fe,\odot} \over f_{\rm b}} \;  { \bigintssss f_{\star}(M_{\rm h}) \; M_{\rm h} \; {{\rm d}n / {\rm d}M_{\rm h}}  \; {\rm d}M_{\rm h} \over  \bigintssss  M_{\rm h} \; {{\rm d}n / {\rm d}M_{\rm h}}  \; {\rm d}M_{\rm h}} \, ,
        \label{eq:z_u2} 
\end{equation}
 where we estimate the uncertainty on ${ r_{o} \mathcal{Y}_{\rm Fe,\odot}} $ from Fig. \ref{fig:z_fe_vs_ms_mb}, by accounting for limits on the stellar Fe abundance, $ Z^{\star}_{{\rm Fe},\odot} = 1.2\pm 0.1$, and by augmenting the relative uncertainty from 20\% to 35\% to account for eventual variations of the efficiency with halo mass.  Furthermore, by adopting the Tinker HMF \citep{Tinker:2008} and taking $f_{\star}(M_{\rm h})$ from \citet{Coupon:2015}, see also Fig. \ref{fig:shmr} and Sect. \ref{sec:missing},  we get:  
\begin{equation}
 0.35 <  \, \langle Z^{\rm b}_{\rm Fe,\odot} \rangle \, < 0.77  \, .
        \label{eq:z_u3} 
\end{equation}

Ours is not the first estimate of the metallicity of the Universe. We are aware of estimates in \cite{Fukugita:2004},
\cite{Shull:2014} and \cite{Maoz:2017}. Converted into \cite{Asplund:2009} solar abundance units, the measurements presented in the above papers are $\langle Z^{\rm b}_{\rm Fe,\odot} \rangle \, \sim 0.1$, with uncertainties, largely dominated by systematics which we estimate to be $\sim$ 50\% (see Table \ref{tab:un_abund}). This implies a discrepancy of a factor between 2 and 8 with ours. Unsurprisingly, the mismatch is explained by the difference in metal yield. Indeed, \cite{Shull:2014} and \cite{Maoz:2017}  adopted yields from measured SN rates and nucleo-synthesis models which are a factor of a few smaller than the one we use here. Comparison with the metallicity of the Universe estimated by \cite{Fukugita:2004} provides further insight. These authors perform their estimate by taking stock of the metal content of different environments, with major contributions coming from: stars in galaxies, cool gas, hot gas in groups and clusters and warm gas in IGM and CGM (see their Table 3). 
A critical point is that the warm/hot gas in IGM and CGM, which accounts for a sizable fraction of baryons in the Universe, is assumed to feature a very modest metallicity $\sim 3\%$ of solar. A metallicity of $\sim$1/3 would take their estimate from a value that is comparable to the one estimated by \cite{Shull:2014} and \cite{Maoz:2017} to another that is consisted with ours. In other words, we have two competing scenarios: the first is characterized by a small yield ($\mathcal{Y}_{{\rm Fe},\odot} \sim 1$), a metallicity of the Universe of $\sim$ 0.1, and requires the bulk of metals to be in stars; the second features a substantially larger yield ($\mathcal{Y}_{{\rm Fe},\odot} \sim 3 - 6$), a metallicity of the Universe in the range 0.3-0.8 and the bulk of metals are in the warm or hot gas. 

\begin{table}
        \centering
        \caption{Metal abundance of the Universe}
        \resizebox{\columnwidth}{!}{%
                \begin{tabular}{|c| c | c |c |c |}      
                \hline
                                                                 & Fu04$\mathrm{^{(a)}}$ & Sh14 &  Ma17    &  Tw \\
                \hline
                $\langle Z^{\rm b}_{\rm Fe,\odot} \rangle $  & $0.08 \pm 0.04$       & $0.11 \pm 0.07$ &  $0.08\pm 0.01$ & $0.56\pm0.21$ \\
                \hline
                \end{tabular}%
        }
        \begin{list}{}{}
                \item[Notes.]
                $\mathrm{^{(a)}}$ When no uncertainty has been provided in original work, we assume a  50\% error.
                \item[References.] (Fu04) \cite{Fukugita:2004}; (Sh14) \cite{Shull:2014}; (Ma17) \cite{Maoz:2017}; (Tw) this work.
        \end{list}
        \label{tab:un_abund}
\end{table}

Having derived the mean metal abundance of the Universe, we now estimate how Fe is shared between different constituents.
The fraction of metals locked in stars, $\phi_\star$, assuming the stellar fraction derived in \cite{Coupon:2015} and including the augmented uncertainty on the yield described above, is  $\phi_\star = 0.15-0.28$, the remainder has to be in gas, either bound to halos or not. 
We point out that we favor the stellar fraction estimated in \citet{Coupon:2015}  over the one in \cite{Shuntov:2022}, because it plays an important part in a self consistent description of the distribution of metals in clusters discussed at the end of Sect. \ref{sec:missing}. That said, we note that by adopting the \cite{Shuntov:2022} stellar fraction we get an even smaller fraction of metals in stars: $\phi_\star = 0.06-0.20$, see Sect. \ref{sec:coutskirts}.

We estimate the fraction of metals in groups and clusters, $\phi^{ICM+IGrM}_{\rm Fe}$, as:
\begin{equation}
        \phi_{ICM+IGrM}  =   \, { Z^{\rm ICM+IGrM}_{\rm Fe,\odot} \over \langle Z^{\rm b}_{\rm Fe,\odot} \rangle  } \; { F_{\rm gas}(M_{\rm h} > 5 \times 10^{13}{\rm M_\odot}) \over f_{\rm b} }\, ,
        \label{eq:f_fe} 
\end{equation}
where $F_{\rm gas}(M_{\rm h} > 5 \times 10^{13}{\rm M_\odot})$ is the fraction of the gas fraction associated to halos with $M_{\rm h} > 5 \times 10^{13}{\rm M_\odot}$, it is defined as:
\begin{multline}
          F_{\rm gas}(M_{\rm h} > 5 \times 10^{13}{\rm M_\odot}) \equiv   \\ 
          { \bigintssss_{5 \times 10^{13}{\rm M_\odot}}^{2 \times 10^{15}{\rm M_\odot}} f_{\rm gas}(M_{\rm h}) \; M_{\rm h} \; {{\rm d}n / {\rm d}M_{\rm h}}  \; {\rm d}M_{\rm h} \over  \bigintssss_{10^{11}{\rm M_\odot}}^{2 \times 10^{15}{\rm M_\odot}}  M_{\rm h} \; {{\rm d}n / {\rm d}M_{\rm h}} \; {\rm d}M_{\rm h}} \, .
        \label{eq:f_icgm} 
\end{multline}
Formally, the upper integration limit for numerator and denominator and the lower integration limit for the denominator should be set to infinity and zero respectively, we selected high and low values and made sure that $F_{\rm gas}(M_{\rm h} > 5 \times 10^{13}{\rm M_\odot})$ depends very weakly on them. Although the boundary for galaxy group masses is typically set at $10^{13}{\rm M_\odot}$, in light of the dearth of gas fraction measurements at the low mass end, we halted our integration at $5 \times 10^{13}{\rm M_\odot}$. We take $f_{\rm gas}$ from Eq. \ref{eq:f_gas} and, from Eq. \ref{eq:f_icgm}, estimate $F_{\rm gas}(M_{\rm h} > 5 \times 10^{13}{\rm M_\odot}) = 0.19-0.23$. We plug this into Eq. \ref{eq:f_fe},  along with $Z^{\rm ICM+IGrM}_{\rm Fe,\odot} = 0.33 \pm 0.07$,  a weighted mean of the cluster \citep{Ghizzardi:2021} and group \citep{Lovisari:2019} measurements, and $\langle Z^{\rm b}_{\rm Fe,\odot} \rangle$  from Eq. \ref{eq:z_u3}  and derive  $\phi_{ICM+IGrM} = 0.08-0.19$.

\begin{figure}
        \centerline{\includegraphics[angle=0,width=8.8cm]{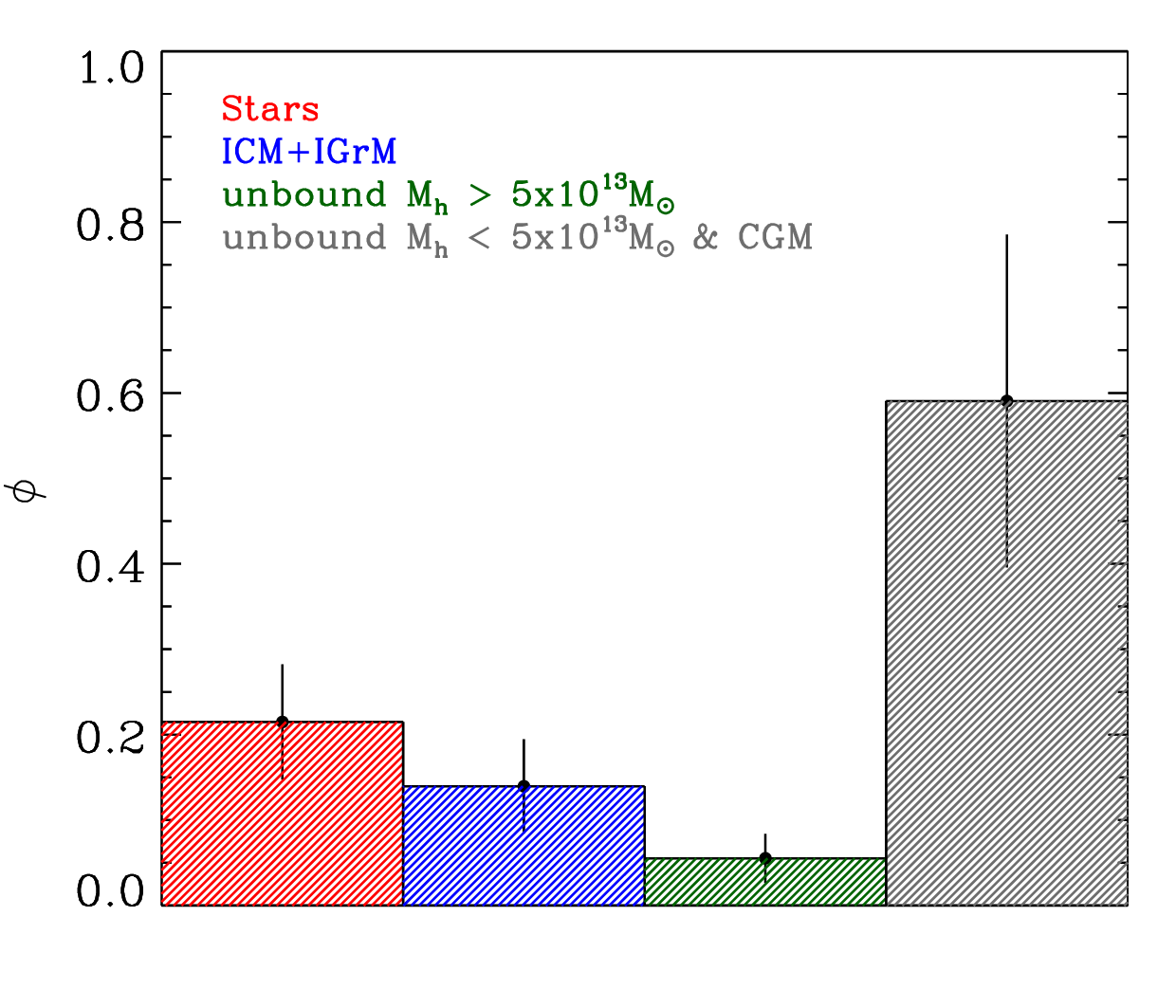}}
        \caption{Cosmic iron budget: the fraction of Fe locked in stars and in the hot gas in massive halos are shown in red and blue respectively. The fraction associated to gas required to reach the baryon fraction in massive halos ($M_{\rm h} > 5 \times 10^{13}{\rm M_\odot}$) is plotted in green. In gray we show the unaccounted fraction, this Fe must reside in gas either in the halos of low mass systems (CGM) or outside them.
        }
        \label{fig:fe_share}
\end{figure}
As discussed in Sect. \ref{sec:goutskirts}, comparison of cluster with group metal abundances allows us to infer that the unbound gas must be enriched with metals to a degree similar to the bound one. By assuming that the missing gas on group and cluster scales, i.e. the gas required to reach the cosmic baryon fraction on such scales, has a metal abundance comparable to that of gas bound to those systems, we estimate that a small fraction of Fe resides in gas outside halos, i.e. $\phi_{\rm l}(M_{\rm h} > 5 \times 10^{13}{\rm M_\odot})=0.02-0.08$. This is likely a very conservative lower limit to the Fe content of gas outside halos as we do not include the contribution of the unbound gas required to reach the baryon fraction for halos smaller than $  5 \times 10^{13}$M$_\odot$. This leaves us with anywhere between two-fifths and four-fifths of  the cosmic Fe to be divided between low mass halos and the unbound gas required to reach the cosmic baryon fraction for them,  $\phi_{\rm l}(M_{\rm h} < 5 \times 10^{13}{\rm M_\odot}) + \phi_{\rm CGM}=0.40-0.79$.  Although we cannot say which of these two components dominates, since the metallicity of the bound and free gas are not likely to be dissimilar, the issue of whether the bulk of metals in the Universe is bound or not is intimately connected to the one of whether most of the gas in the Universe is bound or not. In Fig. \ref{fig:fe_share} we show, in graphic form, the split of the cosmic Fe between different constituents. 

There are two main conclusions to be drawn from the analysis presented in this section. First: about half of the metals in the Universe have yet to be measured, they are most likely in a tenuous, ionized and gaseous medium either in or between galaxies; second, our first conclusion  provides much needed observational support to one of the most common assumptions that have been made to get to this gas, namely that  metal lines, either in emission \citep[see][and refs. therein]{Parimbelli:2022}, or absorption  \citep[see][and refs. therein]{Nicastro:2023}, are associated to it. The lines have to be there because the metals are there. On average metallicities need to be at about one-third solar in the unbound gas and higher, likely close to  solar, in the CGM, see Sect. \ref{sec:coutskirts}.

It is of some interest to compare our estimate of the cosmic metal budget with others. 
\citet{Peeples:2014}, by performing a detailed evaluation of the metal yields for different sources and elements at $z \sim 0 $, estimate the fraction of metals in stars, for stellar masses ranging from $10^{9}{\rm M_\odot}$ to $3 \times 10^{11}{\rm M_\odot}$,  to be between 10\% and 40\% (see their Fig. 5). Recently, \cite{Sanders:2023} found a similar result at  $z \sim 2 $ albeit with a somewhat improved statistics.
Quite remarkably, these estimates are consistent with ours, which is based on an entirely independent approach and focuses on one element only, namely, Fe. 
  
\citet{Peroux:2020}, by adopting the yields derived by \cite{Peeples:2014}, came up with an estimate of the fraction of one-half for the metal content in stars  (see their Fig. 9), which is more than twice the value inferred by \cite{Peeples:2014}. 
What is also puzzling is that \citet{Peroux:2020} simply stated that:  "at $z \sim 0.1$, half of the metals are in stars," without providing any estimate on the uncertainties. \citet{Lehner:2019}, by characterizing the metallicities and physical properties of cool, photo-ionized gas presumably located in and around halos, estimated that this gas contains no more than a few percent of the metals expelled from galaxies. By comparing this estimate with ours,  namely, about one-half of the metals in gas in and around low mass halos, we conclude, as other authors have done previously \citep[e.g.][]{Tumlinson:2017,Lehner:2019},  that the missing metals likely reside in a low density and high ionization gas, which cannot be traced through optical/UV spectroscopy. We do want to highlight that, as far as we know,  this is the first time such a conclusion has been reached on the basis of observational, albeit indirect, evidence, rather than on simulations or theoretical considerations.

\section{Future prospects}\label{sec:prospects}
In this work we have made a connection between stellar assembly in DM halos and observed properties of the Fe abundance to derive a description of the enrichment process in apex accretors. The construct we have built is simple and novel, however, as such, it is susceptible to confirmation and, of course, falsification.
There are several lines of research worthy of further investigation and we take a look at the most promising ones below.

As discussed in Sect. \ref{sec:scatter} the scatter on the metal abundance in the outer regions of massive clusters is smaller than 15\%. This number comes from the analysis of the XCOP sample \citep{Ghizzardi:2021}; more specifically, it is a measure of the total scatter. In that paper we did not provide an assessment of the intrinsic scatter because we estimated that the systematic errors on our measurements could bias it substantially.
A  major contributor to these systematics is the limited knowledge of the instrumental background. Current activities within the framework of the CHEX-MATE heritage program \citep{CM:2021} are leading to a significant reduction in background associated systematics on surface brightness \citep{Bartalucci:2023} and temperature (Rossetti et al., in prep) measurements. If, as we hope, these improvements will have an impact on abundance measurements as well, it may be possible to revisit the XCOP cluster measurements and provide an actual estimate of the scatter. This will allow us to provide better estimates on the dispersion of parameters such as $f_{\star}$, $f_{\rm b}$, and $r_{o} \mathcal{Y}_{\rm Fe} $ and, in turn, to  constraint  the properties of the low mass halos where the bulk of metals are synthesized, see Sect. \ref{sec:scatter}.

Further progress will become possible through high resolution spectroscopy over a large ($\sim$1 sq. deg.) solid angle, something that in the foreseeable future can only be attempted through missions like ATHENA that combine a high-resolution spectrometer \citep[XIFU,][]{Barret:2022} and a wide-field imager \citep[WFI,][]{Meidinger:2020}. The XIFU will allow for the thermal structure to characterized, thereby permitting the construction of a detailed and largely unbiased spectral model of the ICM. The WFI can then be used to apply such a model to larger regions. As for current instrumentation, background associated systematics will need to be kept under tight control \citep[see][]{Molendi:2017}. Unlike current instrumentation, both XIFU \citep{Cucchetti:2018,Lotti:2021}and WFI \citep{Miller:2022} have been designed from the start to maximize background reproducibility. Unfortunately, it appears that, within the current ATHENA reformulation process, a key element in the background characterization strategy, the Athena High Energy Particle Monitor (AHEPaM) may be canceled. This would 
result in a substantial reduction in background reproducibility impacting this and many other science cases.

As discussed in Sect. \ref{sec:goutskirts},  the  metal abundance in less massive systems, namely, poor clusters and groups, is known to a lesser extent than in more massive ones. In this case, improvements will come from a better characterization of the foregrounds and by breaking the degeneracy between normalization and L-shell derived abundance \citep[e.g.,][]{Buote:2000,MG:2001}. While some progress on the first issue can be made with currently available data (Riva et al in prep.), the second can only be addressed through higher spectral resolution observations afforded by the soon-to-be-launched XRISM \citep{Tashiro_XRISM:2018} and XIFU \citep{Barret:2022}. A key point that will have to be addressed is whether the metal abundance profiles of groups in outer regions are flat, as those observed in clusters, or feature a sudden and significant decrement, as observed in a few systems \citep{Lovisari:2019,Sarkar:2022}. Flat profiles would be in agreement with the model presented here while steeply declining ones would probably not.
 
Possibly the most powerful and at the same time easily tested prediction of the model presented in this paper is the "Universal abundance" prediction, that is that metal abundance in the outskirts is roughly the same for all masses and all redshifts respectively above and below a given threshold. This can be tested by comparing the abundance  of local poor clusters with that of high redshift ($z\sim 1$) massive clusters.

In Sect. \ref{sec:budget}, we have shown, for the first time on the basis of  observations, that the ionized gas in and between galaxies contains roughly half of all the metals in the Universe. This result highlights the need for instruments such as XIFU \citep{Barret:2022}  and the line emission mapper probe \citep[LEM][]{Kraft:2022} capable of performing the first direct measurements of these metals. 

From the stellar side, improved constraints on the SHMR for centrals and satellites at $z > 1$ would be quite important. As noted in \cite{Shuntov:2022}, current measurements of ex-situ component are not very constraining. Better data would help improve constraints on  the enrichment process of high redshift halos. Direct measurements of stellar mass in massive halos as a function of redshift  would also be very useful. They would allow confirmation of the work by \cite{Chiu:2018}  and hopefully help test the limits of the simple construct we provide here.

\section{Summary}\label{sec:summary}
In this work, we present a construct connecting stellar assembly in DM halos with observed properties of the Fe abundance of the hot gas in massive systems. Our main findings may be summarized as follows.
\begin{itemize}
                
        \item By requiring consistency between predicted and measured Fe abundances in massive clusters, we derive constraints on the efficiency with which stars produce iron. This, in turn, allows us to estimate that only between a 1/7 and 1/4 of the Fe produced in stars is still locked within them, the bulk is dispersed in gas. 
        
        \item The similarity between metal abundance in clusters and groups implies that gas accreted on the latter as they grow into the former cannot be pristine. That is to say, it must have been previously accreted and enriched in  lower mass halos, namely, of $\sim 10^{12}$M$_{\odot}$. This leads to the unavoidable conclusion that the gas not bound to halos must have a metal abundance that is similar to the one measured in clusters.
        
        \item Although quantities such as $f_\star$ and $r_{o} \mathcal{Y}_{\rm Fe} $, from which metal abundance in massive clusters have been estimated, are currently characterized by significant measurement uncertainties, the small scatter observed in the Fe abundance suggests that they too must be characterized by a modest intrinsic dispersion. This result underscores the validity of an approach such as ours which is based on simple equations relating population averaged properties. 
        
        \item The small scatter in metal abundance in clusters arises from the averaging of hundreds of independent enrichment events originally occurring in halos with masses in the $10^{12}$M$_\odot$ , $10^{13}$ M$_\odot$ range.
        
        \item We provide the first explanation for the Fe abundance versus entropy anti-correlation observed in cluster and groups. The high-entropy, low-abundance gas in the outskirts was previously enriched in smaller halos and acquired its high entropy as it was shock-heated during the accretion onto the main halo. The high-metallicity, low-entropy gas originates from the enrichment process within the progenitor and  a few other star-forming halos. 
        This gas did not suffer shock heating, either because it was enriched in the progenitor halo or because it was originally within the core of an accreted sub-halo that was not shock heated by the merging process.
        
        \item Lack of redshift evolution in cluster outskirt abundance is explained in terms of the self similarity of star formation and the associated enrichment process over a broad redshift range. The proportionality between stellar mass fraction and Fe abundance, on which we base  much of our work, implies that the lack of variation of  $Z^{\rm ICM}_{\rm Fe}$ on $z$ should also hold true for $f_\star$ on  $z$. This has indeed been observed.
        
        \item If, as accumulating evidence suggests, mixing and conduction are heavily suppressed in clusters, we can look at abundance radial profiles very much like geologists look at layers of rock. Within this framework, the absence of substantial radial abundance gradients beyond core regions in massive clusters results from the lack of halo mass and redshift dependence of the metal abundance. 
        
        \item Lack of substantial abundance ratio differences between core and circum-core regions are the consequence of the relatively rapid expulsion of both SNcc and SNIa ejecta. Moreover,  since the metal mass at the center originates from a much smaller number of star-forming halos then in the outskirts, we expect the scatter in abundance ratios to decrease as we move out from the core. 
        
        \item We provide a solution to the so called Fe conundrum, namely, the difference between measured and predicted Fe yield at cluster scales, by requiring that: 1) the stellar mass fraction in massive halos is underestimated;
        2) the gas mass fraction of less massive halos ($<$ a few $10^{14}$ M$_\odot$) is also  underestimated;
and        3) the yield is only marginally consistent with predictions from synthesis models and SN rates and is largely independent of halo mass. 
        
        \item Several arguments suggest that the efficiency of Fe production measured in massive halos also applies to less massive ones. By following this approach we find that in the Universe, no more than $\sim 1/3$ of the Fe is locked in stars, while about 1/2 has yet to be measured. It likely resides in a tenuous warm or hot gaseous medium either in or between galaxies.
        
        \item We highlight the connection between metal yield, universe abundance and enrichment of the warm and hot gaseous medium. A small yield ($\mathcal{Y}_{{\rm Fe},\odot} \sim 1$), assumed in previous studies, is associated  to a metallicity of the Universe of $\sim$ 0.1 and the bulk of metals  in stars; the larger yield ($\mathcal{Y}_{{\rm Fe},\odot} \sim 3 - 6$), which we propose, is associated to a metallicity of the Universe in the range 0.3-0.8 and the bulk of metals in warm and hot gas.
                
\end{itemize}

The connection, that we have built between stellar assembly in small mass halos and Fe abundance of apex accretors is rich in terms of its predictive power. As such, it is susceptible to confirmation (or, conversely, invalidation) through further observational work at X-ray and longer wavelengths. On the X-ray side, although some progress can be made with  currently available data, key measurements (particularly for groups) will only become possible with the advent of high resolution spectrometers such as those on board XRISM and ATHENA.  

In closing this paper, we go back to the questions formulated in the introduction. With respect to\ where the feedback-driven self similarity comes from, the answer lies in the efficiency of the enrichment process which, within the bounds in redshift and halo mass that current data allows us to explore, is consistent with being constant.  
With respect to the mechanism that produces constant and low scatter abundance profiles in cluster and groups, we find that the lack of radial gradients in cluster and group profiles is likely to be associated with the roughly constant abundance of the accreted gas. The low scatter follows from the large ratio between the mass of the halos under investigation and that of those responsible for the bulk of star formation.
\begin{acknowledgements}
We are grateful to Dan Maoz, Alvio Renzini, Aurora Simionescu and Francois Mernier for useful comments. We  thank the referee for valid suggestion that helped improve our work. SM thanks Nico Molendi for the artwork. We acknowledge financial support from INAF mainstream project No.1.05.01.86.13. We acknowledge use of the ACCEPT Archive.
\end{acknowledgements}

\bibliography{biblio_icm_enrichment}

\begin{thebibliography}{108}
\expandafter\ifx\csname natexlab\endcsname\relax\def\natexlab#1{#1}\fi

\bibitem[{{Anders} \& {Grevesse}(1989)}]{AG:1989}
{Anders}, E. \& {Grevesse}, N. 1989, \gca, 53, 197

\bibitem[{{Andreon}(2012)}]{Andreon:2012}
{Andreon}, S. 2012, \aap, 548, A83

\bibitem[{{Artale} {et~al.}(2022){Artale}, {Haider}, {Montero-Dorta},
  {Vogelsberger}, {Martizzi}, {Torrey}, {Bird}, {Hernquist}, \&
  {Marinacci}}]{Artale:2022}
{Artale}, M.~C., {Haider}, M., {Montero-Dorta}, A.~D., {et~al.} 2022, \mnras,
  510, 399

\bibitem[{{Ascasibar} \& {Markevitch}(2006)}]{AM06}
{Ascasibar}, Y. \& {Markevitch}, M. 2006, \apj, 650, 102

\bibitem[{{Asplund} {et~al.}(2009){Asplund}, {Grevesse}, {Sauval}, \&
  {Scott}}]{Asplund:2009}
{Asplund}, M., {Grevesse}, N., {Sauval}, A.~J., \& {Scott}, P. 2009, \araa, 47,
  481

\bibitem[{{Baldi} {et~al.}(2012){Baldi}, {Ettori}, {Molendi}, {Balestra},
  {Gastaldello}, \& {Tozzi}}]{Baldi:2012}
{Baldi}, A., {Ettori}, S., {Molendi}, S., {et~al.} 2012, \aap, 537, A142

\bibitem[{{Balestra} {et~al.}(2007){Balestra}, {Tozzi}, {Ettori}, {Rosati},
  {Borgani}, {Mainieri}, {Norman}, \& {Viola}}]{Balestra:2007}
{Balestra}, I., {Tozzi}, P., {Ettori}, S., {et~al.} 2007, \aap, 462, 429

\bibitem[{{Barret} {et~al.}(2022){Barret}, {Albouys}, {den Herder}, {Piro},
  {Cappi}, {Huovelin}, {Kelley}, {Mas-Hesse}, {Paltani}, {Rauw}, {Rozanska},
  {Svoboda}, {Wilms}, {Yamasaki}, {Audard}, {Bandler}, {Barbera}, {Barcons},
  {Bozzo}, {Ceballos}, {Charles}, {Costantini}, {Dauser}, {Decourchelle},
  {Duband}, {Duval}, {Fiore}, {Gatti}, {Goldwurm}, {den Hartog}, {Jackson},
  {Jonker}, {Kilbourne}, {Korpela}, {Macculi}, {Mendez}, {Mitsuda}, {Molendi},
  {Pajot}, {Pointecouteau}, {Porter}, {Pratt}, {Pr{\^e}le}, {Ravera}, {Sato},
  {Schaye}, {Shinozaki}, {Skup}, {Soucek}, {Thibert}, {Vink}, {Webb}, {Chaoul},
  {Raulin}, {Simionescu}, {Torrejon}, {Acero}, {Branduardi-Raymont}, {Ettori},
  {Finoguenov}, {Grosso}, {Kaastra}, {Mazzotta}, {Miller}, {Miniutti},
  {Nicastro}, {Sciortino}, {Yamaguchi}, {Beaumont}, {Cucchetti}, {D'Andrea},
  {Eckart}, {Ferrando}, {Kammoun}, {Lotti}, {Mesnager}, {Natalucci}, {Peille},
  {de Plaa}, {Ardellier}, {Argan}, {Bellouard}, {Carron}, {Cavazzuti},
  {Fiorini}, {Khosropanah}, {Martin}, {Perry}, {Pinsard}, {Pradines}, {Rigano},
  {Roelfsema}, {Schwander}, {Torrioli}, {Ullom}, {Vera}, {Medinaceli Villegas},
  {Zuchniak}, {Brachet}, {Lo Cicero}, {Doriese}, {Durkin}, {Fioretti},
  {Geoffray}, {Jacques}, {Kirsch}, {Smith}, {Adams}, {Gloaguen}, {Hoogeveen},
  {van der Hulst}, {Kiviranta}, {van der Kuur}, {Ledot}, {van Leeuwen}, {van
  Loon}, {Lyautey}, {Parot}, {Sakai}, {van Weers}, {Abdoelkariem}, {Adam},
  {Adami}, {Aicardi}, {Akamatsu}, {Eleazar Merino Alonso}, {Amato},
  {Andr{\'e}}, {Angelinelli}, {Anon-Cancela}, {Anvar}, {Atienza}, {Attard},
  {Auricchio}, {Balado}, {Bancel}, {Ferrari Barusso}, {Bernard}, {Berrocal},
  {Blin}, {Bonino}, {Bonnet}, {Bonny}, {Boorman}, {Boreux}, {Bounab},
  {Boutelier}, {Boyce}, {Brienza}, {Bruijn}, {Bulgarelli}, {Calarco},
  {Callanan}, {Camus}, {Canourgues}, {Capobianco}, {Cardiel}, {Castellani},
  {Cheatom}, {Chervenak}, {Chiarello}, {Clerc}, {Clerc}, {Cobo}, {Coeur-Joly},
  {Coleiro}, {Colonges}, {Corcione}, {Coriat}, {Coynel}, {Cuttaia}, {D'Ai},
  {D'anca}, {Dadina}, {Daniel}, {DeNigris}, {Dercksen}, {DiPirro}, {Doumayrou},
  {Dubbeldam}, {Dupieux}, {Dupourqu{\'e}}, {Durand}, {Eckert}, {Eiriz},
  {Ercolani}, {Etcheverry}, {Finkbeiner}, {Fiocchi}, {Fossecave}, {Franssen},
  {Frericks}, {Gabici}, {Gant}, {Gao}, {Gastaldello}, {Genolet}, {Ghizzardi},
  {Alcacera Gil}, {Giovannini}, {Godet}, {Gomez-Elvira}, {Gonzalez},
  {Gonzalez}, {Gottardi}, {Granat}, {Gros}, {Guignard}, {Hieltjes}, {Hurtado},
  {Irwin}, {Jacquey}, {Janiuk}, {Jaubert}, {Jim{\'e}nez}, {Jolly}, {Jourdan},
  {Julien}, {Kedziora}, {Korb}, {Kreykenbohm}, {K{\"o}nig}, {Langer}, {Laudet},
  {Laurent}, {Laurenza}, {Lesrel}, {Ligori}, {Lorenz}, {Luminari}, {Maffei},
  {Maisonnave}, {Marelli}, {Massonet}, {Maussang}, {Gonzalo Melchor}, {Le Mer},
  {Michalski}, {Millerioux}, {Mineo}, {Minervini}, {Molin}, {Monestes},
  {Montinaro}, {Mot}, {Murat}, {Nagayoshi}, {Naz{\'e}}, {Nogu{\`e}s}, {Pailot},
  {Panessa}, {Parodi}, {Petit}, {Piconcelli}, {Pinto}, {Encinas Plaza},
  {Poyatos}, {Prouv{\'e}}, {Ptak}, {Puccetti}, {Puccio}, {Ramon}, {Reina},
  {Rioland}, {Rodriguez}, {Roig}, {Rollet}, {Roncarelli}, {Roudil}, {Rudnicki},
  {Sanisidro}, {Sciortino}, {Silva}, {Sordet}, {Soto-Aguilar}, {Spizzi},
  {Surace}, {Fern{\'a}ndez S{\'a}nchez}, {Taralli}, {Terrasa}, {Terrier},
  {Todaro}, {Ubertini}, {Uslenghi}, {Geralt Bij de Vaate}, {Vaccaro},
  {Varisco}, {Varni{\`e}re}, {Vibert}, {Vidriales}, {Villa}, {Vodopivec},
  {Volpe}, {de Vries}, {Wakeham}, {Walmsley}, {Wise}, {de Wit}, \&
  {Wo{\'z}niak}}]{Barret:2022}
{Barret}, D., {Albouys}, V., {den Herder}, J.-W., {et~al.} 2022, arXiv
  e-prints, arXiv:2208.14562

\bibitem[{{Bartalucci} {et~al.}(2023){Bartalucci}, {Molendi}, {Rasia}, {Pratt},
  {Arnaud}, {Rossetti}, {Gastaldello}, {Eckert}, {Balboni}, {Borgani},
  {Bourdin}, {Campitiello}, {De Grandi}, {De Petris}, {Duffy}, {Ettori},
  {Ferragamo}, {Gaspari}, {Gavazzi}, {Ghizzardi}, {Iqbal}, {Kay}, {Lovisari},
  {Mazzotta}, {Maughan}, {Pointecouteau}, {Riva}, \&
  {Sereno}}]{Bartalucci:2023}
{Bartalucci}, I., {Molendi}, S., {Rasia}, E., {et~al.} 2023, arXiv e-prints,
  arXiv:2305.03082

\bibitem[{{Behroozi} {et~al.}(2019){Behroozi}, {Wechsler}, {Hearin}, \&
  {Conroy}}]{Behroozi:2019}
{Behroozi}, P., {Wechsler}, R.~H., {Hearin}, A.~P., \& {Conroy}, C. 2019,
  \mnras, 488, 3143

\bibitem[{{Behroozi} {et~al.}(2010){Behroozi}, {Conroy}, \&
  {Wechsler}}]{Behroozi:2010}
{Behroozi}, P.~S., {Conroy}, C., \& {Wechsler}, R.~H. 2010, \apj, 717, 379

\bibitem[{{Behroozi} {et~al.}(2013){Behroozi}, {Wechsler}, \&
  {Conroy}}]{Behroozi:2013}
{Behroozi}, P.~S., {Wechsler}, R.~H., \& {Conroy}, C. 2013, \apj, 770, 57

\bibitem[{{Biffi} {et~al.}(2018){Biffi}, {Planelles}, {Borgani}, {Rasia},
  {Murante}, {Fabjan}, \& {Gaspari}}]{Biffi:2018}
{Biffi}, V., {Planelles}, S., {Borgani}, S., {et~al.} 2018, \mnras, 476, 2689

\bibitem[{{Blackwell} {et~al.}(2022){Blackwell}, {Bregman}, \&
  {Snowden}}]{Blackwell:2022}
{Blackwell}, A.~E., {Bregman}, J.~N., \& {Snowden}, S.~L. 2022, \apj, 927, 104

\bibitem[{{Bregman} {et~al.}(2010){Bregman}, {Anderson}, \&
  {Dai}}]{Bregman:2010}
{Bregman}, J.~N., {Anderson}, M.~E., \& {Dai}, X. 2010, \apjl, 716, L63

\bibitem[{{Buote}(2000)}]{Buote:2000}
{Buote}, D.~A. 2000, \mnras, 311, 176

\bibitem[{{Cavagnolo} {et~al.}(2009){Cavagnolo}, {Donahue}, {Voit}, \&
  {Sun}}]{Cavagnolo:2009}
{Cavagnolo}, K.~W., {Donahue}, M., {Voit}, G.~M., \& {Sun}, M. 2009, \apjs,
  182, 12

\bibitem[{{CHEX-MATE Collaboration} {et~al.}(2021){CHEX-MATE Collaboration},
  {Arnaud}, {Ettori}, {Pratt}, {Rossetti}, {Eckert}, {Gastaldello}, {Gavazzi},
  {Kay}, {Lovisari}, {Maughan}, {Pointecouteau}, {Sereno}, {Bartalucci},
  {Bonafede}, {Bourdin}, {Cassano}, {Duffy}, {Iqbal}, {Maurogordato}, {Rasia},
  {Sayers}, {Andrade-Santos}, {Aussel}, {Barnes}, {Barrena}, {Borgani},
  {Burkutean}, {Clerc}, {Corasaniti}, {Cuillandre}, {De Grandi}, {De Petris},
  {Dolag}, {Donahue}, {Ferragamo}, {Gaspari}, {Ghizzardi}, {Gitti}, {Haines},
  {Jauzac}, {Johnston-Hollitt}, {Jones}, {K{\'e}ruzor{\'e}}, {Le Brun},
  {Mayet}, {Mazzotta}, {Melin}, {Molendi}, {Nonino}, {Okabe}, {Paltani},
  {Perotto}, {Pires}, {Radovich}, {Rubino-Martin}, {Salvati}, {Saro},
  {Sartoris}, {Schellenberger}, {Streblyanska}, {Tarr{\'\i}o}, {Tozzi},
  {Umetsu}, {van der Burg}, {Vazza}, {Venturi}, {Yepes}, \&
  {Zarattini}}]{CM:2021}
{CHEX-MATE Collaboration}, {Arnaud}, M., {Ettori}, S., {et~al.} 2021, \aap,
  650, A104

\bibitem[{{Chiu} {et~al.}(2018){Chiu}, {Mohr}, {McDonald}, {Bocquet}, {Desai},
  {Klein}, {Israel}, {Ashby}, {Stanford}, {Benson}, {Brodwin}, {Abbott},
  {Abdalla}, {Allam}, {Annis}, {Bayliss}, {Benoit-L{\'e}vy}, {Bertin}, {Bleem},
  {Brooks}, {Buckley-Geer}, {Bulbul}, {Capasso}, {Carlstrom}, {Rosell},
  {Carretero}, {Castander}, {Cunha}, {D'Andrea}, {da Costa}, {Davis}, {Diehl},
  {Dietrich}, {Doel}, {Drlica-Wagner}, {Eifler}, {Evrard}, {Flaugher},
  {Garc{\'\i}a-Bellido}, {Garmire}, {Gaztanaga}, {Gerdes}, {Gonzalez}, {Gruen},
  {Gruendl}, {Gschwend}, {Gupta}, {Gutierrez}, {Hlavacek-L}, {Honscheid},
  {James}, {Jeltema}, {Kraft}, {Krause}, {Kuehn}, {Kuhlmann}, {Kuropatkin},
  {Lahav}, {Lima}, {Maia}, {Marshall}, {Melchior}, {Menanteau}, {Miquel},
  {Murray}, {Nord}, {Ogando}, {Plazas}, {Rapetti}, {Reichardt}, {Romer},
  {Roodman}, {Sanchez}, {Saro}, {Scarpine}, {Schindler}, {Schubnell}, {Sharon},
  {Smith}, {Smith}, {Soares-Santos}, {Sobreira}, {Stalder}, {Stern},
  {Strazzullo}, {Suchyta}, {Swanson}, {Tarle}, {Vikram}, {Walker}, {Weller}, \&
  {Zhang}}]{Chiu:2018}
{Chiu}, I., {Mohr}, J.~J., {McDonald}, M., {et~al.} 2018, \mnras, 478, 3072

\bibitem[{{Chu} {et~al.}(2021){Chu}, {Durret}, \& {M{\'a}rquez}}]{Chu:2021}
{Chu}, A., {Durret}, F., \& {M{\'a}rquez}, I. 2021, \aap, 649, A42

\bibitem[{{Chu} {et~al.}(2022){Chu}, {Sarron}, {Durret}, \&
  {M{\'a}rquez}}]{Chu:2022}
{Chu}, A., {Sarron}, F., {Durret}, F., \& {M{\'a}rquez}, I. 2022, \aap, 666,
  A54

\bibitem[{{Comparat} {et~al.}(2022){Comparat}, {Truong}, {Merloni},
  {Pillepich}, {Ponti}, {Driver}, {Bellstedt}, {Liske}, {Aird}, {Br{\"u}ggen},
  {Bulbul}, {Davies}, {Villalba}, {Georgakakis}, {Haberl}, {Liu}, {Maitra},
  {Nandra}, {Popesso}, {Predehl}, {Robotham}, {Salvato}, {Thorne}, \&
  {Zhang}}]{Comparat:2022}
{Comparat}, J., {Truong}, N., {Merloni}, A., {et~al.} 2022, \aap, 666, A156

\bibitem[{{Contini}(2021)}]{Contini:2021}
{Contini}, E. 2021, Galaxies, 9, 60

\bibitem[{{Coupon} {et~al.}(2015){Coupon}, {Arnouts}, {van Waerbeke},
  {Moutard}, {Ilbert}, {van Uitert}, {Erben}, {Garilli}, {Guzzo}, {Heymans},
  {Hildebrandt}, {Hoekstra}, {Kilbinger}, {Kitching}, {Mellier}, {Miller},
  {Scodeggio}, {Bonnett}, {Branchini}, {Davidzon}, {De Lucia}, {Fritz}, {Fu},
  {Hudelot}, {Hudson}, {Kuijken}, {Leauthaud}, {Le F{\`e}vre}, {McCracken},
  {Moscardini}, {Rowe}, {Schrabback}, {Semboloni}, \& {Velander}}]{Coupon:2015}
{Coupon}, J., {Arnouts}, S., {van Waerbeke}, L., {et~al.} 2015, \mnras, 449,
  1352

\bibitem[{{Coupon} {et~al.}(2012){Coupon}, {Kilbinger}, {McCracken}, {Ilbert},
  {Arnouts}, {Mellier}, {Abbas}, {de la Torre}, {Goranova}, {Hudelot}, {Kneib},
  \& {Le F{\`e}vre}}]{Coupon:2012}
{Coupon}, J., {Kilbinger}, M., {McCracken}, H.~J., {et~al.} 2012, \aap, 542, A5

\bibitem[{{Cowley} {et~al.}(2018){Cowley}, {Caputi}, {Deshmukh}, {Ashby},
  {Fazio}, {Le F{\`e}vre}, {Fynbo}, {Ilbert}, {McCracken}, {Milvang-Jensen}, \&
  {Somerville}}]{Cowley:2018}
{Cowley}, W.~I., {Caputi}, K.~I., {Deshmukh}, S., {et~al.} 2018, \apj, 853, 69

\bibitem[{{Cucchetti} {et~al.}(2018){Cucchetti}, {Pointecouteau}, {Barret},
  {Lotti}, {Macculi}, {Molendi}, {Pajot}, {Peille}, {Piro}, \&
  {Pratt}}]{Cucchetti:2018}
{Cucchetti}, E., {Pointecouteau}, E., {Barret}, D., {et~al.} 2018, in Society
  of Photo-Optical Instrumentation Engineers (SPIE) Conference Series, Vol.
  10699, Space Telescopes and Instrumentation 2018: Ultraviolet to Gamma Ray,
  ed. J.-W.~A. {den Herder}, S.~{Nikzad}, \& K.~{Nakazawa}, 106994N

\bibitem[{{De Grandi} {et~al.}(2016){De Grandi}, {Eckert}, {Molendi},
  {Girardi}, {Roediger}, {Gaspari}, {Gastaldello}, {Ghizzardi}, {Nonino}, \&
  {Rossetti}}]{Degrandi:2016}
{De Grandi}, S., {Eckert}, D., {Molendi}, S., {et~al.} 2016, \aap, 592, A154

\bibitem[{{De Grandi} {et~al.}(2004){De Grandi}, {Ettori}, {Longhetti}, \&
  {Molendi}}]{Degrandi:2004}
{De Grandi}, S., {Ettori}, S., {Longhetti}, M., \& {Molendi}, S. 2004, \aap,
  419, 7

\bibitem[{{De Grandi} \& {Molendi}(2009)}]{DGM:2009}
{De Grandi}, S. \& {Molendi}, S. 2009, \aap, 508, 565

\bibitem[{{Dekel} \& {Birnboim}(2006)}]{Dekel:2006}
{Dekel}, A. \& {Birnboim}, Y. 2006, \mnras, 368, 2

\bibitem[{{Eckert} {et~al.}(2016){Eckert}, {Ettori}, {Coupon}, {Gastaldello},
  {Pierre}, {Melin}, {Le Brun}, {McCarthy}, {Adami}, {Chiappetti}, {Faccioli},
  {Giles}, {Lavoie}, {Lef{\`e}vre}, {Lieu}, {Mantz}, {Maughan}, {McGee},
  {Pacaud}, {Paltani}, {Sadibekova}, {Smith}, \& {Ziparo}}]{Eckert:2016}
{Eckert}, D., {Ettori}, S., {Coupon}, J., {et~al.} 2016, \aap, 592, A12

\bibitem[{{Eckert} {et~al.}(2021){Eckert}, {Gaspari}, {Gastaldello}, {Le Brun},
  \& {O'Sullivan}}]{Eckert:2021}
{Eckert}, D., {Gaspari}, M., {Gastaldello}, F., {Le Brun}, A. M.~C., \&
  {O'Sullivan}, E. 2021, Universe, 7, 142

\bibitem[{{Eckert} {et~al.}(2017){Eckert}, {Gaspari}, {Owers}, {Roediger},
  {Molendi}, {Gastaldello}, {Paltani}, {Ettori}, {Venturi}, {Rossetti}, \&
  {Rudnick}}]{Eckert:2017}
{Eckert}, D., {Gaspari}, M., {Owers}, M.~S., {et~al.} 2017, \aap, 605, A25

\bibitem[{{Eckert} {et~al.}(2019){Eckert}, {Ghirardini}, {Ettori}, {Rasia},
  {Biffi}, {Pointecouteau}, {Rossetti}, {Molendi}, {Vazza}, {Gastaldello},
  {Gaspari}, {De Grandi}, {Ghizzardi}, {Bourdin}, {Tchernin}, \&
  {Roncarelli}}]{Eckert_non_th_XCOP:2019}
{Eckert}, D., {Ghirardini}, V., {Ettori}, S., {et~al.} 2019, \aap, 621, A40

\bibitem[{{Eckert} {et~al.}(2014){Eckert}, {Molendi}, {Owers}, {Gaspari},
  {Venturi}, {Rudnick}, {Ettori}, {Paltani}, {Gastaldello}, \&
  {Rossetti}}]{Eckert:2014}
{Eckert}, D., {Molendi}, S., {Owers}, M., {et~al.} 2014, \aap, 570, A119

\bibitem[{{Edwards} {et~al.}(2020){Edwards}, {Salinas}, {Stanley}, {Holguin
  West}, {Trierweiler}, {Alpert}, {Coelho}, {Koppaka}, {Tremblay}, {Martel}, \&
  {Li}}]{Edwards:2020}
{Edwards}, L. O.~V., {Salinas}, M., {Stanley}, S., {et~al.} 2020, \mnras, 491,
  2617

\bibitem[{{Ettori} {et~al.}(2015){Ettori}, {Baldi}, {Balestra}, {Gastaldello},
  {Molendi}, \& {Tozzi}}]{Ettori:2015}
{Ettori}, S., {Baldi}, A., {Balestra}, I., {et~al.} 2015, \aap, 578, A46

\bibitem[{{Evrard} \& {Henry}(1991)}]{Evrard:1991}
{Evrard}, A.~E. \& {Henry}, J.~P. 1991, \apj, 383, 95

\bibitem[{{Flores} {et~al.}(2021){Flores}, {Mantz}, {Allen}, {Morris},
  {Canning}, {Bleem}, {Calzadilla}, {Floyd}, {McDonald}, \&
  {Ruppin}}]{Flores:2021}
{Flores}, A.~M., {Mantz}, A.~B., {Allen}, S.~W., {et~al.} 2021, \mnras, 507,
  5195

\bibitem[{{Freundlich} \& {Maoz}(2021)}]{Freundlich:2021}
{Freundlich}, J. \& {Maoz}, D. 2021, \mnras, 502, 5882

\bibitem[{{Fukugita} \& {Peebles}(2004)}]{Fukugita:2004}
{Fukugita}, M. \& {Peebles}, P.~J.~E. 2004, \apj, 616, 643

\bibitem[{{Gallazzi} {et~al.}(2014){Gallazzi}, {Bell}, {Zibetti}, {Brinchmann},
  \& {Kelson}}]{Gallazzi:2014}
{Gallazzi}, A., {Bell}, E.~F., {Zibetti}, S., {Brinchmann}, J., \& {Kelson},
  D.~D. 2014, \apj, 788, 72

\bibitem[{{Gastaldello} {et~al.}(2021){Gastaldello}, {Simionescu}, {Mernier},
  {Biffi}, {Gaspari}, {Sato}, \& {Matsushita}}]{Gasta:2021}
{Gastaldello}, F., {Simionescu}, A., {Mernier}, F., {et~al.} 2021, Universe, 7,
  208

\bibitem[{{Ghizzardi} {et~al.}(2014){Ghizzardi}, {De Grandi}, \&
  {Molendi}}]{Ghizzardi:2014}
{Ghizzardi}, S., {De Grandi}, S., \& {Molendi}, S. 2014, \aap, 570, A117

\bibitem[{{Ghizzardi} {et~al.}(2021{\natexlab{a}}){Ghizzardi}, {Molendi}, {van
  der Burg}, {De Grandi}, {Bartalucci}, {Gastaldello}, {Rossetti}, {Biffi},
  {Borgani}, {Eckert}, {Ettori}, {Gaspari}, {Ghirardini}, \&
  {Rasia}}]{Ghizzardi:2021}
{Ghizzardi}, S., {Molendi}, S., {van der Burg}, R., {et~al.}
  2021{\natexlab{a}}, \aap, 646, A92

\bibitem[{{Ghizzardi} {et~al.}(2021{\natexlab{b}}){Ghizzardi}, {Molendi}, {van
  der Burg}, {De Grandi}, {Bartalucci}, {Gastaldello}, {Rossetti}, {Biffi},
  {Borgani}, {Eckert}, {Ettori}, {Gaspari}, {Ghirardini}, \&
  {Rasia}}]{Ghizzardi:2021b}
{Ghizzardi}, S., {Molendi}, S., {van der Burg}, R., {et~al.}
  2021{\natexlab{b}}, \aap, 652, C3

\bibitem[{{Girelli} {et~al.}(2020){Girelli}, {Pozzetti}, {Bolzonella},
  {Giocoli}, {Marulli}, \& {Baldi}}]{Girelli:2020}
{Girelli}, G., {Pozzetti}, L., {Bolzonella}, M., {et~al.} 2020, \aap, 634, A135

\bibitem[{{Greggio} \& {Renzini}(2011)}]{Greggio:2011}
{Greggio}, L. \& {Renzini}, A. 2011, {Stellar Populations. A User Guide from
  Low to High Redshift}

\bibitem[{{Grylls} {et~al.}(2019){Grylls}, {Shankar}, {Zanisi}, \&
  {Bernardi}}]{Grylls:2019}
{Grylls}, P.~J., {Shankar}, F., {Zanisi}, L., \& {Bernardi}, M. 2019, \mnras,
  483, 2506

\bibitem[{{Hoyle}(1946)}]{Hoyle:1946}
{Hoyle}, F. 1946, \mnras, 106, 343

\bibitem[{{Kaiser}(1986)}]{Kaiser:1986}
{Kaiser}, N. 1986, \mnras, 222, 323

\bibitem[{{Kluge} \& {Bender}(2023)}]{Kluge:2023}
{Kluge}, M. \& {Bender}, R. 2023, arXiv e-prints, arXiv:2304.03527

\bibitem[{{Kraft} {et~al.}(2022){Kraft}, {Markevitch}, {Kilbourne}, {Adams},
  {Akamatsu}, {Ayromlou}, {Bandler}, {Barbera}, {Bennett}, {Bhardwaj}, {Biffi},
  {Bodewits}, {Bogdan}, {Bonamente}, {Borgani}, {Branduardi-Raymont},
  {Bregman}, {Burchett}, {Cann}, {Carter}, {Chakraborty}, {Churazov}, {Crain},
  {Cumbee}, {Dave}, {DiPirro}, {Dolag}, {Bertrand Doriese}, {Drake}, {Dunn},
  {Eckart}, {Eckert}, {Ettori}, {Forman}, {Galeazzi}, {Gall}, {Gatuzz}, {Hell},
  {Hodges-Kluck}, {Jackman}, {Jahromi}, {Jennings}, {Jones}, {Kaaret},
  {Kavanagh}, {Kelley}, {Khabibullin}, {Kim}, {Koutroumpa}, {Kovacs}, {Kuntz},
  {Lau}, {Lee}, {Leutenegger}, {Lin}, {Lisse}, {Lo Cicero}, {Lovisari},
  {McCammon}, {McEntee}, {Mernier}, {Miller}, {Nagai}, {Negro}, {Nelson},
  {Ness}, {Nulsen}, {Ogorzalek}, {Oppenheimer}, {Oskinova}, {Patnaude},
  {Pfeifle}, {Pillepich}, {Plucinsky}, {Pooley}, {Porter}, {Randall}, {Rasia},
  {Raymond}, {Ruszkowski}, {Sakai}, {Sarkar}, {Sasaki}, {Sato},
  {Schellenberger}, {Schaye}, {Simionescu}, {Smith}, {Steiner}, {Stern}, {Su},
  {Sun}, {Tremblay}, {Truong}, {Tutt}, {Ursino}, {Veilleux}, {Vikhlinin},
  {Vladutescu-Zopp}, {Vogelsberger}, {Walker}, {Weaver}, {Weigt}, {Werk},
  {Werner}, {Wolk}, {Zhang}, {Zhang}, {Zhuravleva}, \& {ZuHone}}]{Kraft:2022}
{Kraft}, R., {Markevitch}, M., {Kilbourne}, C., {et~al.} 2022, arXiv e-prints,
  arXiv:2211.09827

\bibitem[{{Leauthaud} {et~al.}(2012){Leauthaud}, {Tinker}, {Bundy}, {Behroozi},
  {Massey}, {Rhodes}, {George}, {Kneib}, {Benson}, {Wechsler}, {Busha},
  {Capak}, {Cort{\^e}s}, {Ilbert}, {Koekemoer}, {Le F{\`e}vre}, {Lilly},
  {McCracken}, {Salvato}, {Schrabback}, {Scoville}, {Smith}, \&
  {Taylor}}]{Leauthaud:2012}
{Leauthaud}, A., {Tinker}, J., {Bundy}, K., {et~al.} 2012, \apj, 744, 159

\bibitem[{{Leccardi} {et~al.}(2010){Leccardi}, {Rossetti}, \&
  {Molendi}}]{Leccardi:2010}
{Leccardi}, A., {Rossetti}, M., \& {Molendi}, S. 2010, \aap, 510, A82

\bibitem[{{Legrand} {et~al.}(2019){Legrand}, {McCracken}, {Davidzon}, {Ilbert},
  {Coupon}, {Aghanim}, {Douspis}, {Capak}, {Le F{\`e}vre}, \&
  {Milvang-Jensen}}]{Legrand:2019}
{Legrand}, L., {McCracken}, H.~J., {Davidzon}, I., {et~al.} 2019, \mnras, 486,
  5468

\bibitem[{{Lehner} {et~al.}(2019){Lehner}, {Wotta}, {Howk}, {O'Meara},
  {Oppenheimer}, \& {Cooksey}}]{Lehner:2019}
{Lehner}, N., {Wotta}, C.~B., {Howk}, J.~C., {et~al.} 2019, \apj, 887, 5

\bibitem[{{Lin} {et~al.}(2012){Lin}, {Stanford}, {Eisenhardt}, {Vikhlinin},
  {Maughan}, \& {Kravtsov}}]{Lin:2012}
{Lin}, Y.-T., {Stanford}, S.~A., {Eisenhardt}, P. R.~M., {et~al.} 2012, \apjl,
  745, L3

\bibitem[{{Liu} {et~al.}(2020){Liu}, {Tozzi}, {Ettori}, {De Grandi},
  {Gastaldello}, {Rosati}, \& {Norman}}]{Liu:2020}
{Liu}, A., {Tozzi}, P., {Ettori}, S., {et~al.} 2020, \aap, 637, A58

\bibitem[{{Loewenstein}(2013)}]{Lowenstein13}
{Loewenstein}, M. 2013, \apj, 773, 52

\bibitem[{{Lotti} {et~al.}(2021){Lotti}, {D'Andrea}, {Molendi}, {Macculi},
  {Minervini}, {Fioretti}, {Laurenza}, {Jacquey}, \& {Piro}}]{Lotti:2021}
{Lotti}, S., {D'Andrea}, M., {Molendi}, S., {et~al.} 2021, \apj, 909, 111

\bibitem[{{Lovisari} \& {Reiprich}(2019)}]{Lovisari:2019}
{Lovisari}, L. \& {Reiprich}, T.~H. 2019, \mnras, 483, 540

\bibitem[{{Madau} \& {Dickinson}(2014)}]{Madau:2014}
{Madau}, P. \& {Dickinson}, M. 2014, \araa, 52, 415

\bibitem[{{Mantz} {et~al.}(2017){Mantz}, {Allen}, {Morris}, {Simionescu},
  {Urban}, {Werner}, \& {Zhuravleva}}]{Mantz:2017}
{Mantz}, A.~B., {Allen}, S.~W., {Morris}, R.~G., {et~al.} 2017, \mnras, 472,
  2877

\bibitem[{{Maoz} \& {Graur}(2017)}]{Maoz:2017}
{Maoz}, D. \& {Graur}, O. 2017, \apj, 848, 25

\bibitem[{{Maraston}(2005)}]{Maraston:2005}
{Maraston}, C. 2005, \mnras, 362, 799

\bibitem[{{McDonald} {et~al.}(2016){McDonald}, {Bulbul}, {de Haan}, {Miller},
  {Benson}, {Bleem}, {Brodwin}, {Carlstrom}, {Chiu}, {Forman},
  {Hlavacek-Larrondo}, {Garmire}, {Gupta}, {Mohr}, {Reichardt}, {Saro},
  {Stalder}, {Stark}, \& {Vieira}}]{McDonald:2016}
{McDonald}, M., {Bulbul}, E., {de Haan}, T., {et~al.} 2016, \apj, 826, 124

\bibitem[{{Meidinger} {et~al.}(2020){Meidinger}, {Albrecht}, {Beitler},
  {Bonholzer}, {Emberger}, {Frank}, {Lederhuber}, {M{\"u}ller-Seidlitz},
  {Nandra}, {Oser}, {Ott}, {Plattner}, \& {Strecker}}]{Meidinger:2020}
{Meidinger}, N., {Albrecht}, S., {Beitler}, C., {et~al.} 2020, in Society of
  Photo-Optical Instrumentation Engineers (SPIE) Conference Series, Vol. 11444,
  Society of Photo-Optical Instrumentation Engineers (SPIE) Conference Series,
  114440T

\bibitem[{{Mernier} \& {Biffi}(2022)}]{Mernier:2022}
{Mernier}, F. \& {Biffi}, V. 2022, in Handbook of X-ray and Gamma-ray
  Astrophysics. Edited by Cosimo Bambi and Andrea Santangelo, 12

\bibitem[{{Mernier} {et~al.}(2017){Mernier}, {de Plaa}, {Kaastra}, {Zhang},
  {Akamatsu}, {Gu}, {Kosec}, {Mao}, {Pinto}, {Reiprich}, {Sanders},
  {Simionescu}, \& {Werner}}]{Mernier:2017}
{Mernier}, F., {de Plaa}, J., {Kaastra}, J.~S., {et~al.} 2017, \aap, 603, A80

\bibitem[{{Miller} {et~al.}(2022){Miller}, {Grant}, {Bautz}, {Molendi},
  {Kraft}, {Nulsen}, {Bulbul}, {Allen}, {Burrows}, {Eraerds}, {Fioretti},
  {Gastaldello}, {Hall}, {Hubbard}, {Keelan}, {Meidinger}, {Perinati}, {Rau},
  \& {Wilkins}}]{Miller:2022}
{Miller}, E.~D., {Grant}, C.~E., {Bautz}, M.~W., {et~al.} 2022, Journal of
  Astronomical Telescopes, Instruments, and Systems, 8, 018001

\bibitem[{{Mitchell} \& {Schaye}(2022)}]{Mitchell:2022}
{Mitchell}, P.~D. \& {Schaye}, J. 2022, \mnras, 511, 2600

\bibitem[{{Molendi}(2017)}]{Molendi:2017}
{Molendi}, S. 2017, Experimental Astronomy, 44, 263

\bibitem[{{Molendi} {et~al.}(2023){Molendi}, {De Grandi}, {Rossetti},
  {Bartalucci}, {Gastaldello}, {Ghizzardi}, \& {Gaspari}}]{Molendi:2023}
{Molendi}, S., {De Grandi}, S., {Rossetti}, M., {et~al.} 2023, \aap, 670, A104

\bibitem[{{Molendi} {et~al.}(2016){Molendi}, {Eckert}, {De Grandi}, {Ettori},
  {Gastaldello}, {Ghizzardi}, {Pratt}, \& {Rossetti}}]{Molendi:2016}
{Molendi}, S., {Eckert}, D., {De Grandi}, S., {et~al.} 2016, \aap, 586, A32

\bibitem[{{Molendi} \& {Gastaldello}(2001)}]{MG:2001}
{Molendi}, S. \& {Gastaldello}, F. 2001, \aap, 375, L14

\bibitem[{{Molendi} \& {Pizzolato}(2001)}]{Molendi:2001}
{Molendi}, S. \& {Pizzolato}, F. 2001, \apj, 560, 194

\bibitem[{{Moster} {et~al.}(2013){Moster}, {Naab}, \& {White}}]{Moster:2013}
{Moster}, B.~P., {Naab}, T., \& {White}, S. D.~M. 2013, \mnras, 428, 3121

\bibitem[{{Nandra} {et~al.}(2013){Nandra}, {Barret}, {Barcons}, {Fabian}, {den
  Herder}, {Piro}, {Watson}, {Adami}, {Aird}, {Afonso}, {Alexander},
  {Argiroffi}, {Amati}, {Arnaud}, {Atteia}, {Audard}, {Badenes}, {Ballet},
  {Ballo}, {Bamba}, {Bhardwaj}, {Stefano Battistelli}, {Becker}, {De Becker},
  {Behar}, {Bianchi}, {Biffi}, {B{\^\i}rzan}, {Bocchino}, {Bogdanov}, {Boirin},
  {Boller}, {Borgani}, {Borm}, {Bouch{\'e}}, {Bourdin}, {Bower}, {Braito},
  {Branchini}, {Branduardi-Raymont}, {Bregman}, {Brenneman}, {Brightman},
  {Br{\"u}ggen}, {Buchner}, {Bulbul}, {Brusa}, {Bursa}, {Caccianiga},
  {Cackett}, {Campana}, {Cappelluti}, {Cappi}, {Carrera}, {Ceballos},
  {Christensen}, {Chu}, {Churazov}, {Clerc}, {Corbel}, {Corral}, {Comastri},
  {Costantini}, {Croston}, {Dadina}, {D'Ai}, {Decourchelle}, {Della Ceca},
  {Dennerl}, {Dolag}, {Done}, {Dovciak}, {Drake}, {Eckert}, {Edge}, {Ettori},
  {Ezoe}, {Feigelson}, {Fender}, {Feruglio}, {Finoguenov}, {Fiore}, {Galeazzi},
  {Gallagher}, {Gandhi}, {Gaspari}, {Gastaldello}, {Georgakakis},
  {Georgantopoulos}, {Gilfanov}, {Gitti}, {Gladstone}, {Goosmann}, {Gosset},
  {Grosso}, {Guedel}, {Guerrero}, {Haberl}, {Hardcastle}, {Heinz}, {Alonso
  Herrero}, {Herv{\'e}}, {Holmstrom}, {Iwasawa}, {Jonker}, {Kaastra}, {Kara},
  {Karas}, {Kastner}, {King}, {Kosenko}, {Koutroumpa}, {Kraft}, {Kreykenbohm},
  {Lallement}, {Lanzuisi}, {Lee}, {Lemoine-Goumard}, {Lobban}, {Lodato},
  {Lovisari}, {Lotti}, {McCharthy}, {McNamara}, {Maggio}, {Maiolino}, {De
  Marco}, {de Martino}, {Mateos}, {Matt}, {Maughan}, {Mazzotta}, {Mendez},
  {Merloni}, {Micela}, {Miceli}, {Mignani}, {Miller}, {Miniutti}, {Molendi},
  {Montez}, {Moretti}, {Motch}, {Naz{\'e}}, {Nevalainen}, {Nicastro}, {Nulsen},
  {Ohashi}, {O'Brien}, {Osborne}, {Oskinova}, {Pacaud}, {Paerels}, {Page},
  {Papadakis}, {Pareschi}, {Petre}, {Petrucci}, {Piconcelli}, {Pillitteri},
  {Pinto}, {de Plaa}, {Pointecouteau}, {Ponman}, {Ponti}, {Porquet}, {Pounds},
  {Pratt}, {Predehl}, {Proga}, {Psaltis}, {Rafferty}, {Ramos-Ceja}, {Ranalli},
  {Rasia}, {Rau}, {Rauw}, {Rea}, {Read}, {Reeves}, {Reiprich}, {Renaud},
  {Reynolds}, {Risaliti}, {Rodriguez}, {Rodriguez Hidalgo}, {Roncarelli},
  {Rosario}, {Rossetti}, {Rozanska}, {Rovilos}, {Salvaterra}, {Salvato}, {Di
  Salvo}, {Sanders}, {Sanz-Forcada}, {Schawinski}, {Schaye}, {Schwope},
  {Sciortino}, {Severgnini}, {Shankar}, {Sijacki}, {Sim}, {Schmid}, {Smith},
  {Steiner}, {Stelzer}, {Stewart}, {Strohmayer}, {Str{\"u}der}, {Sun}, {Takei},
  {Tatischeff}, {Tiengo}, {Tombesi}, {Trinchieri}, {Tsuru}, {Ud-Doula},
  {Ursino}, {Valencic}, {Vanzella}, {Vaughan}, {Vignali}, {Vink}, {Vito},
  {Volonteri}, {Wang}, {Webb}, {Willingale}, {Wilms}, {Wise}, {Worrall},
  {Young}, {Zampieri}, {In't Zand}, {Zane}, {Zezas}, {Zhang}, \&
  {Zhuravleva}}]{Nandra_Athena:2013}
{Nandra}, K., {Barret}, D., {Barcons}, X., {et~al.} 2013, arXiv e-prints,
  arXiv:1306.2307

\bibitem[{{Nicastro} {et~al.}(2023){Nicastro}, {Krongold}, {Fang},
  {Fraternali}, {Mathur}, {Bianchi}, {De Rosa}, {Piconcelli}, {Zappacosta},
  {Bischetti}, {Feruglio}, \& {Gupta}}]{Nicastro:2023}
{Nicastro}, F., {Krongold}, Y., {Fang}, T., {et~al.} 2023, arXiv e-prints,
  arXiv:2302.04247

\bibitem[{{Parimbelli} {et~al.}(2022){Parimbelli}, {Branchini}, {Viel},
  {Villaescusa-Navarro}, \& {ZuHone}}]{Parimbelli:2022}
{Parimbelli}, G., {Branchini}, E., {Viel}, M., {Villaescusa-Navarro}, F., \&
  {ZuHone}, J. 2022, arXiv e-prints, arXiv:2209.00657

\bibitem[{{Peeples} {et~al.}(2014){Peeples}, {Werk}, {Tumlinson},
  {Oppenheimer}, {Prochaska}, {Katz}, \& {Weinberg}}]{Peeples:2014}
{Peeples}, M.~S., {Werk}, J.~K., {Tumlinson}, J., {et~al.} 2014, \apj, 786, 54

\bibitem[{{P{\'e}roux} \& {Howk}(2020)}]{Peroux:2020}
{P{\'e}roux}, C. \& {Howk}, J.~C. 2020, \araa, 58, 363

\bibitem[{{Pratt} {et~al.}(2022){Pratt}, {Arnaud}, {Maughan}, \&
  {Melin}}]{Pratt:2022}
{Pratt}, G.~W., {Arnaud}, M., {Maughan}, B.~J., \& {Melin}, J.~B. 2022, \aap,
  665, A24

\bibitem[{{Rauscher} \& {Patk{\'o}s}(2011)}]{Rauscher:2011}
{Rauscher}, T. \& {Patk{\'o}s}, A. 2011, in Handbook of Nuclear Chemistry, ed.
  A.~{V{\'e}rtes}, S.~{Nagy}, Z.~{Klencs{\'a}r}, R.~G. {Lovas}, \&
  F.~{R{\"o}sch}, 611

\bibitem[{{Renzini} \& {Andreon}(2014)}]{RA14}
{Renzini}, A. \& {Andreon}, S. 2014, \mnras, 444, 3581

\bibitem[{{Rossetti} \& {Molendi}(2010)}]{Rossetti:2010}
{Rossetti}, M. \& {Molendi}, S. 2010, \aap, 510, 83

\bibitem[{{Sanders} {et~al.}(2023){Sanders}, {Shapley}, {Jones}, {Shivaei},
  {Popping}, {Reddy}, {Dav{\'e}}, {Price}, {Mobasher}, {Kriek}, {Coil}, \&
  {Siana}}]{Sanders:2023}
{Sanders}, R.~L., {Shapley}, A.~E., {Jones}, T., {et~al.} 2023, \apj, 942, 24

\bibitem[{{Saracco} {et~al.}(2023){Saracco}, {Barbera}, {De Propris},
  {Bevacqua}, {Marchesini}, {De Lucia}, {Fontanot}, {Hirschmann}, {Nonino},
  {Pasquali}, {Spiniello}, \& {Tortora}}]{Saracco:2023}
{Saracco}, P., {Barbera}, F.~L., {De Propris}, R., {et~al.} 2023, \mnras, 520,
  3027

\bibitem[{{Sarkar} {et~al.}(2022){Sarkar}, {Su}, {Truong}, {Randall},
  {Mernier}, {Gastaldello}, {Biffi}, \& {Kraft}}]{Sarkar:2022}
{Sarkar}, A., {Su}, Y., {Truong}, N., {et~al.} 2022, \mnras, 516, 3068

\bibitem[{{Sheardown} {et~al.}(2018){Sheardown}, {Roediger}, {Su}, {Kraft},
  {Fish}, {ZuHone}, {Forman}, {Jones}, {Churazov}, \&
  {Nulsen}}]{Sheardown:2018}
{Sheardown}, A., {Roediger}, E., {Su}, Y., {et~al.} 2018, \apj, 865, 118

\bibitem[{{Shull} {et~al.}(2014){Shull}, {Danforth}, \& {Tilton}}]{Shull:2014}
{Shull}, J.~M., {Danforth}, C.~W., \& {Tilton}, E.~M. 2014, \apj, 796, 49

\bibitem[{{Shuntov} {et~al.}(2022){Shuntov}, {McCracken}, {Gavazzi}, {Laigle},
  {Weaver}, {Davidzon}, {Ilbert}, {Kauffmann}, {Faisst}, {Dubois}, {Koekemoer},
  {Moneti}, {Milvang-Jensen}, {Mobasher}, {Sanders}, \& {Toft}}]{Shuntov:2022}
{Shuntov}, M., {McCracken}, H.~J., {Gavazzi}, R., {et~al.} 2022, \aap, 664, A61

\bibitem[{{Sun} {et~al.}(2007){Sun}, {Jones}, {Forman}, {Vikhlinin}, {Donahue},
  \& {Voit}}]{Sun:2007}
{Sun}, M., {Jones}, C., {Forman}, W., {et~al.} 2007, \apj, 657, 197

\bibitem[{{Tashiro} {et~al.}(2018){Tashiro}, {Maejima}, {Toda}, {Kelley},
  {Reichenthal}, {Lobell}, {Petre}, {Guainazzi}, {Costantini}, {Edison},
  {Fujimoto}, {Grim}, {Hayashida}, {den Herder}, {Ishisaki}, {Paltani},
  {Matsushita}, {Mori}, {Sneiderman}, {Takei}, {Terada}, {Tomida}, {Akamatsu},
  {Angelini}, {Arai}, {Awaki}, {Babyk}, {Bamba}, {Barfknecht}, {Barnstable},
  {Bialas}, {Blagojevic}, {Bonafede}, {Brambora}, {Brenneman}, {Brown},
  {Brown}, {Burns}, {Canavan}, {Carnahan}, {Chiao}, {Comber}, {Corrales}, {de
  Vries}, {Dercksen}, {Diaz-Trigo}, {Dillard}, {DiPirro}, {Done}, {Dotani},
  {Ebisawa}, {Eckart}, {Enoto}, {Ezoe}, {Ferrigno}, {Fukazawa}, {Fujita},
  {Furuzawa}, {Gallo}, {Graham}, {Gu}, {Hagino}, {Hamaguchi}, {Hatsukade},
  {Hawes}, {Hayashi}, {Hegarty}, {Hell}, {Hiraga}, {Hodges-Kluck}, {Holland},
  {Hornschemeier}, {Hoshino}, {Ichinohe}, {Iizuka}, {Ishibashi}, {Ishida},
  {Ishikawa}, {Ishimura}, {James}, {Kallman}, {Kara}, {Katsuda}, {Kenyon},
  {Kilbourne}, {Kimball}, {Kitaguti}, {Kitamoto}, {Kobayashi}, {Kohmura},
  {Koyama}, {Kubota}, {Leutenegger}, {Lockard}, {Loewenstein}, {Maeda},
  {Marbley}, {Markevitch}, {Matsumoto}, {Matsuzaki}, {McCammon}, {McNamara},
  {Miko}, {Miller}, {Miller}, {Minesugi}, {Mitsuishi}, {Mizuno}, {Mori},
  {Mukai}, {Murakami}, {Mushotzky}, {Nakajima}, {Nakamura}, {Nakashima},
  {Nakazawa}, {Natsukari}, {Nigo}, {Nishioka}, {Nobukawa}, {Nobukawa}, {Noda},
  {Odaka}, {Ogawa}, {Ohashi}, {Ohno}, {Ohta}, {Okajima}, {Okamoto}, {Onizuka},
  {Ota}, {Ozaki}, {Plucinsky}, {Porter}, {Pottschmidt}, {Sato}, {Sato},
  {Sawada}, {Seta}, {Shelton}, {Shibano}, {Shida}, {Shidatsu}, {Shirron},
  {Simionescu}, {Smith}, {Someya}, {Soong}, {Suagawara}, {Szymkowiak},
  {Takahashi}, {Tamagawa}, {Tamura}, {Tanaka}, {Terashima}, {Tsuboi},
  {Tsujimoto}, {Tsunemi}, {Tsuru}, {Uchida}, {Uchiyama}, {Ueda}, {Uno},
  {Walsh}, {Watanabe}, {Williams}, {Wolfs}, {Wright}, {Yamada}, {Yamaguchi},
  {Yamaoka}, {Yamasaki}, {Yamauchi}, {Yamauchi}, {Yanagase}, {Yaqoob},
  {Yasuda}, {Yoshioka}, {Zabala}, \& {Irina}}]{Tashiro_XRISM:2018}
{Tashiro}, M., {Maejima}, H., {Toda}, K., {et~al.} 2018, in Society of
  Photo-Optical Instrumentation Engineers (SPIE) Conference Series, Vol. 10699,
  \procspie, 1069922

\bibitem[{{Tinker} {et~al.}(2008){Tinker}, {Kravtsov}, {Klypin}, {Abazajian},
  {Warren}, {Yepes}, {Gottl{\"o}ber}, \& {Holz}}]{Tinker:2008}
{Tinker}, J., {Kravtsov}, A.~V., {Klypin}, A., {et~al.} 2008, \apj, 688, 709

\bibitem[{{Tumlinson} {et~al.}(2017){Tumlinson}, {Peeples}, \&
  {Werk}}]{Tumlinson:2017}
{Tumlinson}, J., {Peeples}, M.~S., \& {Werk}, J.~K. 2017, \araa, 55, 389

\bibitem[{{van de Voort} {et~al.}(2011){van de Voort}, {Schaye}, {Booth},
  {Haas}, \& {Dalla Vecchia}}]{vandeVoort:2011}
{van de Voort}, F., {Schaye}, J., {Booth}, C.~M., {Haas}, M.~R., \& {Dalla
  Vecchia}, C. 2011, \mnras, 414, 2458

\bibitem[{{van der Burg} {et~al.}(2015){van der Burg}, {Hoekstra}, {Muzzin},
  {Sif{\'o}n}, {Balogh}, \& {McGee}}]{vdB15}
{van der Burg}, R.~F.~J., {Hoekstra}, H., {Muzzin}, A., {et~al.} 2015, \aap,
  577, A19

\bibitem[{{van Uitert} {et~al.}(2016){van Uitert}, {Cacciato}, {Hoekstra},
  {Brouwer}, {Sif{\'o}n}, {Viola}, {Baldry}, {Bland-Hawthorn}, {Brough},
  {Brown}, {Choi}, {Driver}, {Erben}, {Heymans}, {Hildebrandt}, {Joachimi},
  {Kuijken}, {Liske}, {Loveday}, {McFarland}, {Miller}, {Nakajima}, {Peacock},
  {Radovich}, {Robotham}, {Schneider}, {Sikkema}, {Taylor}, \& {Verdoes
  Kleijn}}]{vanUitert:2016}
{van Uitert}, E., {Cacciato}, M., {Hoekstra}, H., {et~al.} 2016, \mnras, 459,
  3251

\bibitem[{{Weaver} {et~al.}(2022){Weaver}, {Davidzon}, {Toft}, {Ilbert},
  {McCracken}, {Gould}, {Jespersen}, {Steinhardt}, {Lagos}, {Capak}, {Casey},
  {Chartab}, {Faisst}, {Hayward}, {Kartaltepe}, {Kauffmann}, {Koekemoer},
  {Kokorev}, {Laigle}, {Liu}, {Long}, {Magdis}, {McPartland}, {Milvang-Jensen},
  {Mobasher}, {Moneti}, {Peng}, {Sanders}, {Shuntov}, {Sneppen}, {Valentino},
  {Zalesky}, \& {Zamorani}}]{Weaver:2022}
{Weaver}, J.~R., {Davidzon}, I., {Toft}, S., {et~al.} 2022, arXiv e-prints,
  arXiv:2212.02512

\bibitem[{{Werner} {et~al.}(2013){Werner}, {Urban}, {Simionescu}, \&
  {Allen}}]{Werner:2013}
{Werner}, N., {Urban}, O., {Simionescu}, A., \& {Allen}, S.~W. 2013, \nat, 502,
  656

\bibitem[{{Wright} {et~al.}(2020){Wright}, {Lagos}, {Power}, \&
  {Mitchell}}]{Wright:2020}
{Wright}, R.~J., {Lagos}, C. d.~P., {Power}, C., \& {Mitchell}, P.~D. 2020,
  \mnras, 498, 1668

\bibitem[{{Zahid} {et~al.}(2017){Zahid}, {Kudritzki}, {Conroy}, {Andrews}, \&
  {Ho}}]{Zahid:2017}
{Zahid}, H.~J., {Kudritzki}, R.-P., {Conroy}, C., {Andrews}, B., \& {Ho}, I.~T.
  2017, \apj, 847, 18

\bibitem[{{Zhang} {et~al.}(2019){Zhang}, {Yanny}, {Palmese}, {Gruen}, {To},
  {Rykoff}, {Leung}, {Collins}, {Hilton}, {Abbott}, {Annis}, {Avila}, {Bertin},
  {Brooks}, {Burke}, {Carnero Rosell}, {Carrasco Kind}, {Carretero}, {Cunha},
  {D'Andrea}, {da Costa}, {De Vicente}, {Desai}, {Diehl}, {Dietrich}, {Doel},
  {Drlica-Wagner}, {Eifler}, {Evrard}, {Flaugher}, {Fosalba}, {Frieman},
  {Garc{\'\i}a-Bellido}, {Gaztanaga}, {Gerdes}, {Gruendl}, {Gschwend},
  {Gutierrez}, {Hartley}, {Hollowood}, {Honscheid}, {Hoyle}, {James},
  {Jeltema}, {Kuehn}, {Kuropatkin}, {Li}, {Lima}, {Maia}, {March}, {Marshall},
  {Melchior}, {Menanteau}, {Miller}, {Miquel}, {Mohr}, {Ogand o}, {Plazas},
  {Romer}, {Sanchez}, {Scarpine}, {Schubnell}, {Serrano}, {Sevilla-Noarbe},
  {Smith}, {Soares-Santos}, {Sobreira}, {Suchyta}, {Swanson}, {Tarle},
  {Thomas}, {Wester}, \& {DES Collaboration}}]{Zhang_ICL_2019}
{Zhang}, Y., {Yanny}, B., {Palmese}, A., {et~al.} 2019, \apj, 874, 165

\bibitem[{{Zhuravleva} {et~al.}(2019){Zhuravleva}, {Churazov}, {Schekochihin},
  {Allen}, {Vikhlinin}, \& {Werner}}]{Zhuravleva:2019}
{Zhuravleva}, I., {Churazov}, E., {Schekochihin}, A.~A., {et~al.} 2019, Nature
  Astronomy, 3, 832

\bibitem[{{Zu} \& {Mandelbaum}(2015)}]{Zu:2015}
{Zu}, Y. \& {Mandelbaum}, R. 2015, \mnras, 454, 1161

\end{thebibliography}
\end{document}